 \pdfoutput=1
\newcommand*{\ATLASLATEXPATH}{latex/}
\documentclass[cernpreprint,texlive=2016,txfonts,UKenglish]{\ATLASLATEXPATH atlasdoc}

\usepackage{\ATLASLATEXPATH atlaspackage}
\usepackage{\ATLASLATEXPATH atlasbiblatex}

\usepackage{\ATLASLATEXPATH atlascontribute}

\usepackage{\ATLASLATEXPATH atlasphysics}

\graphicspath{{logos/}{figures/}}

\AtlasTitle{Measurement of the prompt $J$/\textbf{$\psi$} pair production cross-section in $pp$ collisions at \textbf{$\sqrt{s}$}~=~8~TeV with the ATLAS detector}

\author{The ATLAS Collaboration}

\AtlasRefCode{BPHY-2015-02}

\PreprintIdNumber{CERN-EP-2016-211}

 \AtlasJournal{EPJC}
 \AtlasJournalRef{Eur. Phys. J. C (2017) 77: 76}
 \AtlasDOI{10.1140/epjc/s10052-017-4644-9}

\AtlasAbstract{The production of two prompt $J/\psi$ mesons, each with transverse momenta $p_{\mathrm{T}}>8.5$~GeV and rapidity $|y| < 2.1$, is studied using a sample of proton-proton collisions at $\sqrt{s} = 8$~TeV, corresponding to an integrated luminosity of 11.4 fb$^{-1}$ collected in 2012 with the ATLAS detector at the LHC. The differential cross-section, assuming unpolarised $J/\psi$ production, is measured as a function of the transverse momentum of the lower-$p_{\mathrm{T}}$ $J/\psi$ meson, di-$J/\psi$ $p_{\mathrm{T}}$ and mass, the difference in rapidity between the two $J/\psi$ mesons, and the azimuthal angle between the two $J/\psi$ mesons.  The fraction of prompt pair events due to double parton scattering is determined by studying kinematic correlations between the two $J/\psi$ mesons.  The total and double parton scattering cross-sections are compared with predictions.  The effective cross-section of double parton scattering is measured to be $\sigma_{\mathrm{eff}} = 6.3 \pm 1.6 \mathrm{(stat)} \pm 1.0 \mathrm{(syst)}$~mb.
}

\hypersetup{pdftitle={ATLAS document},pdfauthor={The ATLAS Collaboration}}

\begin{document}

\maketitle

\tableofcontents

\newcommand{\beq}{\begin{equation}}
\newcommand{\eeq}{\end{equation}}

\clearpage

\section{Introduction}
\label{sec:intro}
The study of the simultaneous production of two prompt $J/\psi$ mesons offers an opportunity to test our understanding of non-perturbative Quantum Chromodynamics (QCD).  These events are also sensitive to next-to-leading-order (NLO) and higher-order perturbative QCD corrections, in addition to providing an opportunity to study and compare $J/\psi$ production models. Di-$J/\psi$ events can be produced from a single gluon--gluon collision via single parton scattering (SPS) or from two independent parton--parton scatters in a single proton--proton collision, known as double parton scattering (DPS). 

In particular, the production of di-$J/\psi$ events via double parton scattering presents a unique insight into the structure of the proton and allows a better comprehension of backgrounds to searches for new phenomena.  Although the di-$J/\psi$ process has a low production rate in hadron collisions, the high luminosity and energy of the LHC allows a more detailed study than previously possible~\cite{JpsiHadron, HumpertJpsi, NA3Jpsi,NA3DiJpsi, InclDiJ,ThDiJandtetra, LHCbDiJ, DiJImpact, PromptDiJLHC,SPSth,SPSth2,FeedDownDPS}.  State-of-the-art techniques have been developed to describe di-$J/\psi$ production in leading-order (LO), NLO, next-to-leading-order colour singlet non-relativistic QCD computations without loops (NLO*), and intrinsic parton transverse momentum calculations.  Contributions of gluon fragmentation and quark fragmentation, which occur at even higher order calculations have been shown to make a large difference in the predictions~\cite{DiJImpact}.

\begin{figure}[h!tb]
\begin{center}
\begin{tabular}{cc}
\subfloat[]{
\includegraphics*[width=0.36\textwidth]{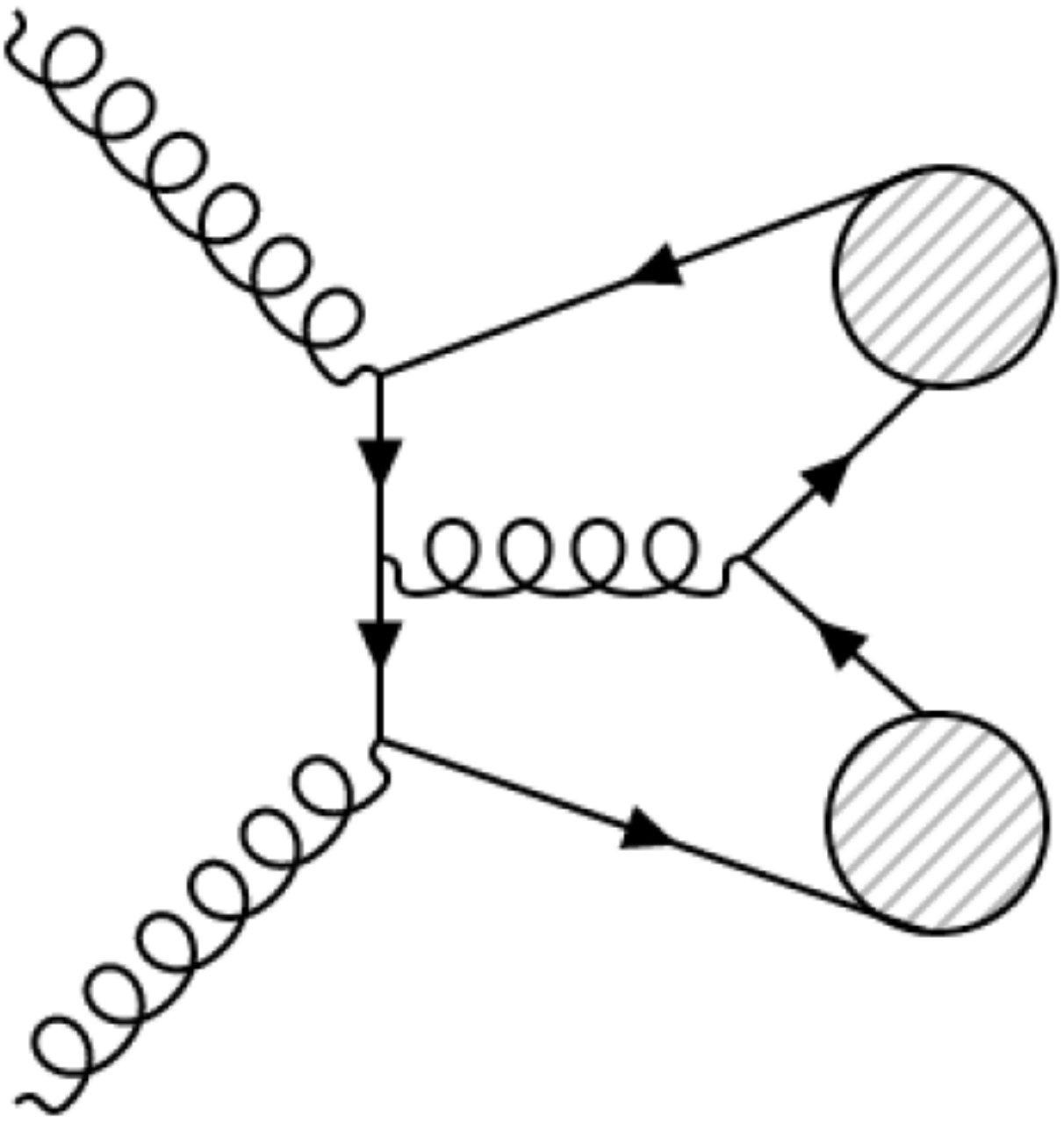}
\label{fig:LOCS}
}&
\subfloat[]{
\includegraphics*[width=0.34\textwidth]{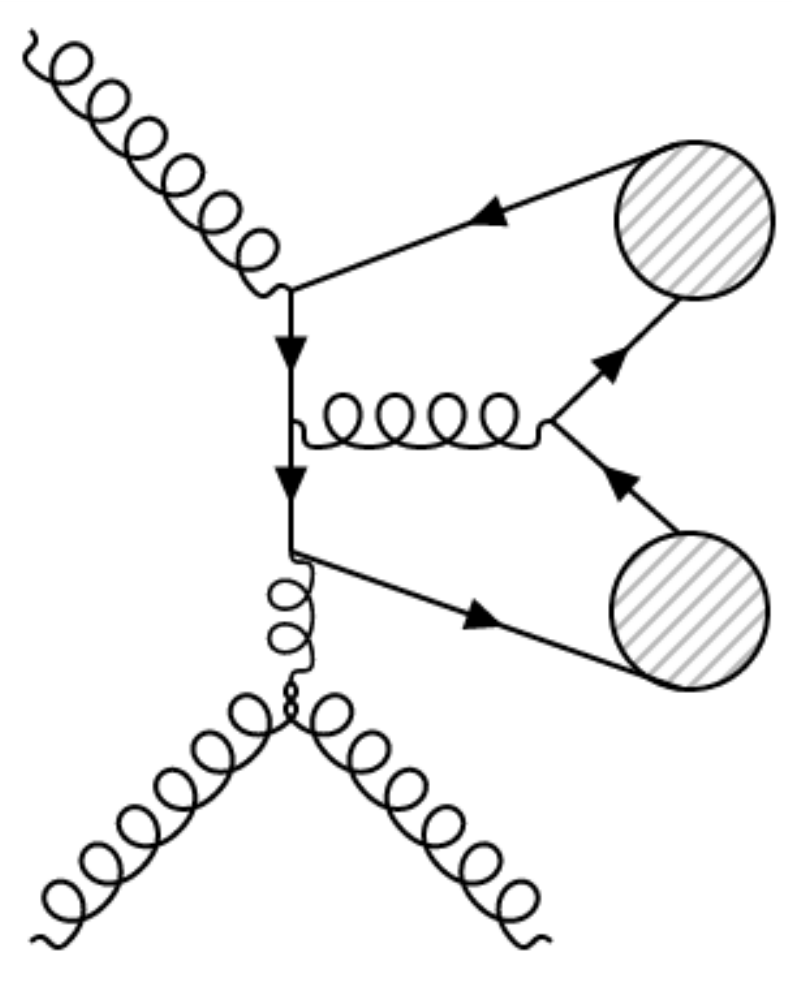}
\label{fig:NLOCS}
}\\
\subfloat[]{
\includegraphics*[width=0.42\textwidth]{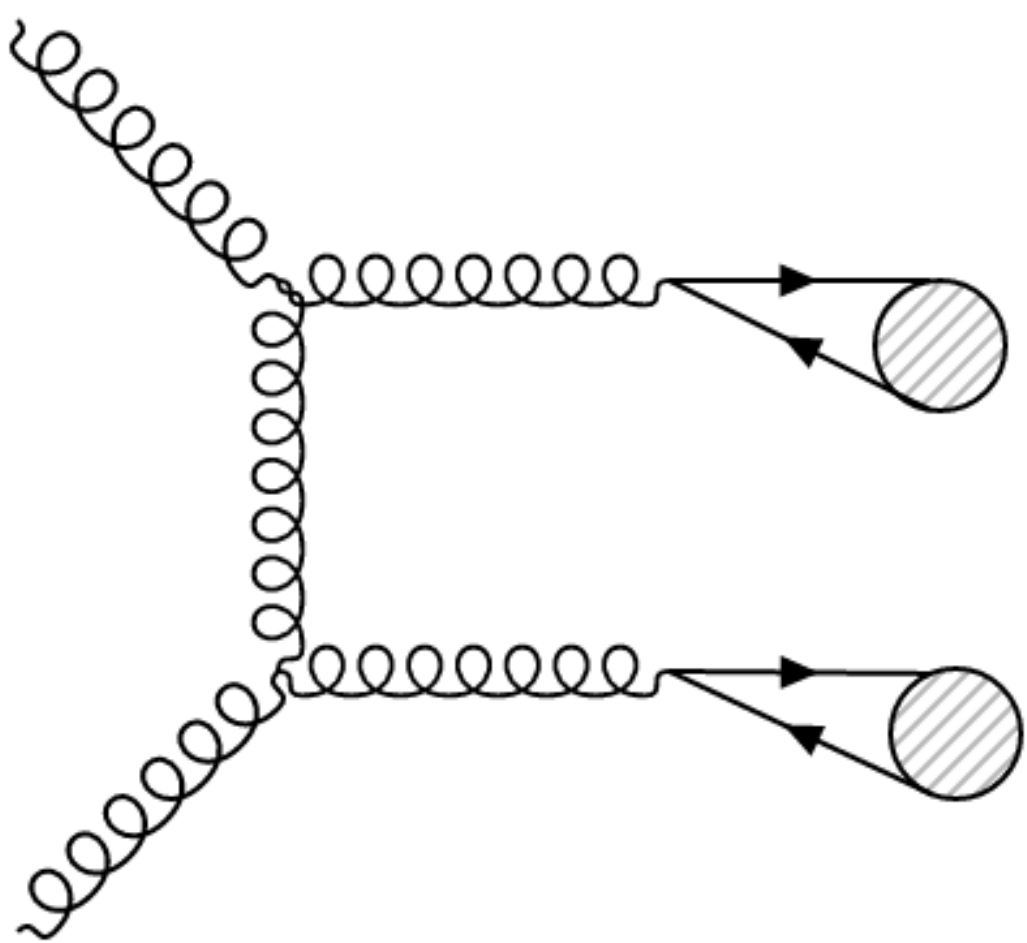}
\label{fig:CO2}
}&
\subfloat[]{
\includegraphics*[width=0.22\textwidth]{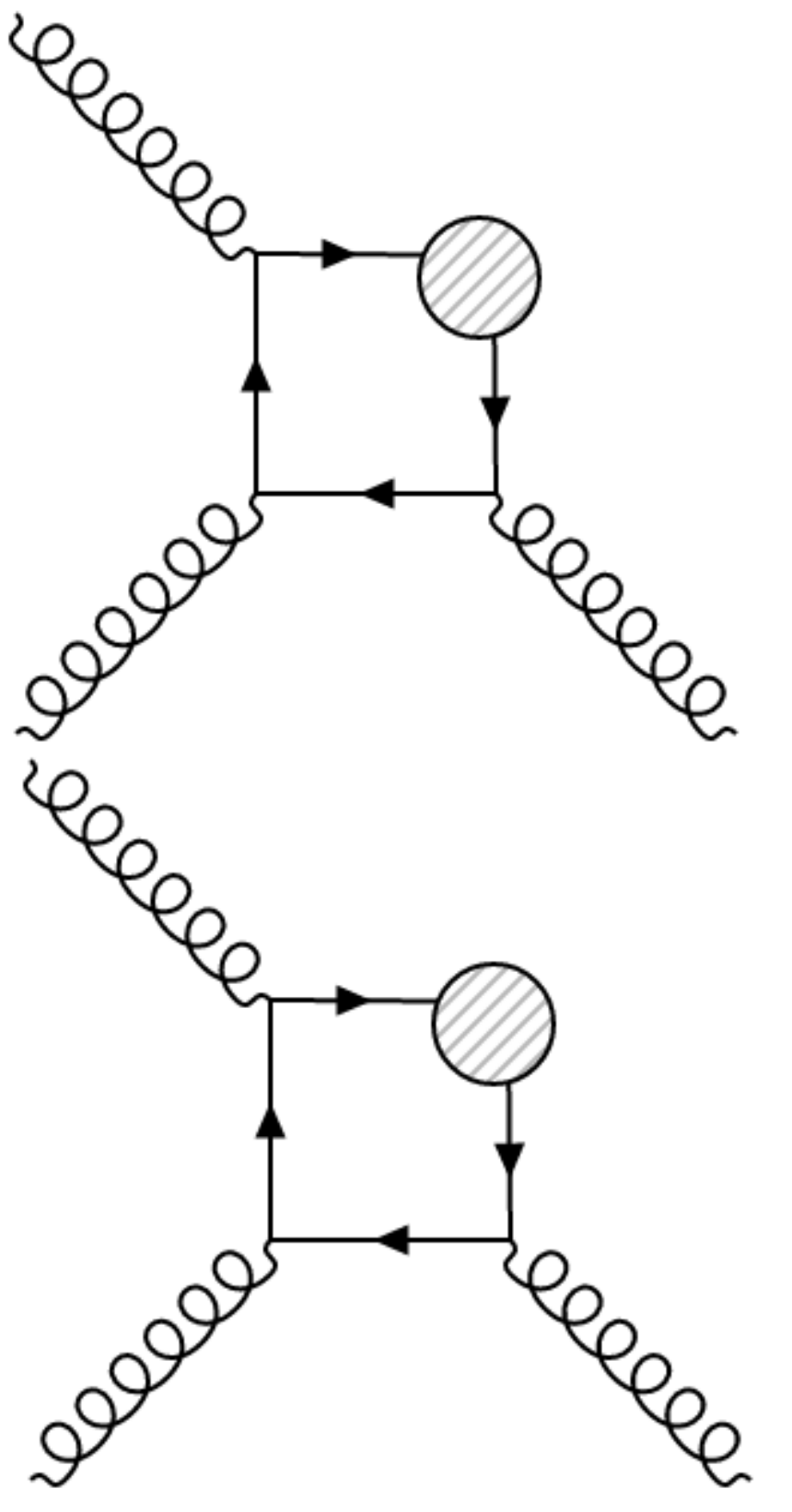}
\label{fig:DPS}
}
\end{tabular}
\caption{Examples of Feynman diagrams of prompt--prompt $J/\psi$ pair production in $pp$ collisions for~\protect\subref{fig:LOCS} leading-order production and~\protect\subref{fig:NLOCS} next-to-leading-order production in which the circle represents the $J/\psi$ meson produced in a colour-singlet state,~\protect\subref{fig:CO2} leading-order production where the circles represent a $c\bar{c}$ pair in a colour-octet state, and~\protect\subref{fig:DPS} double parton scattering.}
\label{fig:fdiag}
\end{center}
\end{figure}

Prompt $J/\psi$ mesons can be produced directly~\cite{ATLASJpsi, ATLASJpsi2, LHCbJpsi, LHCbJpsi2, CMSJpsi, CMSJpsi2, ALICEJpsi,CS2} or via a higher-mass charmonium state, such as $\chi_{c}  \rightarrow J/\psi + X$ or $\psi(\mathrm{2S}) \rightarrow J/\psi + X$.  These feed-down events resemble direct gluon--gluon fusion $J/\psi$ production.  Non-prompt events can be identified by their displaced decay vertex from the decay of $b$-hadrons.  The focus of this paper is on prompt--prompt (PP) di-$J/\psi$ production with the decay $J/\psi$~$\rightarrow$~$\mu^{+}$~$\mu^{-}$.  This decay channel has the advantage of a clean four-muon signal.  Examples of prompt--prompt di-$J/\psi$ production diagrams are shown in Figure~\ref{fig:fdiag}.

It is expected that DPS plays a larger role at high energies and could be increasingly important for $c\bar{c}c\bar{c}$ production~\cite{DPSenergy, LHCbJD}.  DPS can help to explain the cross-section of multi-jet production and the large difference in rapidity\footnote{ATLAS uses a right-handed coordinate system where the nominal interaction point (IP) is defined as the origin, the $z$-axis defines the beam direction, and the $x$--$y$ plane is transverse to the beam direction.  The positive $x$-axis is defined to point from the IP to the centre of the LHC ring, and the positive $y$-axis points upward.  The azimuthal angle, $\phi$, is measured around the beam axis and the polar angle, $\theta$, is measured from the beam axis.  The pseudorapidity is defined as $\eta = -\mathrm{ln[tan}(\theta/2)]$ and for massive particles the rapidity $y = 1/2 \cdot \mathrm{ln}[(E+p_{\mathrm{z}})/(E-p_{\mathrm{z}})]$ is used.} for hard interaction processes~\cite{MPILHC}.  As di-$J/\psi$ production is dominated by gluon interactions, it gives information complementary to recent effective cross-section measurements of quark-dominated interactions.  The effective cross-section is a factor which relates the production cross-section of the two individual interactions to the total interaction.  Due to the low production rate of prompt $J/\psi$ meson pairs, this process has not been studied in as much detail as other DPS processes and can therefore provide a good test of the process dependence of the effective cross-section.  Testing possible correlations of non-perturbative origin between the partons in DPS events could lead to a better comprehension of non-perturbative QCD~\cite{MPISum}.  

DPS has been studied in multiple final-states such as $W(\rightarrow \ell\nu)+\,$2 jets~\cite{WDPS,CMSW2J}, $Z(\rightarrow \ell^{+}\ell^{-})+\,$$J/\psi$~\cite{ATLASZJpsi}, quarkonium plus open charm~\cite{LHCbJD, LHCbYD}, double quarkonium~\cite{D0DiJ,D0JY}, dijet~\cite{D0Dijet}, $\gamma+\,$3 jets~\cite{CDFxF,gamma,CDF2013,CDF4j}, and 4 jets~\cite{AFS,UA2,CDF4j,ATLAS4jEP}.  The production of di-$J/\psi$ events via DPS and SPS is described in Refs.~\cite{DPSth,Jpsi2,JpsiDPSNuc,SPSth2,SPSth,FeedDownDPS,FeedDownDPS2,DPSth2}.

In this paper, the differential cross-sections for PP $J/\psi$ pairs are measured as functions of the transverse momentum of the lower-$p_{\mathrm{T}}$ $J/\psi$ meson, the di-$J/\psi$ transverse momentum, and the di-$J/\psi$ mass.  The lower-$p_{\mathrm{T}}$ (sub-leading) $J/\psi$ meson is denoted as $J/\psi_{2}$ hereafter.  Measurements are presented in two regions of the sub-leading $J/\psi$ meson rapidity both within the muon $p_{\mathrm{T}}$ acceptance and extrapolated to the full acceptance by integrating the muon transverse momentum to zero.  Additionally, the differential cross-section over the full $J/\psi$ rapidity region defined by the muon selection criteria is measured for these variables along with the difference in rapidity and the azimuthal angle between the two $J/\psi$ mesons.  Using the sub-leading $J/\psi$ meson allows the full range of the muon kinematic region to be explored.  As the mass resolution for $J/\psi$ mesons is worse at forward rapidities, the cross-section is measured in two rapidity regions, $|y(J/\psi_{2})|<1.05$ and $1.05\leq|y(J/\psi_{2})|<2.1$, to increase the sensitivity of the results.

A data-driven method is used to produce background-subtracted SPS-weighted and DPS-weighted distributions of several kinematic variables. The measured distributions are compared with both the leading-order (LO) DPS and NLO* SPS predictions in the same fiducial volume. Finally, using the PP di-$J/\psi$ cross-section, the fraction of DPS events, and the prompt $J/\psi$ cross-section~\cite{Single}, the effective cross-section of DPS is measured and compared to previous measurements.

In data collected at $\sqrt{s}$~=~7~TeV during 2011, the CMS experiment measured the cross-section of the pair production of prompt $J/\psi$ mesons extrapolated to a muon $p_{\mathrm{T}}$ of zero assuming unpolarised $J/\psi$ mesons~\cite{CMScross}.  The D0 experiment measured the fiducial prompt $J/\psi$ pair cross-section using data collected at $\sqrt{s}$~=~1.96~TeV~\cite{D0DiJ}.  The LHCb experiment measured the pair production of prompt $J/\psi$ mesons in the forward rapidity region using data collected at $\sqrt{s}$~=~7~TeV~\cite{LHCbDiJ}.  The present measurement of PP di-$J/\psi$ production uses a different kinematic range and a larger data set.

The rest of this document is organised as follows. In Section~\ref{sec:detector}, a brief description of the ATLAS detector and the data samples used in this study is provided.  In Section~\ref{sec:Event}, the methods used in the event reconstruction as well as the selection criteria used in this analysis are reported.  Section~\ref{sec:Sig} focuses on the removal of non-$J/\psi$, non-prompt $J/\psi$, and pile-up backgrounds as well as the calculation of the detector and selection efficiencies used in the extrapolated signal yields.  In Section~\ref{sec:DPSsec}, double parton scattering and the data-driven method to extract the fraction of DPS events from various kinematic variables are discussed.  Section~\ref{sec:Syst} reports the systematic uncertainties.  In Section~\ref{sec:result}, the results of the cross-section measurements and DPS study are presented.  Finally, the findings are summarised in Section~\ref{sec:conclusion}.

\section{ATLAS detector}
\label{sec:detector}

A full and detailed description of the ATLAS detector can be found in Ref.~\cite{ATLAS}.  The inner detector (ID) is composed of the pixel detector, the semiconductor tracker (SCT), and the transition radiation tracker (TRT).  The pixel and SCT detectors cover the range $|\eta| <  2.5$.  The barrel is constructed from concentric cylinders around the beam axis and in the end-caps the disks are oriented perpendicular to the beam axis.  The TRT is made up of straws filled with gas.  It covers a range of $|\eta| \le 2.0$.  The TRT surrounds the SCT and provides $r$--$\phi$ information as well as electron identification information from transition radiation photons.  The ID is surrounded by a solenoid which provides a 2~T axial magnetic field.  The calorimeter has separate electromagnetic and hadronic components.  The muon spectrometer (MS) consists of monitored drift tubes for $|\eta| \le 2.7$ in combination with cathode strip chambers for $2.0<|\eta|<2.7$.  Additionally, there are two types of triggering chambers, the resistive plate chambers (RPC) and the thin gap chambers (TGC).  The MS is designed to provide precise position and momentum measurements in the bending plane and is capable of stand-alone muon reconstruction.  The ATLAS trigger system has three levels (Level-1, Level-2, Event Filter). The Level-1 muon trigger uses information from three layers of RPCs in the barrel region ($|\eta|$~$<$~1.05) and three layers of TGCs in the end-cap regions (1.05~$<$~$|\eta|$~$<$~2.4). The geometric coverage of the Level-1 trigger for single muons is about 99$\%$ in the end-cap regions and about 80$\%$ in the barrel region~\cite{ATLASTrig}. Information from the ID and MS is included in the Level-2 and Event Filter triggers.

\section{Event reconstruction and selection}
\label{sec:Event}
The data set used was collected during 2012 at $\sqrt{s} = 8$ TeV for proton--proton collisions. The total integrated luminosity of the data set used is 14.1~$\pm$~0.3~fb$^{-1}$, and 11.4~$\pm$ 0.3~fb$^{-1}$~\cite{Lum2} after accounting for the prescale factor of the $J/\psi$ dimuon trigger.

The selected events satisfy a $J/\psi$ dimuon trigger requiring two muons with $p_{\mathrm{T}}$~$>$~4~GeV and invariant mass in the range $2.5 < m(\mu\mu) < 4.3$~GeV.  Two $J/\psi$ candidates reconstructed through their decay to a pair of oppositely charged muons are required.  The reconstruction of the muon tracks is described in Ref.~\cite{Reco}.  The offline selection requires that events have at least three muons identified by the MS with matching tracks in the ID.  Due to the ID acceptance, the reconstruction of muons is limited to $|\eta^{\mu}| <$~2.5 and must satisfy the selection criteria described in Ref.~\cite{Reco}.

The two $J/\psi$ candidates in each event are ordered by transverse momentum.  In the event reconstruction, it is permitted that the two $J/\psi$ candidates are associated with two different proton--proton collision vertices.  This is necessary to model the pile-up background.  The signed transverse decay length, $L_{\mathrm{xy}}$, of each $J/\psi$ candidate is defined as the projection of the vector from the closest reconstructed hard-scatter vertex candidate along the beam direction to the $J/\psi$ decay vertex onto the $J/\psi$ transverse momentum vector.  Events with the two $J/\psi$ candidates originating from different vertices are removed later by imposing a limit on the distance along the beam axis between the two vertices and subtracting the pile-up events that make it through this requirement, as described in Section~\ref{sec:pile}.  After subtracting the remaining multiple vertex events, the primary vertex is the common vertex which is closest to each $J/\psi$ candidate along the beam direction.  The following kinematic and geometrical requirements on the muons and $J/\psi$ mesons are applied:  

   \begin{itemize}
       \item $|\eta^{\mu}| <$~2.3 and $p_{\mathrm{T}}^{\mu} >$~2.5~GeV.
	   \item $2.8$~$\leq$~$m(\mu\mu)$~$\leq$~3.4~GeV.
       \item $|y^{J/\psi}| <$~2.1 and $p_{\mathrm{T}}^{J/\psi} >$~8.5~GeV.       
       \item For the triggered $J/\psi$, both of the reconstructed muons must have an ID track matched to a MS track.
       \item For the non-triggered $J/\psi$ candidate, at least one of the reconstructed muons must have an ID track matched to a MS track.
       \item The distance between the two $J/\psi$ decay vertices along the beam direction is required to be $|d_{\mathrm{z}}| <$~1.2~mm.  This requirement aims to select two $J/\psi$ mesons that originate from the same proton--proton collision.
       \item The uncertainty in the measurement of $L_{\mathrm{xy}}$ is required to be less than 0.3~mm.
        
    \end{itemize}

Although the requirement on $|d_{\mathrm{z}}|$ affects events with large decay length, there is negligible bias in the measurement for PP signal events due to the narrow $d_{\mathrm{z}}$ distribution of prompt $J/\psi$ pair production.  This is discussed further in Section~\ref{sec:Syst}.

A total of 1210 events satisfy the above selection criteria.

\section{Signal extraction}
\label{sec:Sig}
The PP differential cross-sections are measured in two rapidity regions based on the sub-leading $J/\psi$ rapidity: the central region, $|y$($J/\psi_{2})|<$1.05, and the forward region, 1.05$\leq|y$($J/\psi_{2})|<$2.1,

\beq
\frac{\Delta \sigma_{i}(pp\rightarrow J/\psi J/\psi  + X)}{ \Delta x } = \frac{N_{\mathrm{sig}}^{i}}{ A_{i} \times \epsilon_{i} \times \mathrm{BF} (J/\psi \rightarrow \mu^{+}\mu^{-})^{2} \times \Delta x  \times \cal{L}}.
\label{eq:sigma}
\eeq

In this equation the differential cross-section in bin $i$, of size $\Delta x$, of the kinematic variable $x$ is a function of the number of PP di-$J/\psi$ signal events in the interval, $N^{i}_{\mathrm{sig}}$; the kinematic acceptance correction, $A_{i}$, which is defined as the probability of a di-$J/\psi$ event in the bin to pass the kinematic requirements; the efficiency, $\epsilon_{i}$, of the trigger, reconstruction, and selection criteria; the branching fraction of a $J/\psi$ meson to two muons, $BF(J/\psi \rightarrow \mu^{+} \mu^{-})$; and the total integrated luminosity of the data set, $\cal{L}$.

The main sources of background to PP di-$J/\psi$ production are non-$J/\psi$ events, non-prompt $J/\psi$ events, and events containing $J/\psi$ mesons originating from two separate proton--proton collisions (called pile-up background).  This analysis uses a sequential extraction of the di-$J/\psi$ PP signal.  First, each event is weighted by the inverse of the trigger, reconstruction and selection efficiencies and the kinematic acceptance.  Next the two-dimensional distribution of the mass of the leading $J/\psi$ candidate against the sub-leading $J/\psi$ candidate is fit using a two-dimensional probability density function (PDF) in a maximum-likelihood fit~\cite{Roofit} to subtract non-$J/\psi$ background and extract the di-$J/\psi$ signal.  The extracted di-$J/\psi$ signal is used to create PP event weights (the probability that the event is prompt--prompt) from the two-dimensional fit of the transverse decay length distribution of the leading $J/\psi$ meson against the sub-leading $J/\psi$ distribution.  The extracted di-$J/\psi$ signal is taken in bins $\Delta x$ of the chosen variable $x$, weighted by the PP event weight and finally the pile-up background is subtracted.

Results are reported for the fiducial cross-section within the acceptance of the muon requirements as well as that corrected for muons produced outside the muon transverse momentum acceptance, described in detail in Ref.~\cite{Single}.  The world-average branching fraction of a $J/\psi$ meson to two muons, 5.96~$\pm$~0.03~$\%$, is used~\cite{PDG}.

\subsection{Efficiency and acceptance}
\label{sec:Eff}
The PP di-$J/\psi$ signal is corrected for the reconstruction, trigger, and event selection efficiencies, $\epsilon_{i}$ in Eq.~(\ref{eq:sigma}).  To obtain the dimuon trigger efficiency, the first step is to calculate the single-muon-trigger efficiency of each muon, multiply the two efficiencies, and then apply a correction term that accounts for correlations between the vertex resolution and opposite-sign requirements, as well as correcting for configurations in which the muons are too close to each other to be resolved by the Level-1 single-muon trigger.  A further correction is applied to account for a bias due to the use of high-$p_{\mathrm{T}}$ single-muon triggers for the efficiency determination.  The correction is determined from the binned ratio of data to Monte Carlo (MC) simulation of an inclusive $J/\psi$ sample generated using \PYTHIA 8.186~\cite{Pythia} with the AU2 set of tuned parameters~\cite{AU2} and CTEQ6L1 parton distribution functions~\cite{CTEQ}. The MC samples are passed through ATLAS detector simulation~\cite{ATLASsim} based on GEANT4~\cite{GEANT4}, and are reconstructed with the same software as the data.  Using the single-muon-trigger efficiencies, the correction term for correlations and the MC correction, the total efficiency for the $J/\psi$ dimuon trigger, $\epsilon$, is then calculated using a modified form of the ``tag and probe'' method presented in Ref.~\cite{trig}.

A correction to the muon reconstruction efficiency is applied using the efficiency scale factors described in Ref.~\cite{Reco}.  Efficiency scale factors have been determined in bins of $q \times \eta$ and $\phi$ separately for muons with and without an ID track that matches an MS track, where $q$ is the charge of the muon.  The scale factors for muons from the triggered $J/\psi$ are taken from the correction that includes ID track matching.  Since the other $J/\psi$ meson only requires one of the muons to have an independent ID track matched to a MS track, a combination of the two efficiency corrections is used.

The kinematic acceptance factor, $A_{i}$ in Eq.~(\ref{eq:sigma}), is determined from simulation which describes the effect of the muon $p_{\mathrm{T}}$ and $\eta$ cuts in the fiducial region definition, and corrects the cross-section for a $J/\psi$ observed in the $J/\psi$ $p_{\mathrm{T}}$ and rapidity fiducial region to the full muon geometric and kinematic acceptance.  The method is described in detail in Ref.~\cite{Single}, and is applied to the fiducial volume of this analysis.  For this correction the $J/\psi$ mesons are assumed to be unpolarised, as the $J/\psi$ polarisation coefficients were found close to zero~\cite{PolarCDF, PolarLHCb, PolarCMS}.  The additional maximum variation of the polarisation assumption is shown in the differential cross-section distributions. 

The final component of the event-weight corrections is the signal efficiency of the selection criteria.  The $d_{\mathrm{z}}$ selection efficiency is 99.9$\%$ and 96.9$\%$ in the central and forward rapidity regions, respectively.  The efficiency of the requirement on the $L_{\mathrm{xy}}$ uncertainty is 91.1$\%$ in the central region and 94.5$\%$ in the forward rapidity region.  The correction is the inverse of the efficiency and is applied to each event.

\subsection{Non-\texorpdfstring{$J/\psi$}~~Background}
The non-$J/\psi$ background comes mostly from semileptonic decays of $b$-hadrons and from dimuon continuum events from Drell--Yan processes.   An unbinned two-dimensional (2-D) maximum-likelihood fit~\cite{unbinML} of the dimuon invariant mass of the leading $J/\psi$ ($J/\psi_{1}$) against the dimuon invariant mass of the sub-leading $J/\psi$ ($J/\psi_{2}$) is used to extract the signal.  To parameterise the mass distribution of $J/\psi$ signal events, a large inclusive $J/\psi$ sample selected from the 2012 $\sqrt{s}$~=~8~TeV ATLAS data is used.  It has the same selection criteria, fiducial volume, and trigger as the di-$J/\psi$ sample with the exception of the cut on $d_{\mathrm{z}}$ which is not applied to the prompt signal.

In the fit of the inclusive $J/\psi$ mass distribution, the signal is modelled by a modified double Crystal Ball function ($CB$)~\cite{CB, CB1, CB2} and the background is modelled by a first-order polynomial.  The modified double Crystal Ball function has a Gaussian core and power-law low-end and high-end tails that are fixed to have the same rate of decrease, described by the parameter $n$ in the references.  The parameter which controls the transition from the core Gaussian to the power-law tails is allowed to be different for each tail.

For the di-$J/\psi$ sample the PDF includes terms for the signal which is parameterised as a product of two normalised $CB$ functions and the normalised background, which is assumed to be constant in the 2-D mass plane.  The values of the $CB$ parameters for each $J/\psi$ candidate are set to the values from the inclusive $J/\psi$ sample in the corresponding rapidity region.  The term for mixed $J/\psi$ and non-$J/\psi$ contributions is not found to be statistically significant within error and is therefore not included in the PDF.  The expression for the PDF used to describe the data is: 

\beq
P = P_{\mathrm{sig}} \times CB\left(m(J/\psi_{1})\right) \times CB(m(J/\psi_{2}))\:+ \: P_{\mathrm{bkg}} \times P_{0},
\label{eq:TrueJ}
\eeq

where $P_{\mathrm{sig}}$ is the fraction of events attributed to signal, $P_{\mathrm{bkg}} = 1 - P_{\mathrm{sig}}$ is the fraction of events attributed to background, and $P_{0}$ is a constant.  The two $J/\psi$ masses are not expected to be correlated and no evidence of a correlation is observed in the data.

The average mass and mass resolution of the reconstructed $J/\psi$ meson depend on $p_{\mathrm{T}}$ (both varying by about 3$\%$ with $p_{\mathrm{T}}$ in the studied region), but in the di-$J/\psi$ sample there are not enough events to let the mean and width float free for each bin of the chosen distribution.  To account for this effect, a correction is applied as a function of $p_{\mathrm{T}}$.  The number of $J/\psi$ signal events obtained from the fit of the inclusive $J/\psi$ sample with and without a fixed mean and width is calculated as a function of $p_{\mathrm{T}}$ and the mass of each $J/\psi$ meson in the di-$J/\psi$ sample is corrected for the mass bias in the corresponding rapidity region.

\begin{figure}[t]
\begin{center}
\begin{tabular}{cc}
\subfloat[]{
\includegraphics*[width=0.43\textwidth]{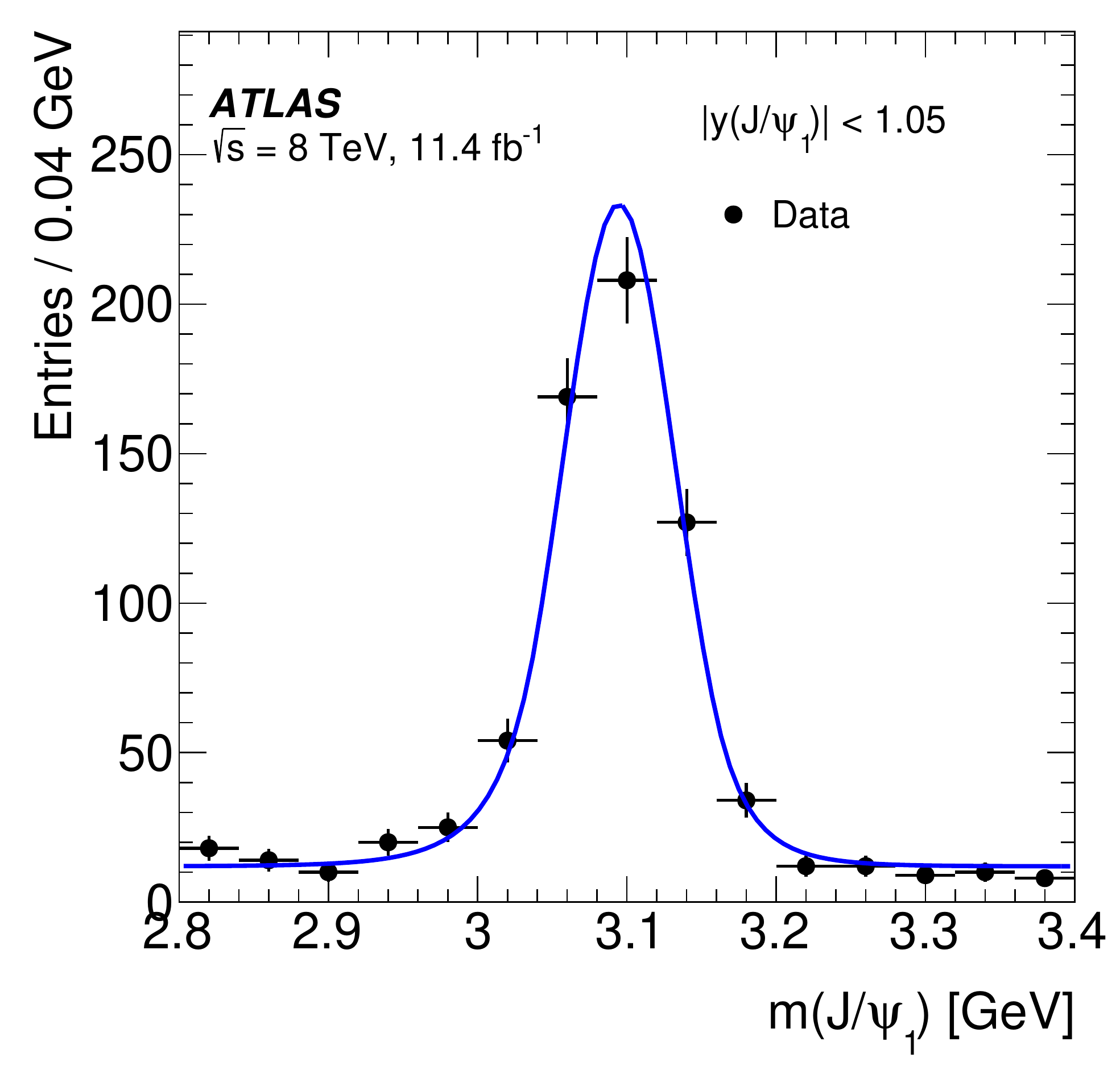}
\label{fig:m11}
}&
\subfloat[]{
\includegraphics*[width=0.43\textwidth]{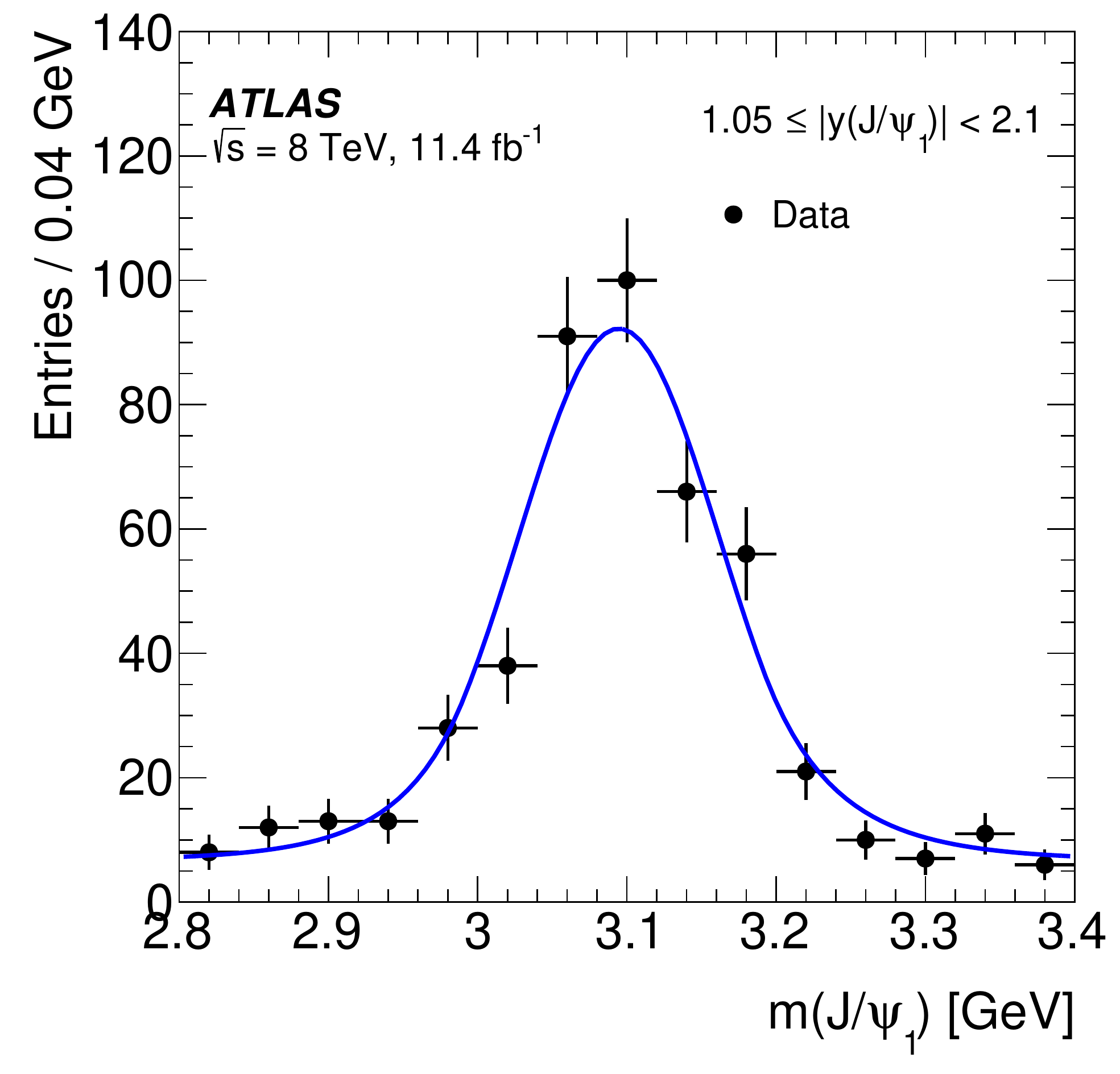}
\label{fig:m12}
}\\
\subfloat[]{
\includegraphics*[width=0.43\textwidth]{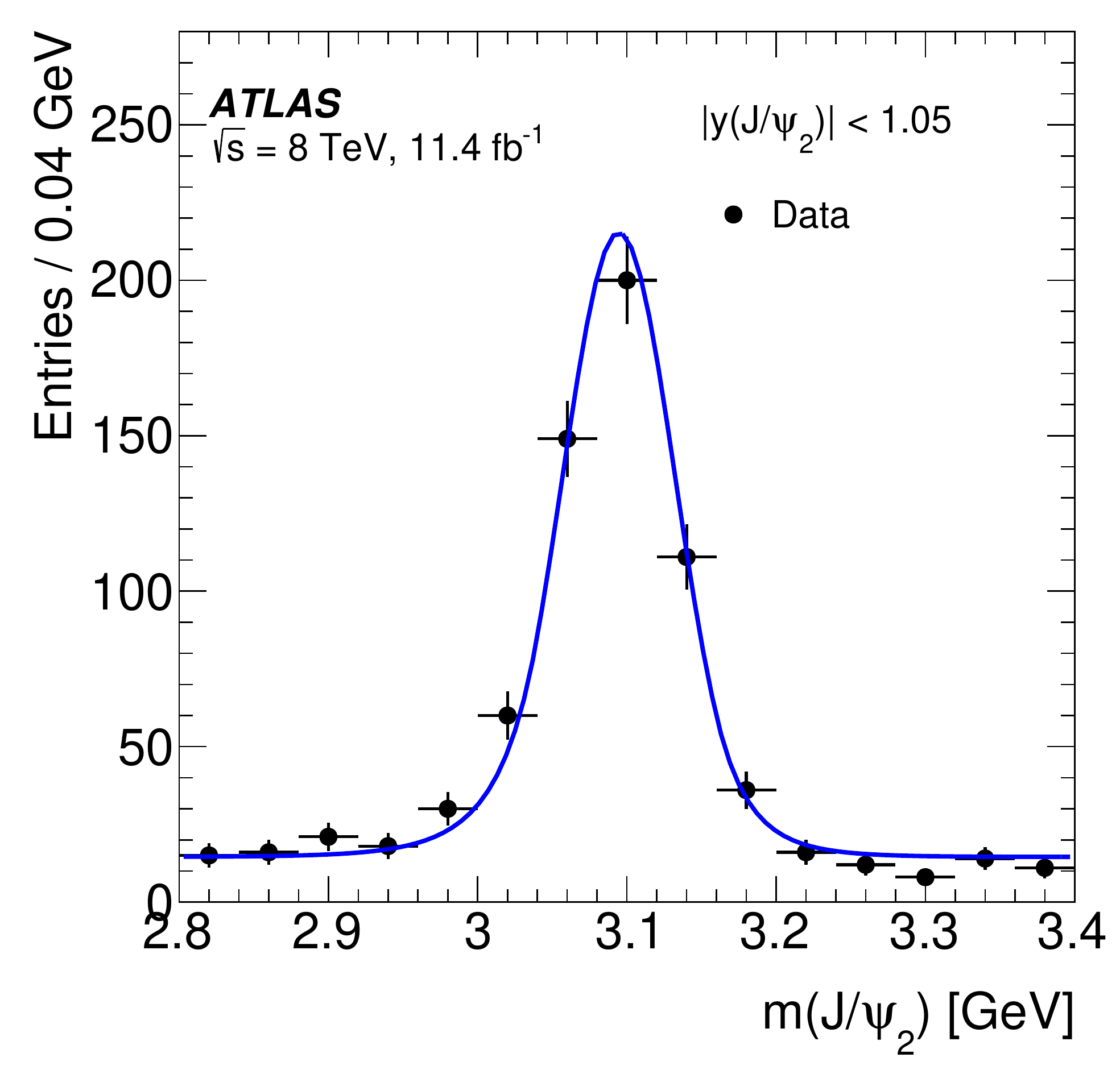}
\label{fig:m21}
}&
\subfloat[]{
\includegraphics*[width=0.43\textwidth]{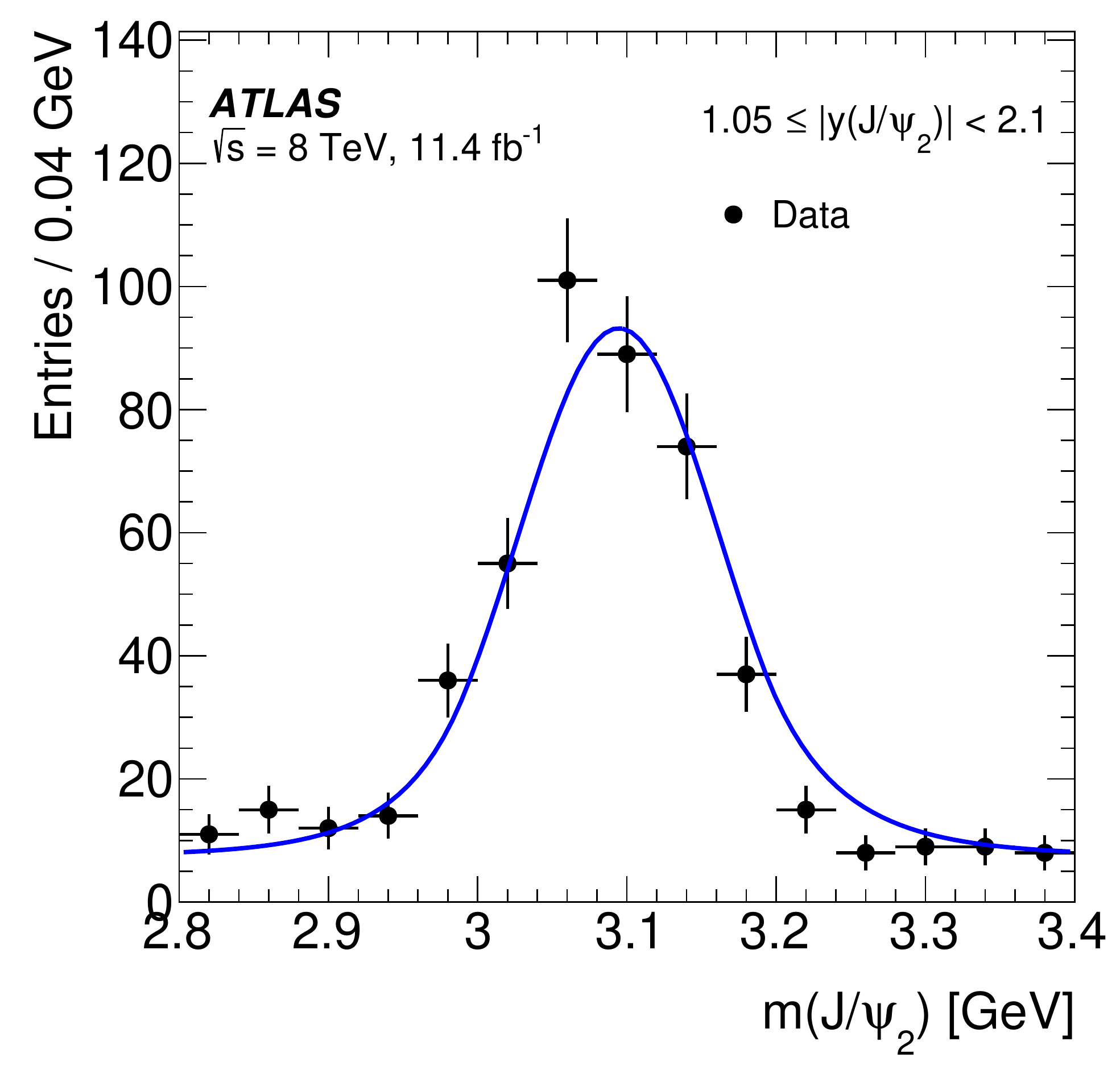}
\label{fig:m22}
}
\end{tabular}
\caption{The 1-D projections of the non-weighted invariant mass spectrum fit of the leading $J/\psi$ in the~\protect\subref{fig:m11} central and~\protect\subref{fig:m12} forward rapidity regions as well as the sub-leading $J/\psi$ in the~\protect\subref{fig:m21} central and~\protect\subref{fig:m22} forward rapidity regions.  The fits use the parameters derived from the inclusive $J/\psi$ sample.}
\label{fig:Mass1D}
\end{center}
\end{figure}

For the extraction of the signal, the data are split into four rapidity regions based on the rapidities of the two $J/\psi$ mesons.  After correcting for the mass bias, the di-$J/\psi$ signal is extracted from the unbinned 2-D maximum-likelihood fit of $m$($J/\psi_{1}$) against $m$($J/\psi_{2}$) in the range 2.8~$\leq$~$m(J/\psi)$~$\leq$~3.4~GeV.  The 1-D projections of the fit onto each $J/\psi$ mass in the central and forward rapidity regions are shown in Figure~\ref{fig:Mass1D}, and are used as an illustration of the shape of the signal and background distributions.  There are 1050~$\pm$~40 non-weighted di-$J/\psi$ events extracted from the 2-D fit of the mass distribution in the fiducial volume.  From the efficiency-weighted unbinned maximum likelihood fit, there are (15.0~$\pm$~0.9)$\times 10^{3}$ di-$J/\psi$ signal events in the full inclusive volume; this uncertainty does not include the uncertainty arising from the extrapolation to the inclusive volume.  The increase is mainly from the transformation to the inclusive volume in which the $p_{\mathrm{T}}$ of the four muons is extrapolated to zero from the fiducial $p_{\mathrm{T}}$ requirements, in addition to the weights from the other efficiency corrections.

\subsection{Non-prompt background}
\label{sec:NP}
After extracting the inclusive di-$J/\psi$ signal, the next step is to extract the PP signal by creating a PP event weight from a fit of the transverse decay length of each $J/\psi$ meson.  The distributions of the transverse decay length $L_{\mathrm{xy}}$ resolution, $R$, the prompt signal, $S$, and the non-prompt background, $N$, are defined as:

\beq
\begin{split}
R &= G_{1}(L_{\mathrm{xy}}) + G_{2}(L_{\mathrm{xy}}) + G_{3}(L_{\mathrm{xy}}) + G_{4}(L_{\mathrm{xy}}) \\
S &= \delta(L_{\mathrm{xy}}) \ast R \\
N &= \frac{1}{\tau}\mathrm{exp}(-L_{\mathrm{xy}}/\tau) \ast R.
\end{split}
\label{eq:LxyS}
\eeq

The resolution function is modelled by the sum of four Gaussian functions, $G_{i}(L_{\mathrm{xy}})$, centred at zero.  This is determined from a study of the $L_{\mathrm{xy}}$ distribution of the inclusive $J/\psi$ sample.  The signal PDF is a delta function convolved with the four-Gaussian resolution function, and the non-prompt background PDF is modelled with a single-sided exponential function with decay constant $\tau$, and is convolved with the four-Gaussian resolution function.  In the di-$J/\psi$ sample, the PP signal is extracted in four subsamples based on the rapidity of each $J/\psi$ meson.  The prompt and non-prompt PDFs are used for both $J/\psi$, with the parameters of the resolution function set to the values from the  inclusive $J/\psi$ sample in the corresponding rapidity region.  A 2-D unbinned maximum-likelihood fit of the mass is performed in bins of $L_{\mathrm{xy}_{1}}$ against $L_{\mathrm{xy}_{2}}$ to get the di-$J/\psi$ signal distribution.

\begin{figure}[ht!b]
\begin{center}
\begin{tabular}{cc}
\subfloat[]{
\includegraphics*[width=0.48\textwidth]{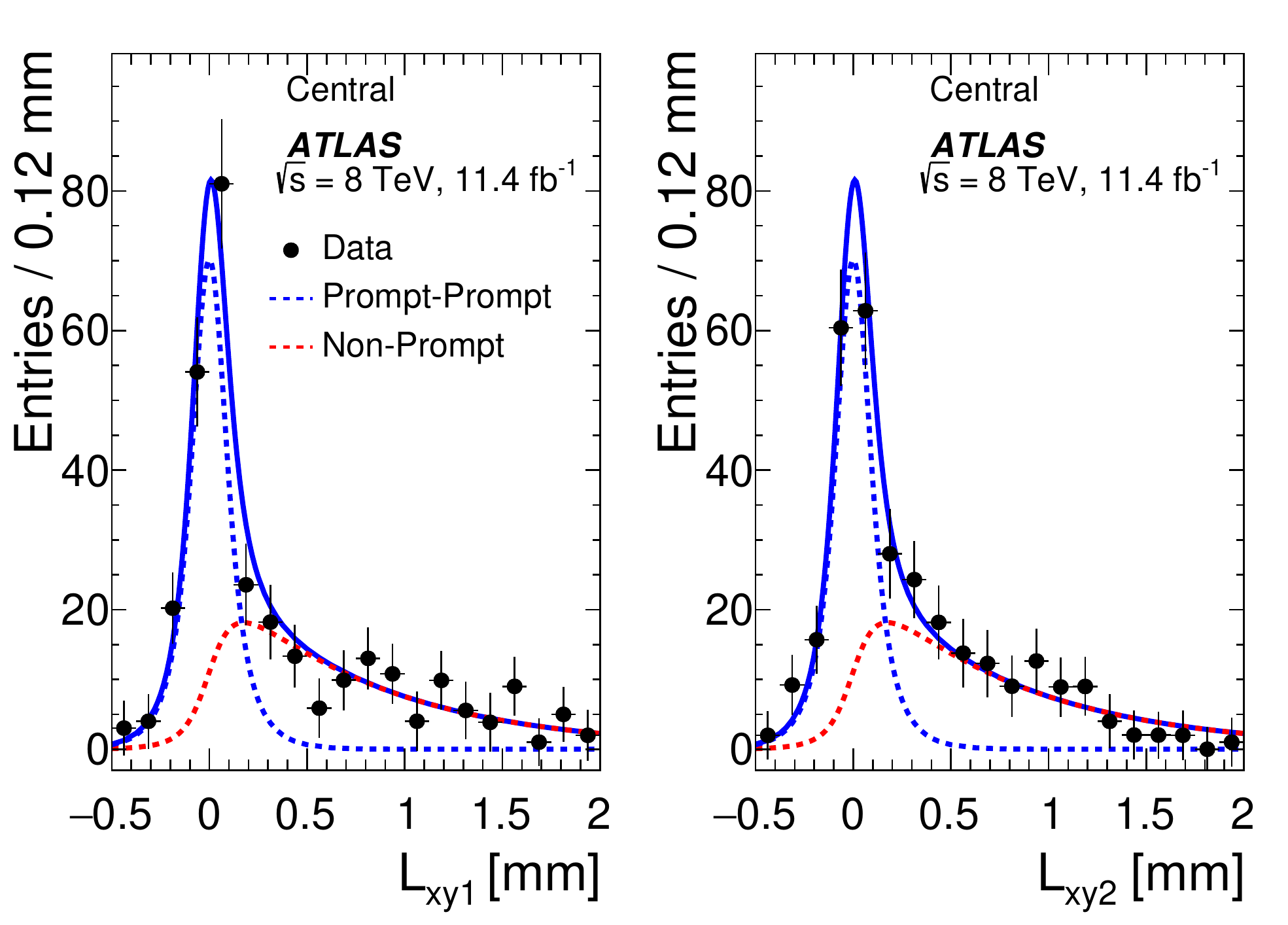}
\label{fig:LxyCC}
}&
\subfloat[]{
\includegraphics*[width=0.48\textwidth]{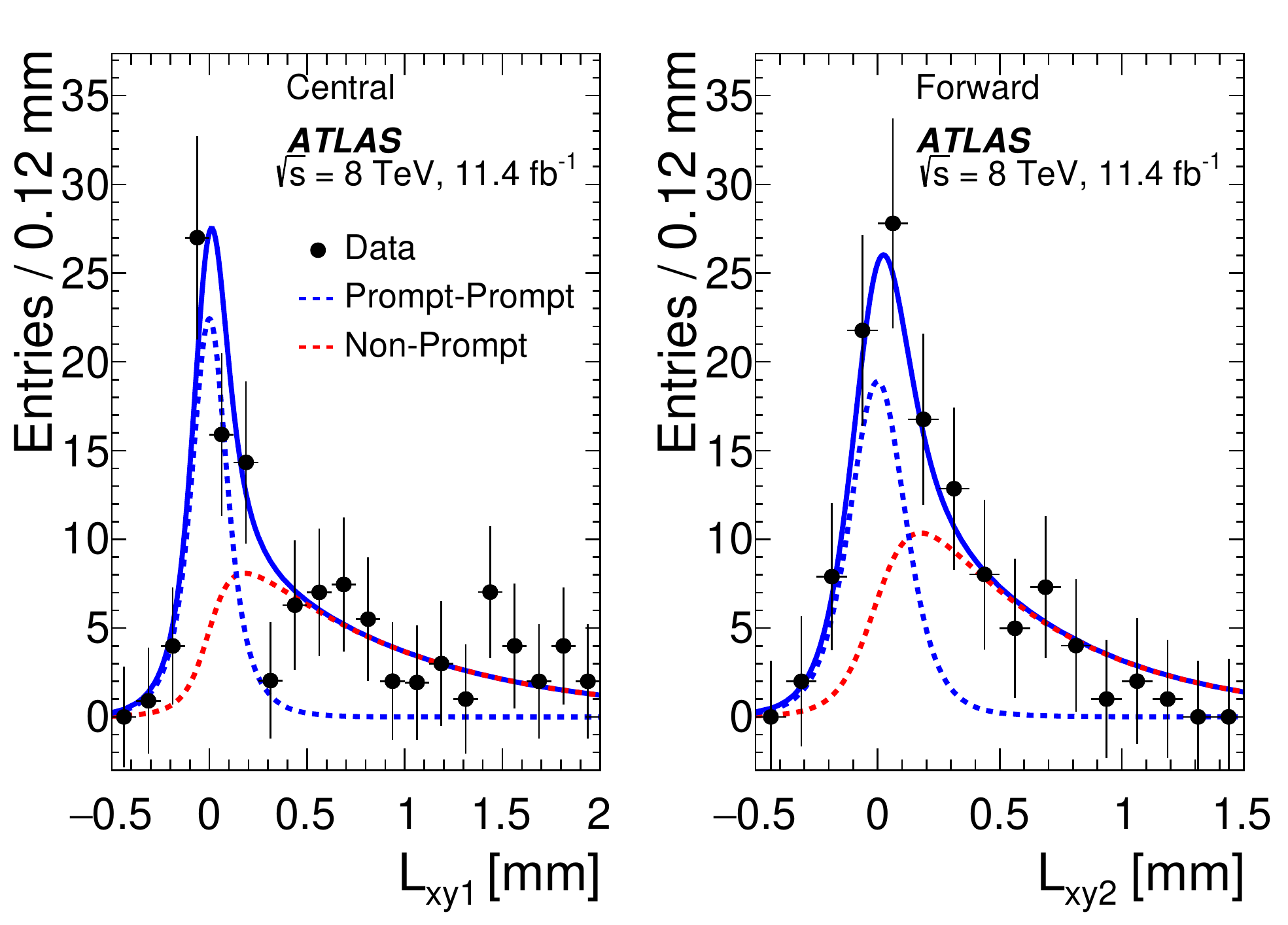}
\label{fig:LxyCF}
}\\
\subfloat[]{
\includegraphics*[width=0.48\textwidth]{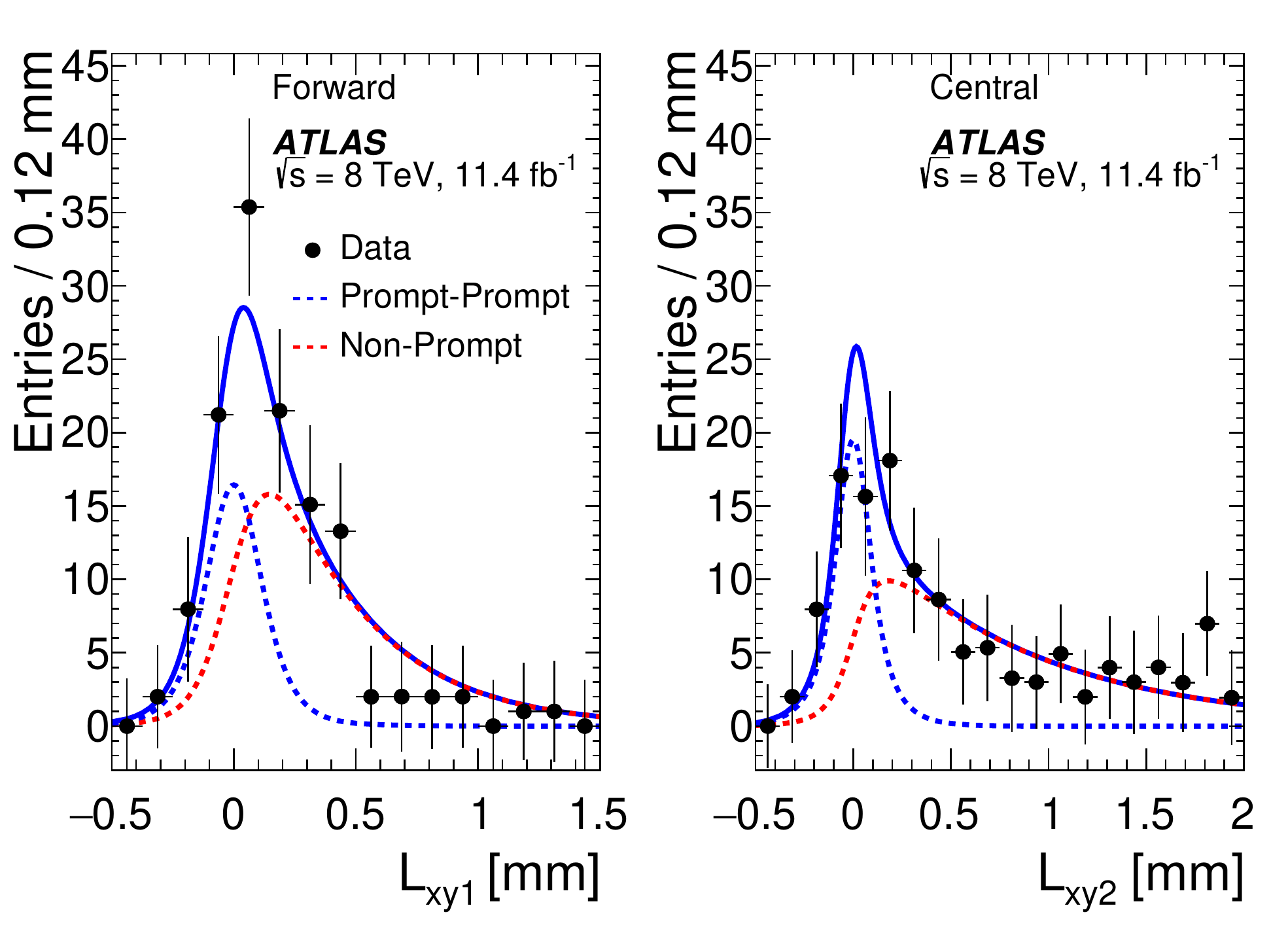}
\label{fig:LxyFC}
}&
\subfloat[]{
\includegraphics*[width=0.48\textwidth]{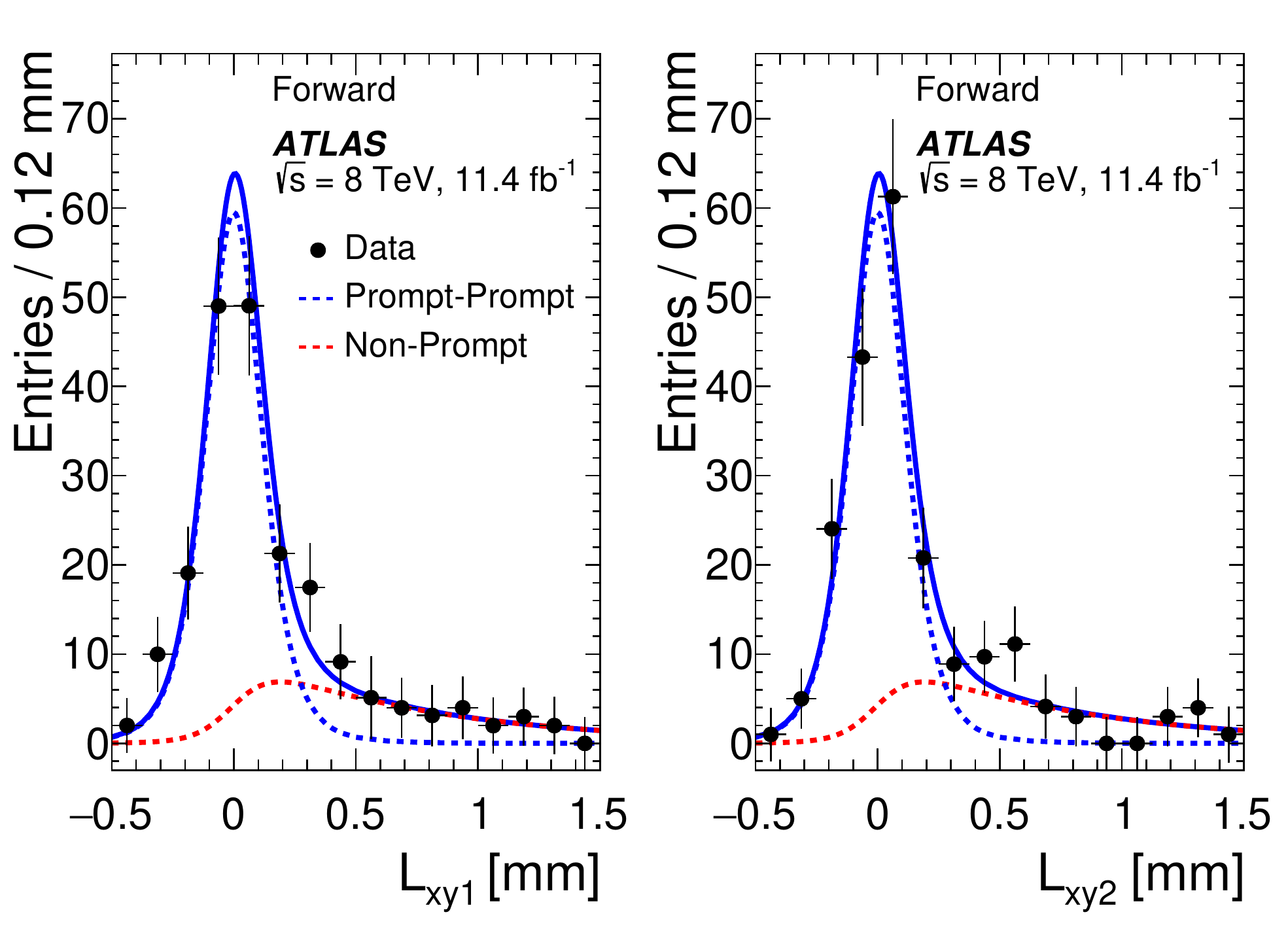}
\label{fig:LxyFF}
}
\end{tabular}
\caption{The background-subtracted non-weighted transverse decay length spectra, $L_{\mathrm{xy}_{1}}$ and $L_{\mathrm{xy}_{2}}$, of the leading and sub-leading $J/\psi$ mesons.  The data are split into four ranges:~\protect\subref{fig:LxyCC} central--central,~\protect\subref{fig:LxyCF} central-forward,~\protect\subref{fig:LxyFC} forward-central, and~\protect\subref{fig:LxyFF} forward--forward.  The prompt--prompt signal component and the non-prompt background component in the fiducial volume are shown.}
\label{fig:LxyD}
\end{center}
\end{figure}

In the central--central and forward--forward cases, when both $J/\psi$ are either central or forward, a single value of the decay constant of the non-prompt exponential function is used to describe both $J/\psi$ candidates.  For the mixed cases, the two $J/\psi$ candidates are allowed to have non-prompt exponential functions with different decay constants.  In these cases, the decay constants of the central--central and forward--forward fits are used as input parameters to a Gaussian penalty function in the fit of the exponential tail of the central and forward $J/\psi$ candidates respectively.  The Gaussian penalty with the mean set to the decay constant of the non-prompt exponential function of either the central--central or forward--forward case is applied to the fit of the $L_{\mathrm{xy}}$ distributions.  The penalty function increases the probability for the decay constant of the exponential tail for the $J/\psi$ candidate to be close to the value of either the central--central or forward--forward case depending on its rapidity. The background-subtracted data are then fit with the product of the prompt PDFs for the two $J/\psi$ mesons and the product of the non-prompt PDFs for the two $J/\psi$ mesons.  The mixed prompt and non-prompt terms are not included in the fit as the contribution is not found to be statistically significant.  Figure~\ref{fig:LxyD} shows the 1-D projections of the results of the fits to data including the projected distributions for the prompt--prompt signal and non-prompt background.  Dividing the PP PDF by the total PDF, shown in Figure~\ref{fig:LxyD}, gives the likelihood for an event to be PP as a function of the transverse decay length and rapidity of the two $J/\psi$ mesons.  By applying the PP probability as a signal weight to each event and then using an unbinned 2-D maximum-likelihood fit of the PP-weighted mass distribution of $J/\psi_{1}$ against $J/\psi_{2}$ in bins of the given kinematic variable, the projected distribution of that variable for the PP di-$J/\psi$ signal is determined.

Because the PP event weight is determined as a function of the transverse decay length of the two $J/\psi$ mesons in four rapidity regions over the full volume, an average value is assumed over the differential distributions.  Since the fraction of PP events, $f_{\mathrm{PP}}$, can vary, an average $f_{\mathrm{PP}}$ leads to a bias of the PP event weight in the differential distributions.  An example of the average $f_{\mathrm{PP}}$ leading to a bias is the $p_{\mathrm{T}}$ spectra of $J/\psi$ mesons, as $f_{\mathrm{PP}}$ decreases with $p_{\mathrm{T}}$.  To determine this bias and to correct for it, MC di-$J/\psi$ samples are used.  Three MC samples (prompt--prompt, prompt--non-prompt, and non-prompt--non-prompt) are produced.  The particle-level MC samples are produced using the second-hard-process mechanism in \PYTHIA 8~\cite{Pythia}.  These scale factors are defined as a function of the reconstructed prompt--prompt fraction, $f_{\mathrm{PP}}$.

\begin{figure}[t]
\begin{center}
\subfloat[]{
\includegraphics*[width=0.42\textwidth]{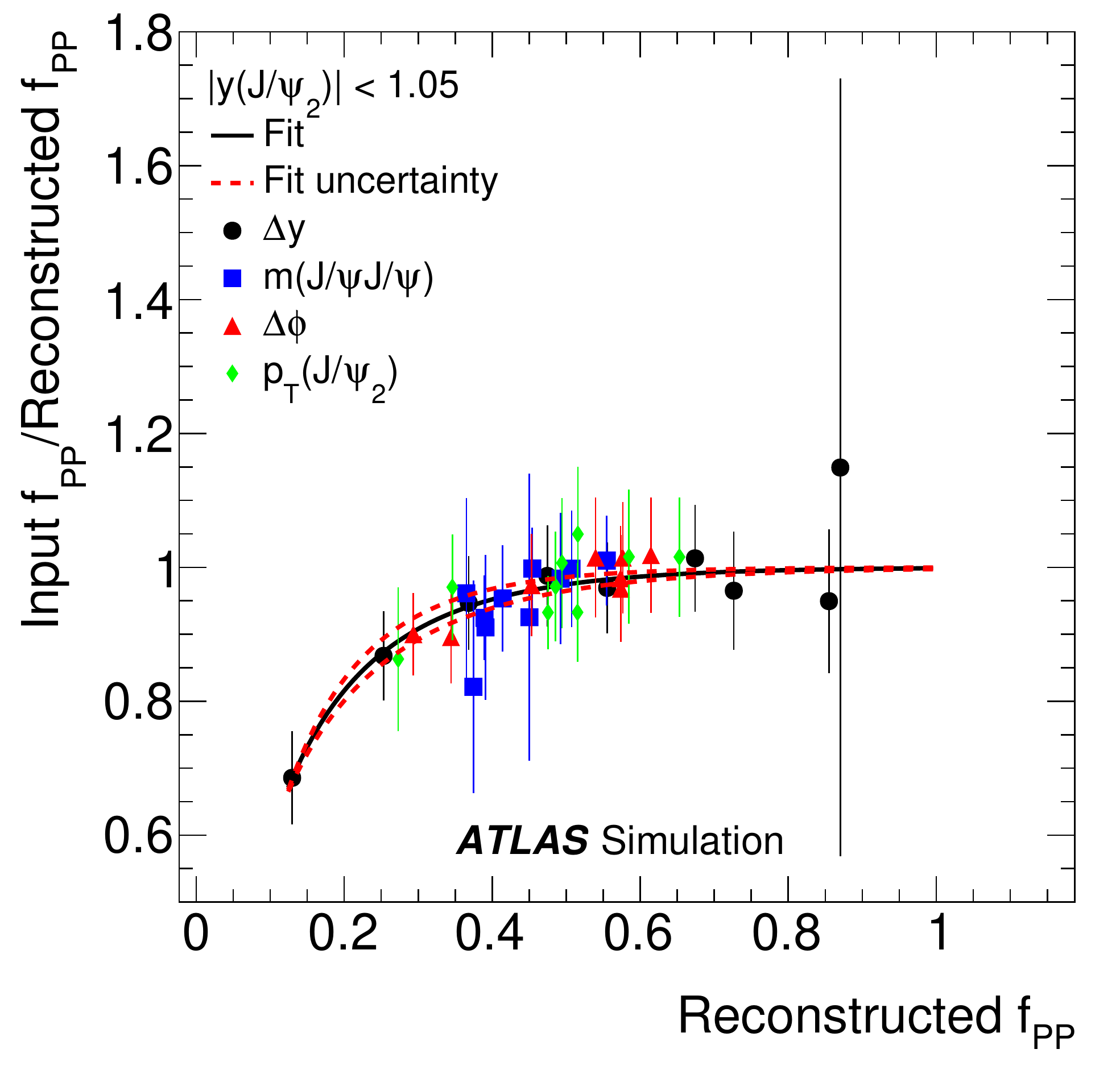}
\label{fig:fc1}
}
\subfloat[]{
\includegraphics*[width=0.42\textwidth]{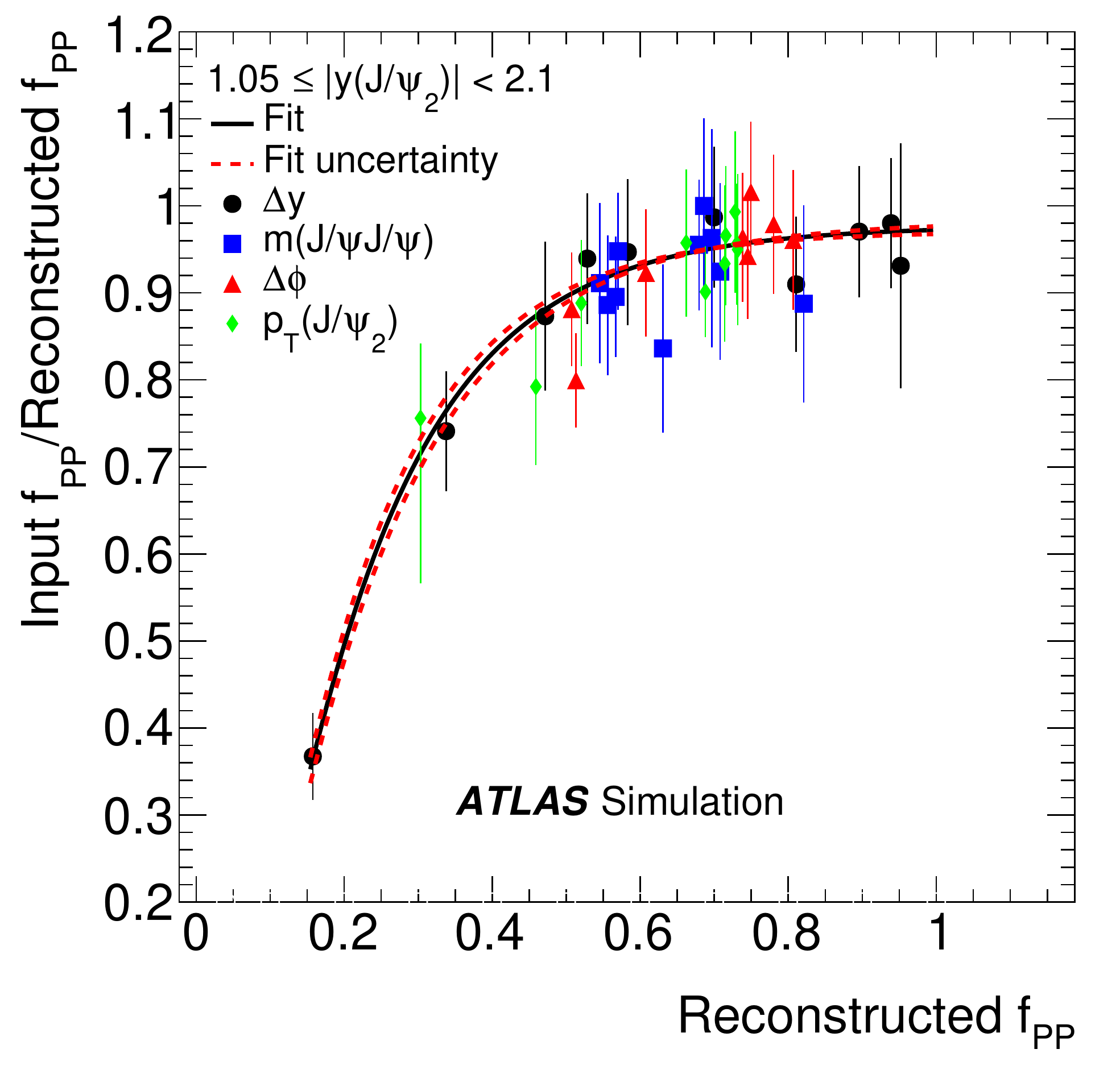}
\label{fig:fc2}
}
\caption{The PP weight bias correction as a function of the reconstructed $f_{\mathrm{PP}}$ in the~\protect\subref{fig:fc1} central and~\protect\subref{fig:fc2} forward rapidity regions for different variables.  The $\Delta y$ distribution is used for the fit.}
\label{fig:fPPCorr}
\end{center}
\end{figure}

The resulting bias correction is displayed as a function of the reconstructed $f_{\mathrm{PP}}$ in Figure~\ref{fig:fPPCorr} for the kinematic variables considered in this analysis.  The lowest reconstructed $f_{\mathrm{PP}}$ is 15$\%$, so the bias correction is only fit above this point.  The correction factor is obtained separately for the central and forward rapidity regions of the sub-leading $J/\psi$ meson.  The bias correction is flat over a wide range of the reconstructed $f_{\mathrm{PP}}$ and drops quickly at low $f_{\mathrm{PP}}$. It is fit with a threshold function defined as $F$$\times$$[1 - \mathrm{erf}(x)]$, where $F$ is a free parameter and erf$(x)$ is the error function obtained by integrating the normal distribution.

\begin{figure}[t]
\begin{center}
\includegraphics*[width=0.6\textwidth]{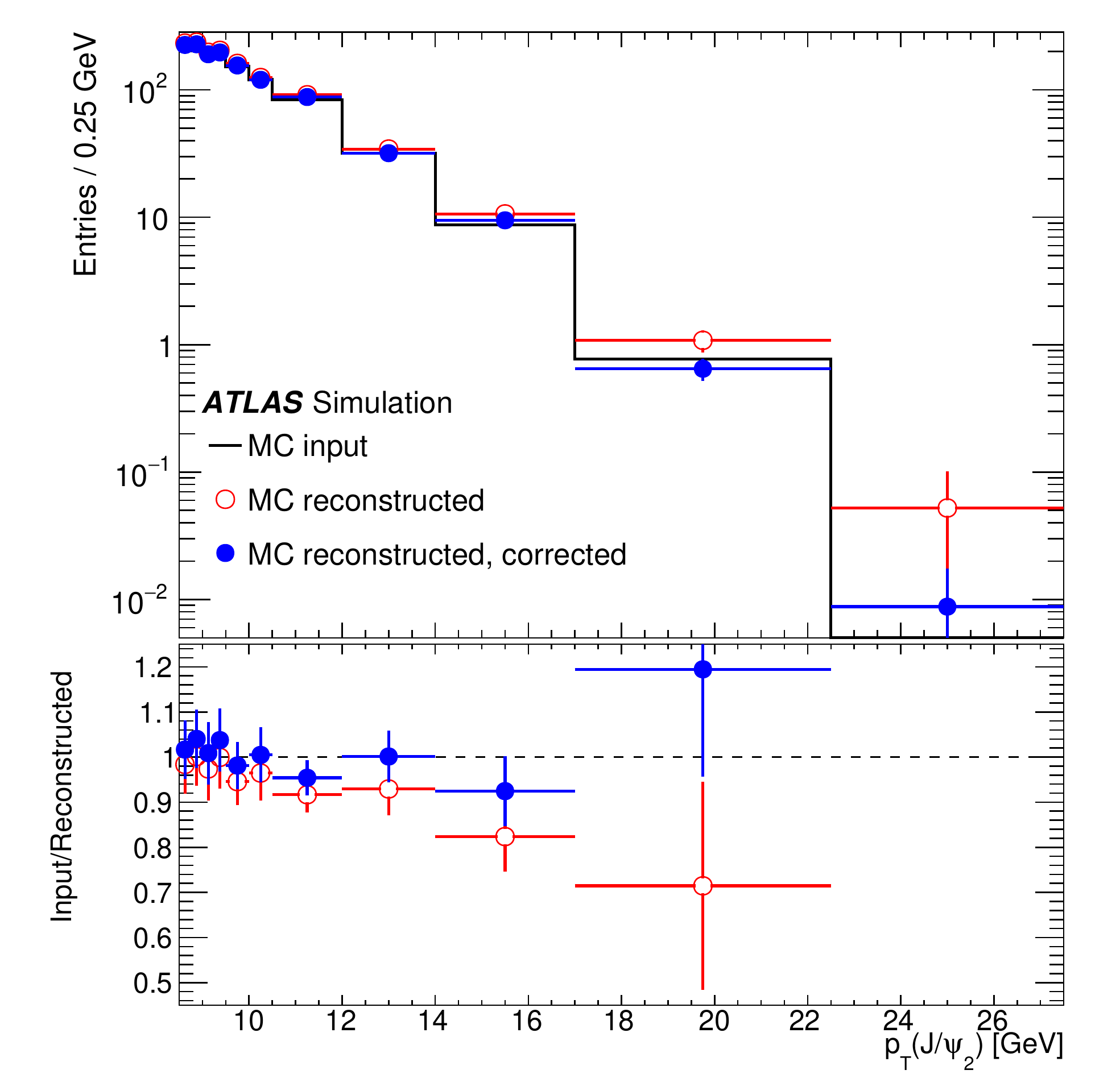}
\caption{A comparison of the PP MC input, MC reconstructed, and MC reconstructed corrected distribution of the sub-leading $J/\psi$ $p_{\mathrm{T}}$, $p_{\mathrm{T}}$($J/\psi_{2}$).  The ratios of the input to the MC reconstructed as well as bias corrected MC reconstructed distributions are shown.}
\label{fig:MCptCorr}
\end{center}
\end{figure}

In Figure~\ref{fig:MCptCorr} the correction is applied to the $p_{\mathrm{T}}$($J/\psi_{2}$) distribution.  The correction factor is found to perform well for each of the variables considered.  The original reconstructed distribution is included for comparison.  Closure tests with MC samples are performed.

\subsection{Pile-up background}
\label{sec:pile}

The remaining background comes from pile-up events, which are multiple uncorrelated collisions in the same beam crossing.  In pile-up events, the two $J/\psi$ mesons originate from two independent proton--proton collisions.  These events have distributions similar to those from DPS.  The requirement on the distance between the trajectories of the two $J/\psi$ mesons along the beam direction, $|d_{\mathrm{z}}|~<$~1.2~mm, is used to remove events that come from two separate primary vertices.

\begin{figure}[t]\begin{center}
\subfloat[]{
\includegraphics*[width=0.47\textwidth]{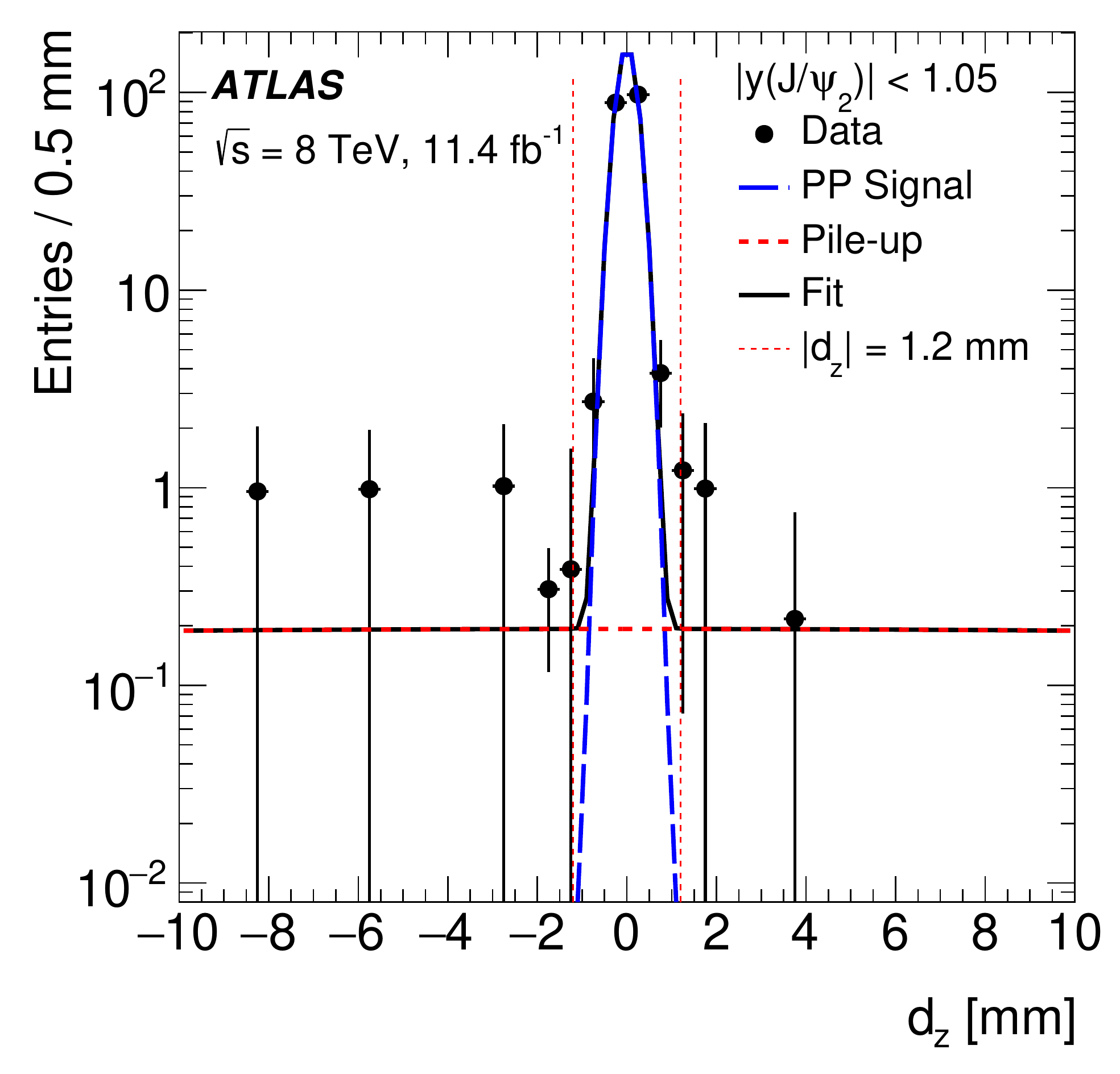}
\label{fig:dz1}
}
\subfloat[]{
\includegraphics*[width=0.47\textwidth]{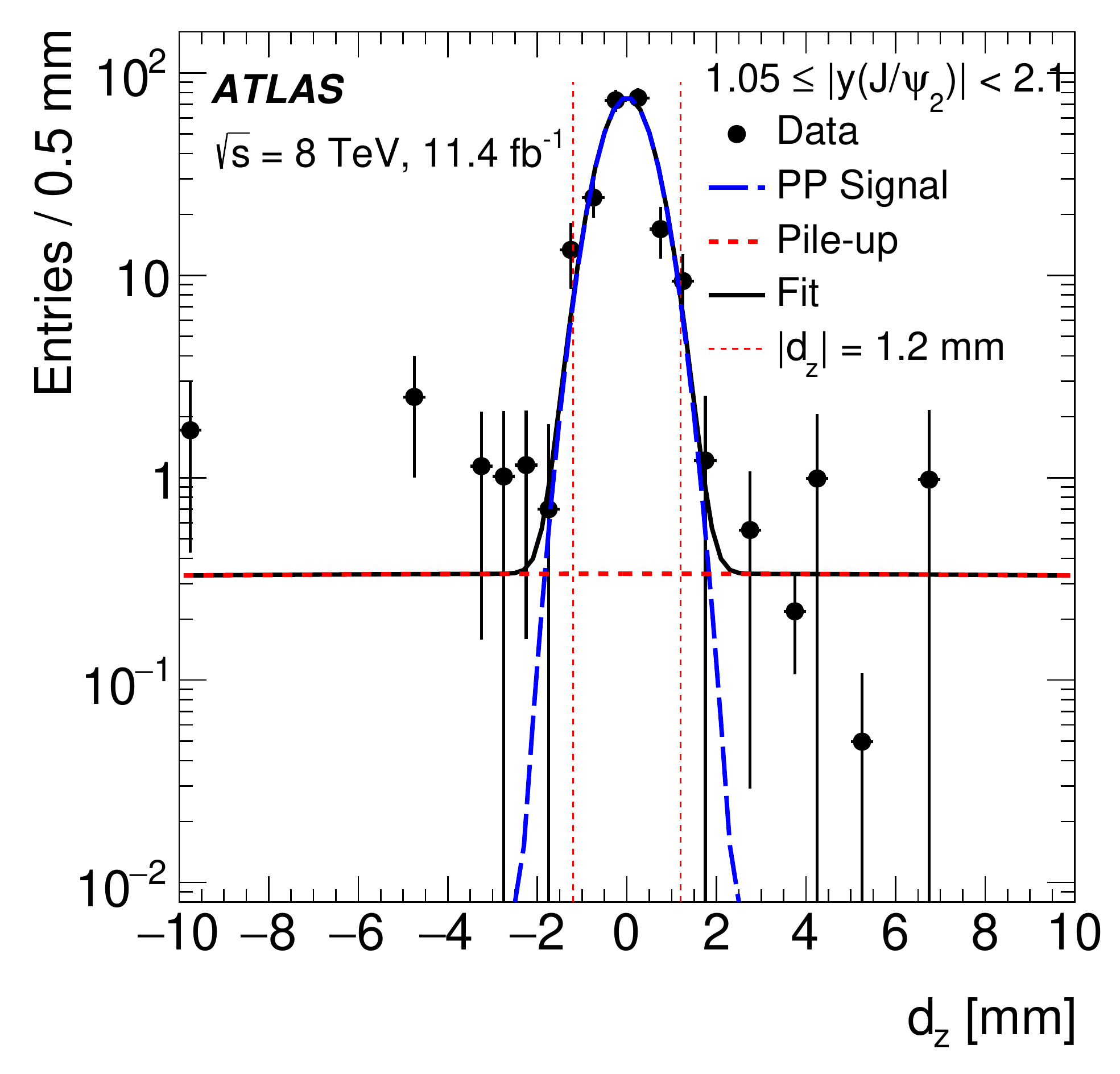}
\label{fig:dz2}
}
\caption{The distribution of the distance between the trajectories of the two $J/\psi$ mesons along the beam direction, $d_{\mathrm{z}}$, after subtraction of the non-$J/\psi$ and non-prompt background. A double Gaussian function is fit to the data in order to determine the fraction of pile-up background events in the~\protect\subref{fig:dz1} central and~\protect\subref{fig:dz2} forward rapidity regions.  A narrow Gaussian describes the prompt $J/\psi$ component and a wider Gaussian describes the component due to pile-up.  Only the relative normalisation is free in the fit. The parameters of both Gaussian functions are determined from a fit to the inclusive $J/\psi$ sample over a wider range of $d_{\mathrm{z}}$.} 
\label{fig:DI}
\end{center}
\end{figure}

The PP background-subtracted $d_{\mathrm{z}}$ distribution is shown in Figure~\ref{fig:DI}.  To determine the amount of pile-up background that passes the $d_{\mathrm{z}}$ selection, a double Gaussian function is fit to the data.  A narrow Gaussian describes the prompt $J/\psi$ component and a wider Gaussian describes the component due to pile-up.  Only the relative normalisation is free in the fit to the di-$J/\psi$ sample.  The other parameters of the Gaussian functions are determined from a fit to the large inclusive $J/\psi$ sample over a wide $d_{\mathrm{z}}$ range.  In the fit of the inclusive $J/\psi$ sample, the pile-up distribution is found to have a Gaussian width of 49.2~$\pm$~0.8~mm.  The background is determined by integrating the fitted function over the selected $d_{\mathrm{z}}$ range of $|d_{\mathrm{z}}|$~$\leq$~1.2~mm.

The amount of pile-up in the accepted sample is $f_{\mathrm{pile-up}}$~=~(0.466~$\pm$~0.034~(stat)~$\pm$~0.004~(syst))$\%$ in the central region and $f_{\mathrm{pile-up}}$~=~(0.802~$\pm$~0.062~(stat)~$\pm$~0.007~(syst))$\%$ in the forward region.  The systematic uncertainty is described in Section~\ref{sec:Syst}.  As a check, the signal PP Gaussian width is allowed to be free in the di-$J/\psi$ $d_{\mathrm{z}}$ fit.  The width is compatible with the value calculated from the  inclusive $J/\psi$ sample but with a much larger uncertainty.  Finally, to remove the pile-up background, the pile-up distributions of the kinematic variables are needed.  This is achieved by reversing the $|d_{\mathrm{z}}|$ requirement to $|d_{\mathrm{z}}|$~$>$~2.0~mm.  The distribution of the chosen variable with this new requirement is then plotted to get the distributions of events coming from two separate primary vertices.  The pile-up distributions are normalised to the correct number of events, $n_{\mathrm{Total}}\times$$f_{\mathrm{pile-up}}$, and subtracted from the PP distributions.

The total number of PP di-$J/\psi$ signal events corrected for the muon acceptance are 3310~$\pm$~330~(central) and  3140~$\pm$~370~(forward) where the uncertainty is extracted from the fit of the weighted data.  In the DPS analysis described in Section~\ref{sec:DPSsec}, the full muon fiducial volume is used.  The number of PP di-$J/\psi$ signal events in the fiducial volume, not corrected for the acceptance, is 1160~$\pm$~70.

\section{Double parton scattering}
\label{sec:DPSsec}
Due to the decrease in the average fraction of the incoming proton momentum carried by a parton at large centre-of-mass energies, the parton densities rapidly increase and therefore DPS phenomena can be of substantial importance at the LHC.  The DPS cross-section is dependent on the transverse distance between partons, and should decrease quickly as a function of transverse energy.  Since at the LHC energies, $J/\psi$ meson production is dominated by gluon--gluon interactions, the DPS cross-section is sensitive to the spatial distribution of gluons in the proton~\cite{gg}.

A simplified ansatz for defining the DPS cross-section in terms of the production cross-sections of the two final states and an effective cross-section is described in Ref.~\cite{ansatz} as:

\beq
\sigma_{\mathrm{eff}} \: = \: \frac{1}{2} \frac{\sigma_{J/\psi}^{2}}{\sigma_{\mathrm{DPS}}^{J/\psi,J/\psi}} \: = \: \frac{1}{2} \frac{\sigma_{J/\psi}^{2}}{f_{\mathrm{DPS}} \times \sigma_{J/\psi J/\psi}},
\label{eq:sigEffDiJpsi}
\eeq

where $f_{\mathrm{DPS}}$ is the fraction of PP di-$J/\psi$ events that are due to DPS.  The factor of 1/2 is because the two final states for di-$J/\psi$ events are identical.

The effective cross-section, $\sigma_{\mathrm{eff}}$, is related to the spatial separation between partons inside the proton.  In the derivation of the effective cross-section ansatz, process and energy independence are assumed to be first-order approximations in perturbative QCD predictions.  There are possible correlations between the fractional momenta of the incoming partons, the fractional momenta of the partons and the impact parameter, as well as spin and colour correlations that are not addressed in this simplified ansatz.  These correlations and a modified effective cross-section ansatz which accounts for these possible correlations are described in Refs.~\cite{corr,corr2,corr3,corr4}.

Completely uncorrelated scatterings and the factorisation of the contributions to the cross-section described by the ansatz would lead to a universal effective cross-section which would be close to the inelastic cross-section.  The measured values of the effective cross-section from multiple experiments range from about 5 to 20 mb~\cite{gamma,WDPS,CMSW2J,UA2,CDF4j,ATLASZJpsi,LHCbJD, LHCbYD, CDF2013,D0JY,D0Dijet} for centre-of-mass energies of 630~GeV to 8~TeV.

\subsection{Data-driven model-independent approach}
\label{sec:DPSmodel}
One of the goals of this analysis is to measure the fraction of DPS events, $f_{\mathrm{DPS}}$, as a function of various parameters such as the mass and $p_{\mathrm{T}}$ of the di-$J/\psi$ system and the difference in rapidity and the azimuthal angle between the two $J/\psi$ mesons in order to probe regions of phase space that are sensitive to different processes.  A second goal of this analysis is to use the di-$J/\psi$ DPS cross-section obtained from the measured $f_{\mathrm{DPS}}$ to determine the effective cross-section of DPS. Additionally the modelling and subtraction of the DPS yield can be useful for studies of SPS quarkonium production models.  

A common method for extracting the DPS contribution involves fitting DPS and SPS templates to the data.  The theoretical predictions for the SPS distributions depend on perturbative QCD corrections of various orders and on $J/\psi$ production models~\cite{CE1,CE2,CS3,CS4,CO,CO1,CO2,JpsiMod1,JpsiMod2,JpsiMod3,JpsiMod4, CS2, JpsiMod5}.  By forming a template based on data, that dependence can be minimised.  

\begin{figure}[t]
\begin{center}
\subfloat[]{
\includegraphics*[width=0.48\textwidth]{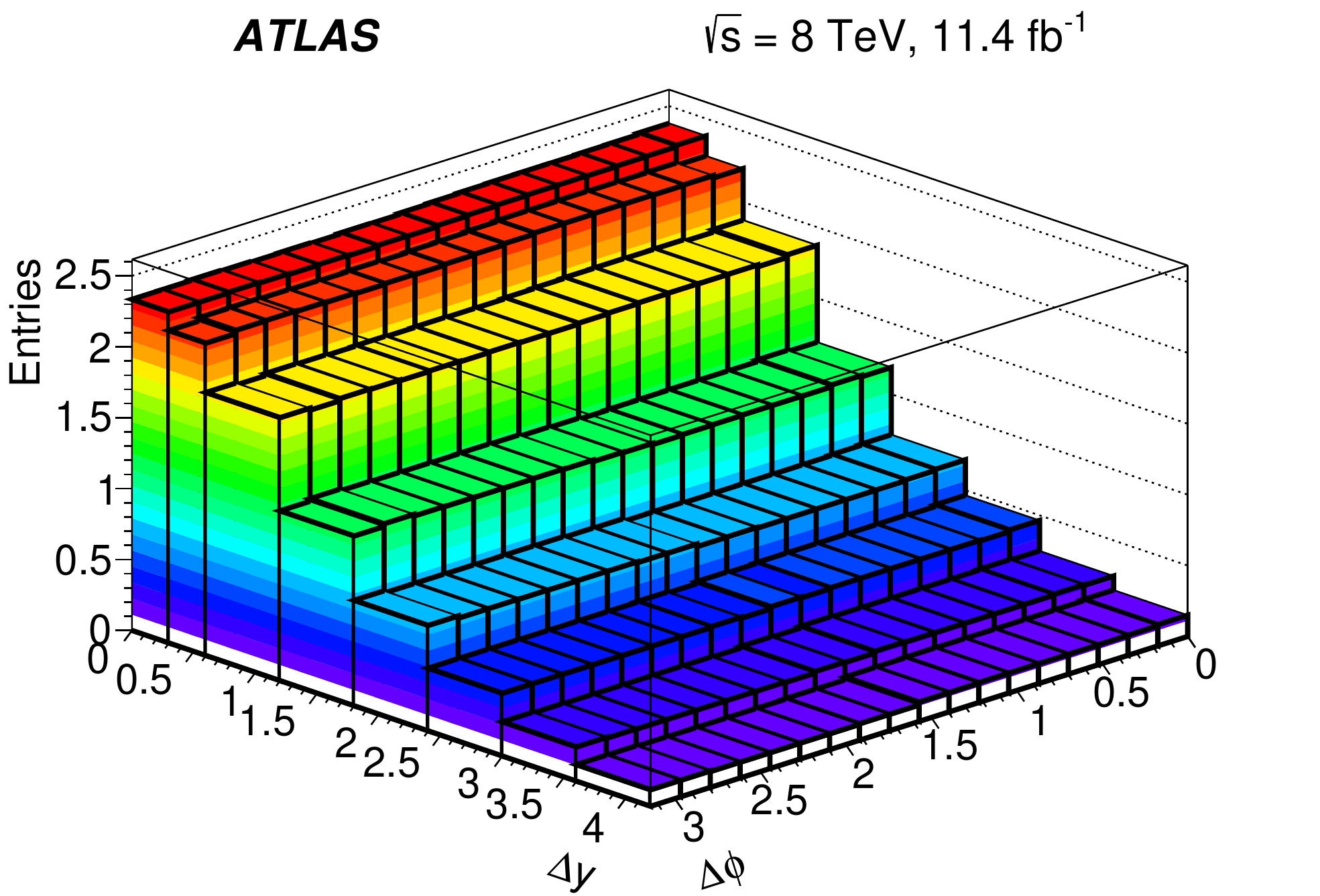}
\label{fig:Dw}
}
\subfloat[]{
\includegraphics*[width=0.48\textwidth]{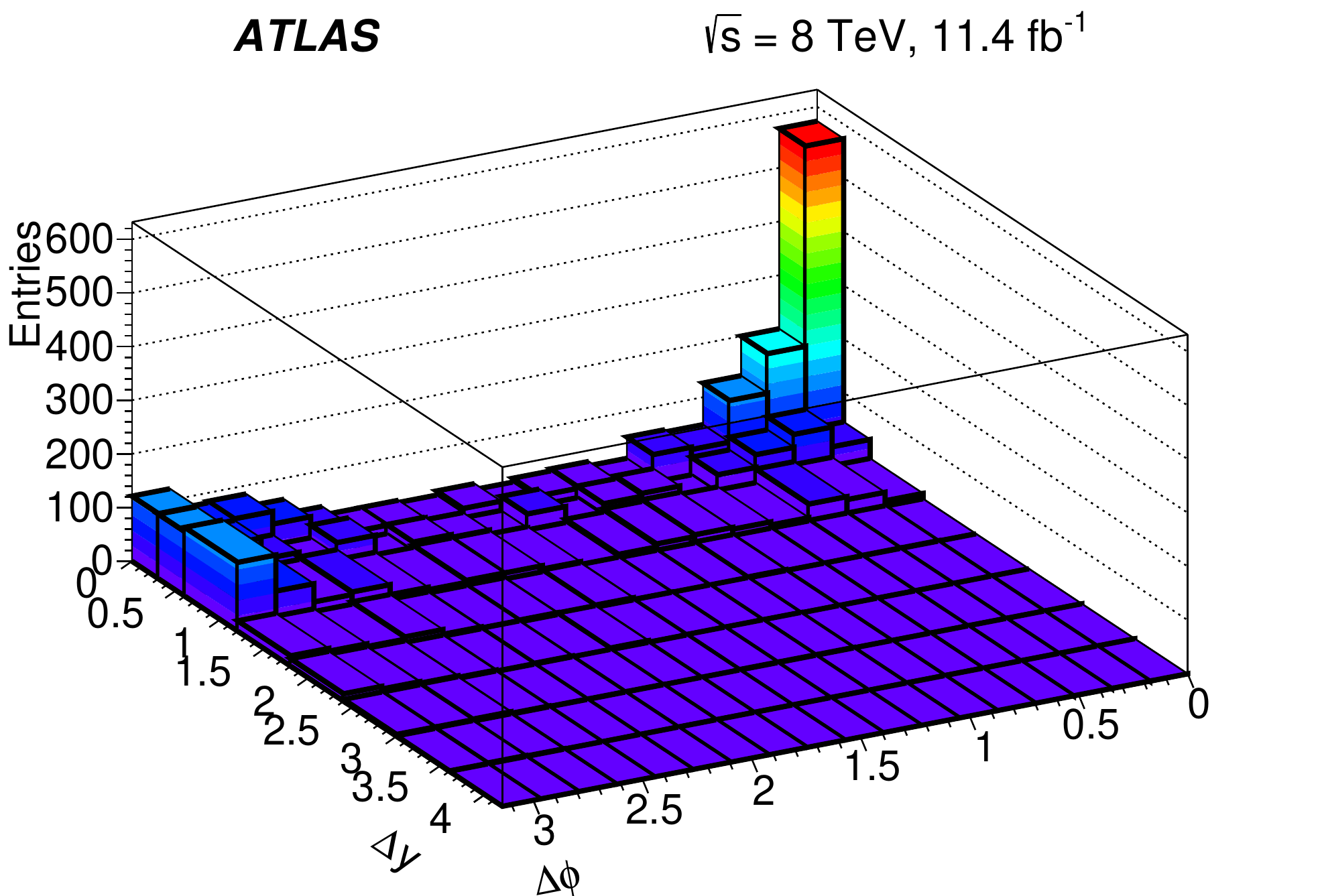}
\label{fig:Sw}
}
\caption{The 2-D data-driven templates of $\Delta y$ against $\Delta \phi$ for~\protect\subref{fig:Dw} DPS obtained by combining $J/\psi$ pairs from different events and normalising to the data and~\protect\subref{fig:Sw} SPS obtained by subtracting the normalised DPS template from the data.  The data-driven templates are used to calculate the DPS and SPS event weights.}
\label{fig:2Dweights}
\end{center}
\end{figure}

In constructing the data-driven DPS template, it is assumed that the two $J/\psi$ candidates are produced independently of each other.  The DPS sample is therefore simulated by combining re-sampled $J/\psi$ mesons from two different random events in the di-$J/\psi$ sample which pass the requirements.  By using events from the di-$J/\psi$ sample, it is ensured that the $J/\psi$ candidates in the DPS sample have the same kinematics as the data.  The distribution of the absolute difference between the rapidities, $\Delta y$, against the absolute difference between the azimuthal angles, $\Delta \phi$, of the two $J/\psi$ candidates for this DPS sample is shown in Figure~\ref{fig:2Dweights}\protect\subref{fig:Dw}.  The template for the SPS component, shown in  Figure~\ref{fig:2Dweights}\protect\subref{fig:Sw}, is obtained by subtracting the DPS template from the $\Delta y$ against $\Delta \phi$ distribution of the background-subtracted data.  The DPS contribution is normalised to the data in the region $\Delta y$~$\geq$~1.8 and $\Delta \phi$~$\leq$~$\pi/2$, where DPS is assumed to dominate and SPS is assumed to be negligible.  The DPS-dominated region is determined after a careful study of the data.  The $\Delta y$ requirement is determined as before this region the data drops off quickly with $\Delta y$ and after it flattens out which is indicative of a dominant DPS contribution.  After examining the data, it is observed that the peak at $\Delta \phi$~$\approx$~$\pi$ has a large tail in $\Delta y$ and therefore an additional requirement is placed to avoid this tail.  Additionally, theoretical predictions~\cite{SPSth,SPSth2} show that SPS is negligible in this region.  The assumption of and sensitivity to the definition of the DPS-dominated region is tested by increasing the $\Delta y$ and varying the $\Delta \phi$ requirements.  By increasing the $\Delta y$ requirement to a smaller region in which SPS is known to be negligible, the possibility of a SPS tail making it into the normalisation region is determined.  Tests of the normalisation are included in Section~\ref{sec:DPSresults}.  At low $\Delta \phi$ and large $\Delta y$, DPS dominates and this validates the choice of region used to normalise the DPS template to the data ($\Delta y$~$\geq$~1.8, $\Delta \phi \leq \pi/2$).

From the 2-D data-driven templates of the SPS and DPS distribution, the DPS and SPS event weights, $w_{\mathrm{DPS}}$ and $w_{\mathrm{SPS}}$, are:

\beq
\begin{split}
w_{\mathrm{DPS}}(\Delta \phi,~\Delta y ) &= \frac{N_{\mathrm{DPS}}(\Delta \phi,~\Delta y )}{N_{\mathrm{Data}}(\Delta \phi,~\Delta y )},\\
w_{\mathrm{SPS}}(\Delta \phi,~\Delta y ) &= \frac{N_{\mathrm{SPS}}(\Delta \phi,~\Delta y )}{N_{\mathrm{Data}}(\Delta \phi,~\Delta y )},
\end{split}
\label{eq:DPSwt}
\eeq

where $N_{\mathrm{Data}}$ is the number of the background-subtracted and bias corrected di-$J/\psi$ data events, and $N_{\mathrm{DPS(SPS)}}$ are the number of background-subtracted and corrected DPS (SPS) events in the normalised template.  

By applying these weights as well as the PP weight, and then extracting the di-$J/\psi$ signal from the 2-D mass fits in bins of the chosen variable one can extract the PP SPS-weighted and DPS-weighted distributions of the kinematic variables studied.  

From these weights, the value of $f_{\mathrm{DPS}}$ is determined.  These weighted distributions are then compared to the sum of the LO DPS and NLO* SPS predicted distributions with $f_{\mathrm{DPS}}$ fixed to the measured experimental value.  Finally, the effective cross-section is calculated and compared to the current measured values.

\section{Systematic uncertainties}
\label{sec:Syst}

Sources of systematic uncertainty and their relative percentage are summarised in Table~\ref{tab:SystCross} for the di-$J/\psi$ cross-section and Table~\ref{tab:SystfDPS} for the $f_{\mathrm{DPS}}$ measurement.  Many of the systematic uncertainties cancel in the $f_{\mathrm{DPS}}$ measurement.

\textbf{Trigger:} The systematic uncertainty due to the trigger selection is estimated by creating one thousand MC templates, varying each bin within the statistical uncertainty of the trigger efficiency and determining the effect on the yield.  Additionally the spatial and vertex correction are varied within their uncertainty.  Finally, a conservative uncertainty for the MC correction is determined by calculating the efficiency weighted yield without the application of the MC correction.  This systematic uncertainty accounts for the use of different low-$p_{\mathrm{T}}$ single-muon triggers in the MC simulation that are not present in data and covers the possible range of trigger corrections.  This is the dominant source of the systematic uncertainty due to the trigger selection. 

\textbf{Muon Reconstruction:} The estimation of the systematic uncertainty due to the two muon reconstruction efficiency correction used in the analysis, described in Section~\ref{sec:Eff}, uses the same MC method as for the trigger efficiency measurement.  The dominant source uncertainty comes from the statistical error in the tag-and-probe fit of $Z \rightarrow \mu \mu$ data for high-$p_{\mathrm{T}}$ muons and $J/\psi \rightarrow \mu \mu$ data for low-$p_{\mathrm{T}}$ muons.

\textbf{Kinematic Acceptance:} Comparing the ratio with and without the acceptance correction for the SPS and DPS MC samples gives the systematic uncertainty for the assumption that the acceptance correction can be applied independently for each $J/\psi$ candidate.  This assumption affects only SPS production.  To measure the systematic uncertainty from bin migration effects due to the detector resolution, the method outlined in Ref.~\cite{Single} is used in which the $J/\psi$ $p_{\mathrm{T}}$ and rapidity spectra with and without convolution the experimental resolution are compared.  The systematic uncertainty as a function of the rapidity is negligible and the systematic uncertainty as a function of the $p_{\mathrm{T}}$ is small relative to the SPS correction. 

\textbf{Mass Model:} To extract signal di-$J/\psi$ events, a 2-D fit of $m$($J/\psi_{1}$) against $m$($J/\psi_{2}$) is used.  The signal parameters are determined from inclusive $J/\psi$ samples in the central and forward rapidity regions.  The mass fit parameters are varied within their uncertainties to estimate the systematic uncertainty due to the choice of the mass model.  Alternative characterisations of the background, such as linear and exponential functions, were tested and were found to have negligible differences.

\textbf{Mass Bias:} The mean and width of the $J/\psi$ mass fit are not constant as a function of $p_{\mathrm{T}}$.  The correction for fixing these values is the ratio of the number of $J/\psi$ from the  inclusive $J/\psi$ sample with fixed mean and width parameters to the number with the mean and width allowed to be free, as a function of $p_{\mathrm{T}}$.  To find the systematic uncertainty of the mass bias correction, the correction is varied within its uncertainties in each bin.  

\textbf{Prompt--Prompt Model:} A 2-D fit of the transverse decay length, $L_{\mathrm{xy}}$, for signal di-$J/\psi$ events is used to extract the PP distributions described in Section~\ref{sec:NP}.  The dependence of the PP model on the resolution function is tested using a triple-Gaussian function and assigning the difference from the default model as a systematic uncertainty.  The systematic uncertainty is larger in the central rapidity region due to the smaller fraction of non-prompt events in the forward region from the event selection requirements.  This allowed for more freedom in the fit of the PP component in the central region and a larger uncertainty from the model.

\textbf{Differential $f_{\mathrm{PP}}$ Correction:} The systematic uncertainty for the correction factor of the differential $f_{\mathrm{PP}}$ bias, described in Section~\ref{sec:NP}, is determined by varying the fit of the bias correction and measuring the difference in the correction factor.  A covariance matrix of the partial derivative of the likelihood function, used to determine the correction, with respect to the free parameters of the fit is used to extract the maximal deviation of the fit of the bias against the measured $f_{\mathrm{PP}}$.  The correction factor is refit with the varied parameters and the difference from the nominal final result is taken as the systematic uncertainty of the correction.

\textbf{Pile-up:} The fit of the $d_{\mathrm{z}}$ distribution is varied by adjusting the pile-up Gaussian width within its uncertainty.   As another test, a constant is used instead of a Gaussian function for the pile-up PDF.  Any bias from the requirement of $|d_{\mathrm{z}}|$~$<$~1.2~mm on PP events is tested by adding a requirement of $|\Delta z_{0}|<\sqrt{2}\times$~1.2~mm for the inclusive $J/\psi$ sample and propagating the change in the prompt PDF to the 2-D fit of the $L_{\mathrm{xy_{1}}}$ against $L_{\mathrm{xy_{2}}}$.  Here, $\Delta z_{0}$ is defined as the difference in the impact parameter of an object with respect to the primary vertex in the $r$--$z$ plane for each muon, and the factor of $\sqrt{2}$ is because the di-$J/\psi$ sample has twice the muons of the inclusive $J/\psi$ sample. 

\textbf{Branching Fraction and Luminosity:} The $J/\psi$ meson to dimuon branching fraction systematic uncertainty is taken from the world-average~\cite{PDG}.  The uncertainty in the integrated luminosity of 1.9$\%$ comes from Ref.~\cite{Lum2} and is propagated through to the cross-section calculation.

\textbf{DPS Model:} The DPS-dominated region of the 2-D template used to create the SPS and DPS event weights, described in Section~\ref{sec:DPSmodel}, is varied in both $\Delta \phi$ and $\Delta y$ to test the dependence of $f_{\mathrm{DPS}}$ on the assumption and definition of the DPS-dominated region.  For $\Delta \phi$ the DPS-dominated region is varied by $\pi/9$, the width of a bin in the DPS-dominated region.  For $\Delta y$, the strictness of the DPS-dominated requirement is only increased to avoid including the SPS tail in the defined DPS-dominated region.  It is increased by a single bin of $\Delta y$ in the DPS-dominated region to 2.4.  In this region, predictions show that SPS is negligible.  The systematic uncertainty due to the contribution from the tail of the SPS distribution extending into the normalisation region has been  determined by reducing the size of the normalisation region.

\textbf{DPS Binning:} As a cross-check, the dependence of $f_{\mathrm{DPS}}$ on the binning of the 2-D template used to create the SPS and DPS event weights is tested.  A finer bin width is used where the distribution falls off more steeply around the NLO SPS peak and the DPS-dominated region is set to avoid the tail of the SPS distribution.  The change in the $f_{\mathrm{DPS}}$ value is well within the uncertainty of $f_{\mathrm{DPS}}$.

\begin{table}[h!tb]
\begin{center}
\caption{The summary of relative systematic uncertainties in the di-$J/\psi$ cross-section in the central and forward rapidity regions of the sub-leading $J/\psi$.  The systematic uncertainties for the branching fraction and luminosity are treated separately.}
\label{tab:SystCross}
\begin{tabular}{l c c }
\hline
\hline
\multicolumn{3}{ c }{Systematic uncertainty: di-$J/\psi$ cross-section $[\%]$} \\
\hline
Source      &    $|y$($J/\psi_{2})|$ $<$ 1.05              &   1.05 $\leq$ $|y$($J/\psi_{2})|$ $<$ 2.1  \\
\hline
\hline
Trigger                       &$\pm$7.5 &$\pm$8.3 \\
Muon reconstruction           &$\pm$1.1 &$\pm$1.3   \\
Kinematic acceptance                    &$\pm$0.4    &$\pm$1.1       \\
Mass model                    &$\pm$0.1 &$\pm$0.1   \\
Mass bias                     &$\pm$0.2 &$\pm$0.2   \\
Prompt--prompt model           &$\pm$0.2    &~~$\pm$0.01       \\
Differential $f_{\mathrm{PP}}$ corr.          &$\pm$0.6    &$\pm$0.3 \\
Pile-up                       &~~$\pm$0.03 &$\pm$0.4   \\
\hline
Total                       &~$\pm$ 7.7     &~$\pm$ 8.5 \\
\hline
\hline
Branching fraction            &  \multicolumn{2}{c} {$\pm$1.1}  \\
Luminosity                    &  \multicolumn{2}{c} {$\pm$1.9}   \\
\hline
\hline
\end{tabular}
\end{center}
\end{table}

\begin{table}[h!tb]
\begin{center}
\caption{The summary of the relative systematic uncertainties for the data-driven $f_{\mathrm{DPS}}$ measurement.  Several systematic uncertainties cancel in the ratio.}
\label{tab:SystfDPS}
\begin{tabular}{l c }
\hline
\hline
\multicolumn{2}{ c }{Systematic uncertainty: $f_{\mathrm{DPS}}$ $[\%]$} \\
\hline
Source      &    Relative uncertainty $[\%]$ \\
\hline
\hline
Trigger                       &$\pm$0.7 \\
Muon reconstruction           &$\pm$0.1   \\
Mass model                    &~~$\pm$0.01   \\
Mass bias                     &~~$\pm$0.02   \\
Prompt--prompt model           &   $\pm$0.1      \\
Differential $f_{\mathrm{PP}}$ corr.          &$\pm$0.1  \\
Pile-up                        &$\pm$0.8   \\
DPS model                    &$\pm$5.6   \\
\hline
\hline
Total                         &~$\pm$ 5.7    \\
\hline
\hline
\end{tabular}
\end{center}
\end{table}

\section{Results}
\label{sec:result}
The PP di-$J/\psi$ and DPS differential cross-sections in the central and forward rapidity regions are measured for the sub-leading $J/\psi$ $p_{\mathrm{T}}$, the di-$J/\psi$ $p_{\mathrm{T}}$, and the di-$J/\psi$ invariant mass corrected for the muon kinematic acceptance.  Also shown are the results over the full $J/\psi$ rapidity range in the muon fiducial volume: the total and DPS cross-sections for the difference in rapidity between the two $J/\psi$ mesons, the azimuthal angle between the two $J/\psi$ mesons, the di-$J/\psi$ invariant mass, and the di-$J/\psi$ $p_{\mathrm{T}}$.  The fraction of DPS events is calculated for each distribution, and the distributions are compared to the LO DPS and NLO* SPS+DPS predictions.   For this comparison, the DPS predictions are normalised to the measured fraction of DPS events and the NLO* SPS predictions are multiplied by the feed-down correction factor from $\psi(\mathrm{2S})$ described in Ref.~\cite{SPSth}, which assumes that feed-down has the same distribution as the NLO* SPS predictions.  For the data, feed-down is part of the PP signal.  Finally, the effective cross-section of DPS is calculated and compared with previous measurements.

\subsection{Cross-section measurement}
\label{sec:sigma}

The fiducial PP cross-section for the region $p_\mathrm{T}$($J/\psi )$~$>$~8.5~GeV,  $|y$($J/\psi )|$~$<$~2.1, $p_{\mathrm{T}}(\mu)$~$\geq$~2.5~GeV and $|\eta(\mu)|$~$<$~2.3 with the two muons from the triggered $J/\psi$ candidate having $p_{\mathrm{T}}(\mu)$~$\geq$~4~GeV is:

{\small $\sigma_{\mathrm Fid}(pp \: \rightarrow \: J/\psi J/\psi \: + \: X ) = \left\{   
     \begin{array}{lr}
     15.6 \pm 1.3 \: (\mathrm{stat}) \pm 1.2 \: (\mathrm{syst}) \: \pm 0.2 \: (\mathrm{BF}) \: \pm 0.3 \: (\mathrm{lumi}) \: \mathrm{pb}, \: \mathrm{for} \: |y| \: < \: 1.05, \\     
     13.5 \pm 1.3 \: (\mathrm{stat}) \pm 1.1 \: (\mathrm{syst}) \: \pm 0.2 \: (\mathrm{BF}) \: \pm 0.3 \: (\mathrm{lumi}) \: \mathrm{pb}, \: \mathrm{for} \: 1.05 \: \leq \: |y| \: < \: 2.1.
     \end{array}
    \right.$
}

The above results are measured in two rapidity regions which are defined in terms of the sub-leading $J/\psi$ meson.  The systematic uncertainties for the branching fraction and luminosity are quoted separately.  The extrapolated cross-section is measured by including the acceptance correction and assuming unpolarised $J/\psi$ production.  This cross-section is measured in the $J/\psi$ fiducial volume $p_{\mathrm{T}}$~$>$~8.5~GeV, $|y|$~$<$~2.1 with no requirement on the kinematics of the muon in the final state.  The total cross-section over the full fiducial $J/\psi$ rapidity is 160~$\pm$~12~(stat)~$\pm$~14~(syst)~$\pm$~2~(BF)~$\pm$~3~(lumi)~pb.  The PP cross-section in the two rapidity regions of the sub-leading $J/\psi$ meson is:

{\small $\sigma(pp \: \rightarrow \: J/\psi J/\psi \: + \: X )= \left\{
     \begin{array}{lr}
     82.2 \pm 8.3 \: (\mathrm{stat}) \pm 6.3 \: (\mathrm{syst}) \: \pm 0.9 \: (\mathrm{BF}) \: \pm 1.6 \: (\mathrm{lumi}) \: \mathrm{pb}, \: \mathrm{for} \: |y| \: < \: 1.05, \\
     78.3 \pm 9.2 \: (\mathrm{stat}) \pm 6.6 \: (\mathrm{syst}) \: \pm 0.9 \: (\mathrm{BF}) \: \pm 1.5 \: (\mathrm{lumi}) \: \mathrm{pb}, \: \mathrm{for} \: 1.05 \: \leq \: |y| \: < \: 2.1.
     \end{array}
    \right.$
}

\begin{figure}[t]
\begin{center}
\subfloat[]{
\includegraphics*[width=0.45\textwidth]{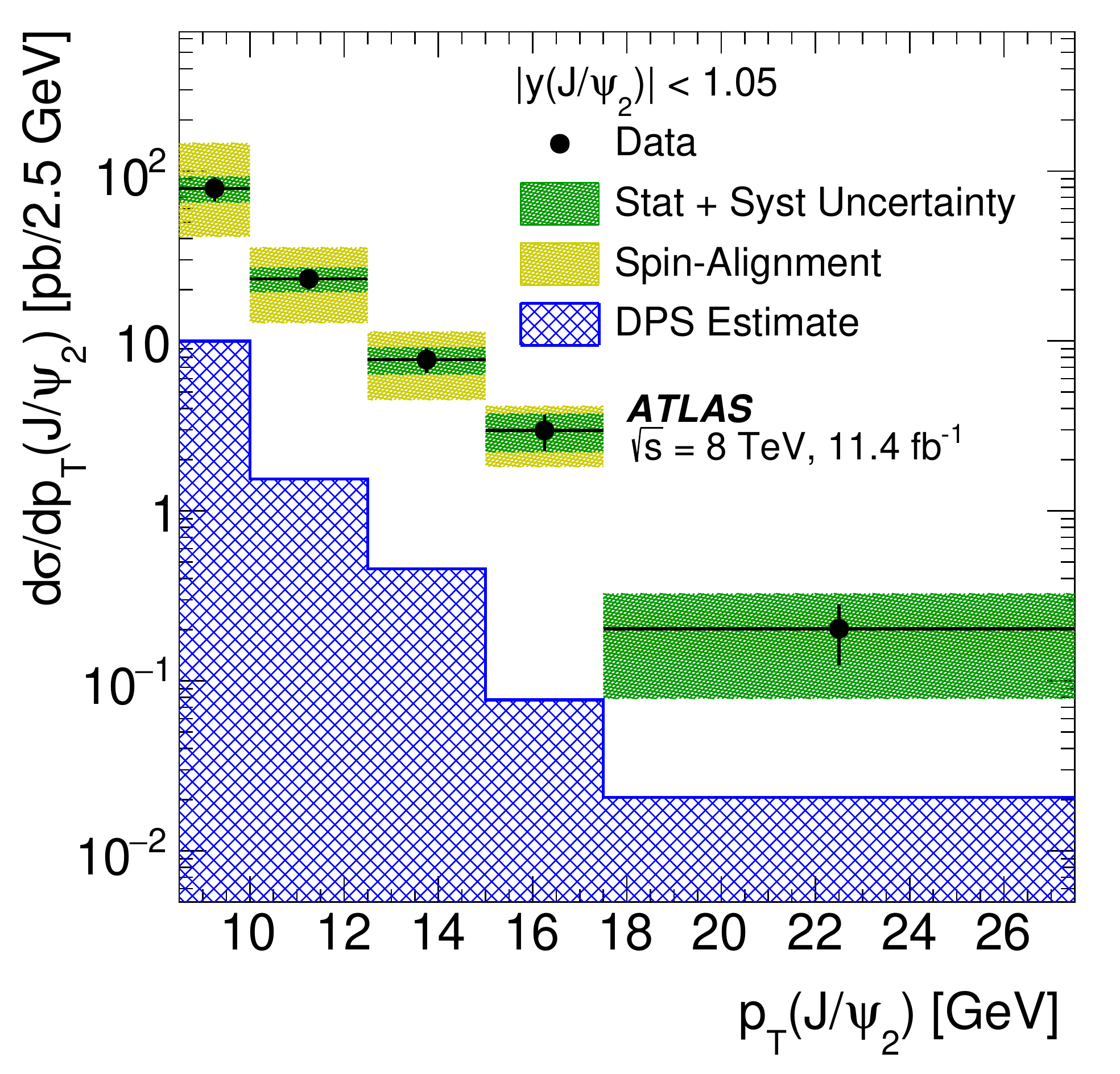}
\label{fig:c1}
}
\subfloat[]{
\includegraphics*[width=0.45\textwidth]{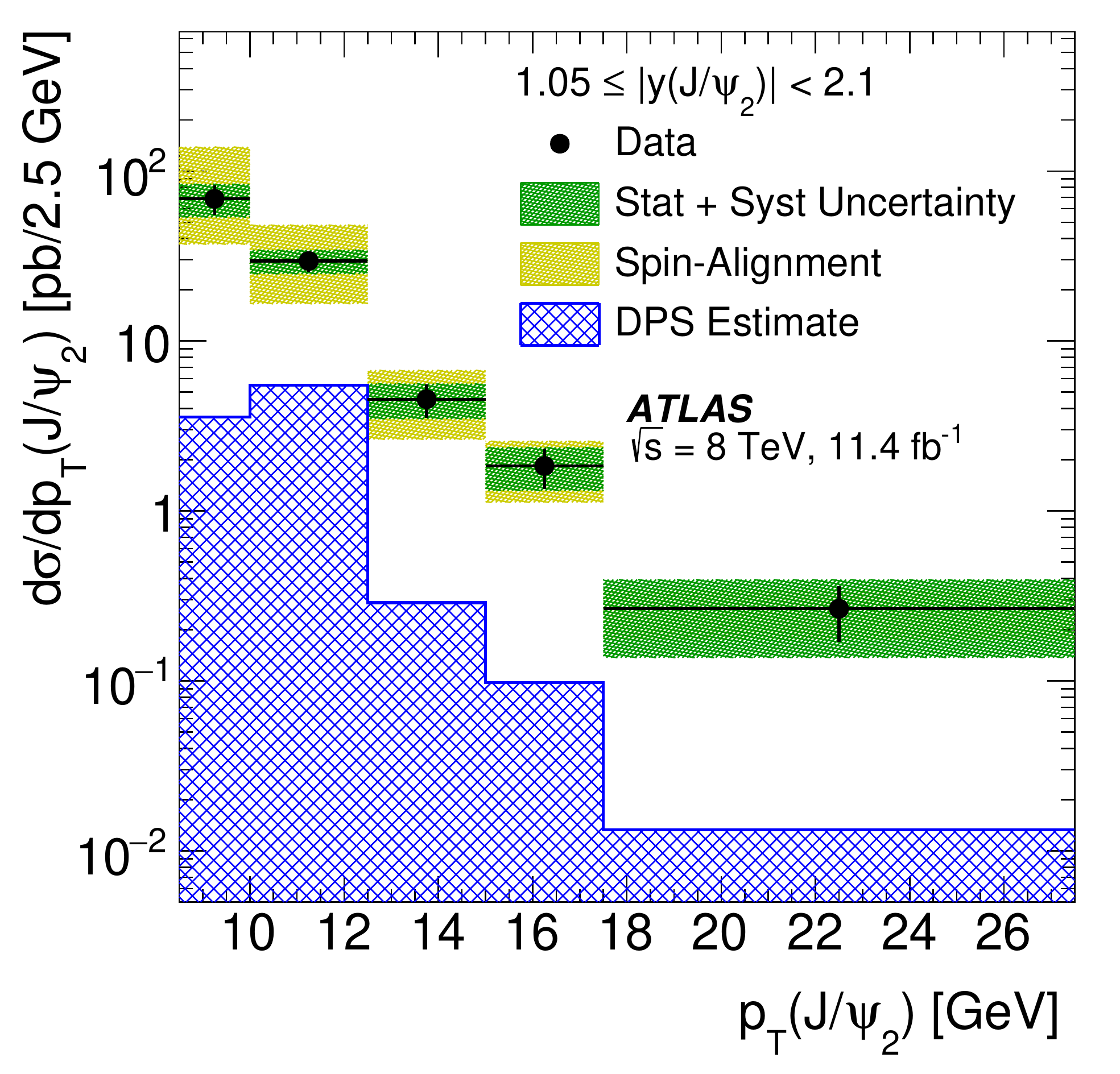}
\label{fig:c2}
}
\caption{The differential cross-section, d$\sigma$/d$p_{\mathrm{T}}(J/\psi_{2})$, in the~\protect\subref{fig:c1} central and~\protect\subref{fig:c2} forward rapidity regions.  The variation due to the choice of $J/\psi$ spin-alignment is shown separately.  The longitudinal polarisation is found to minimise the di-$J/\psi$ cross-section and the positive transverse polarisation is found to maximise the di-$J/\psi$ cross-section.  Also shown is the data-driven DPS distribution derived with the method described in the text.  It is assumed that the DPS weights created within the muon kinematic acceptance can be applied to the acceptance-corrected distributions.}
\label{fig:CrossSyst}
\end{center}
\end{figure}

The differential cross-sections as a function of the sub-leading $J/\psi$ $p_{\mathrm{T}}$ are shown for the central and forward rapidity regions in Figure~\ref{fig:CrossSyst}.  The DPS-weighted distribution created using the data-driven method within the muon kinematic acceptance, which is described in Section~\ref{sec:DPSmodel}, is weighted by the acceptance correction to get the inclusive cross-section and is included in the figure.  It is assumed that the DPS weights created within the muon kinematic acceptance can be applied to the acceptance-corrected distributions.  An in-depth discussion of the DPS-weighted distribution is given in Section~\ref{sec:DPSresults}.

The cross-section results are reported under the assumption of unpolarised $J/\psi$ mesons as the $J/\psi$ polarisation coefficients are close to zero~\cite{PolarCDF, PolarLHCb, PolarCMS}.  As an additional test, the variation of the cross-section has been determined for four extreme cases of $J/\psi$ spin-alignment, one with full longitudinal polarisation and three with different transverse polarisations.  Both $J/\psi$ candidates are assumed to have the same polarisation.  These are maximal polarisations compared to the small possible polarisation at the low-$p_{\mathrm{T}}$ range studied in this analysis.  The maximum deviations from the unpolarised case are given in Table~\ref{tab:SpinTest} and Table~\ref{tab:SpinTestDPS} for the total and DPS di-$J/\psi$ cross-sections, respectively.  The differential variations due to the maximal polarisation scenarios are included separately in the figures for the differential di-$J/\psi$ cross-section.
 
\begin{table}[h!tb]
\begin{center}
\caption{The maximum variation of the di-$J/\psi$ cross-section determined for four extreme cases of $J/\psi$ spin-alignment of maximal polarisation, one with full longitudinal polarisation and three with different full transverse polarisations, relative to the nominal unpolarised assumption.  Both $J/\psi$ candidates are assumed to have the same polarisation.}
\label{tab:SpinTest}
\begin{tabular}{l c c}
\hline
\hline
\multicolumn{3}{ c }{Maximum spin-alignment scenarios: di-$J/\psi$ cross-section} \\
\hline
Scenario      & $|y$($J/\psi_{2})|$ $\leq$ 1.05   & 1.05 $\leq$ $|y$($J/\psi_{2})|$ $<$ 2.1 \\
\hline
\hline
Longitudinal                  &-47$\%$ &-45$\%$ \\
Transverse positive           &+68$\%$ &+82$\%$ \\
Transverse negative           &+39$\%$ &+28$\%$ \\
Transverse zero               &+51$\%$ &+47$\%$ \\
\hline
\hline
\end{tabular}
\end{center}
\end{table}

\begin{table}[h!tb]
\begin{center}
\caption{The maximum variation of the DPS di-$J/\psi$ cross-section determined for four extreme cases of $J/\psi$ spin-alignment of maximal polarisation, one with full longitudinal polarisation and three with different full transverse polarisations, relative to the nominal unpolarised assumption.  Both $J/\psi$ candidates are assumed to have the same polarisation.}
\label{tab:SpinTestDPS}
\begin{tabular}{l c c}
\hline
\hline
\multicolumn{3}{ c }{Maximum spin-alignment scenarios: di-$J/\psi$ DPS cross-section} \\
\hline
Scenario      & $|y$($J/\psi_{2})|$ $\leq$ 1.05   & 1.05 $\leq$ $|y$($J/\psi_{2})|$ $<$ 2.1 \\
\hline
\hline
Longitudinal                  &-47$\%$ &-45$\%$ \\
Transverse positive           &+79$\%$ &+65$\%$ \\
Transverse negative           &+35$\%$ &+35$\%$ \\
Transverse zero               &+54$\%$ &+48$\%$ \\
\hline
\hline
\end{tabular}
\end{center}
\end{table}

\begin{figure}[ht!b]
\begin{center}
\subfloat[]{
\includegraphics*[width=0.45\textwidth]{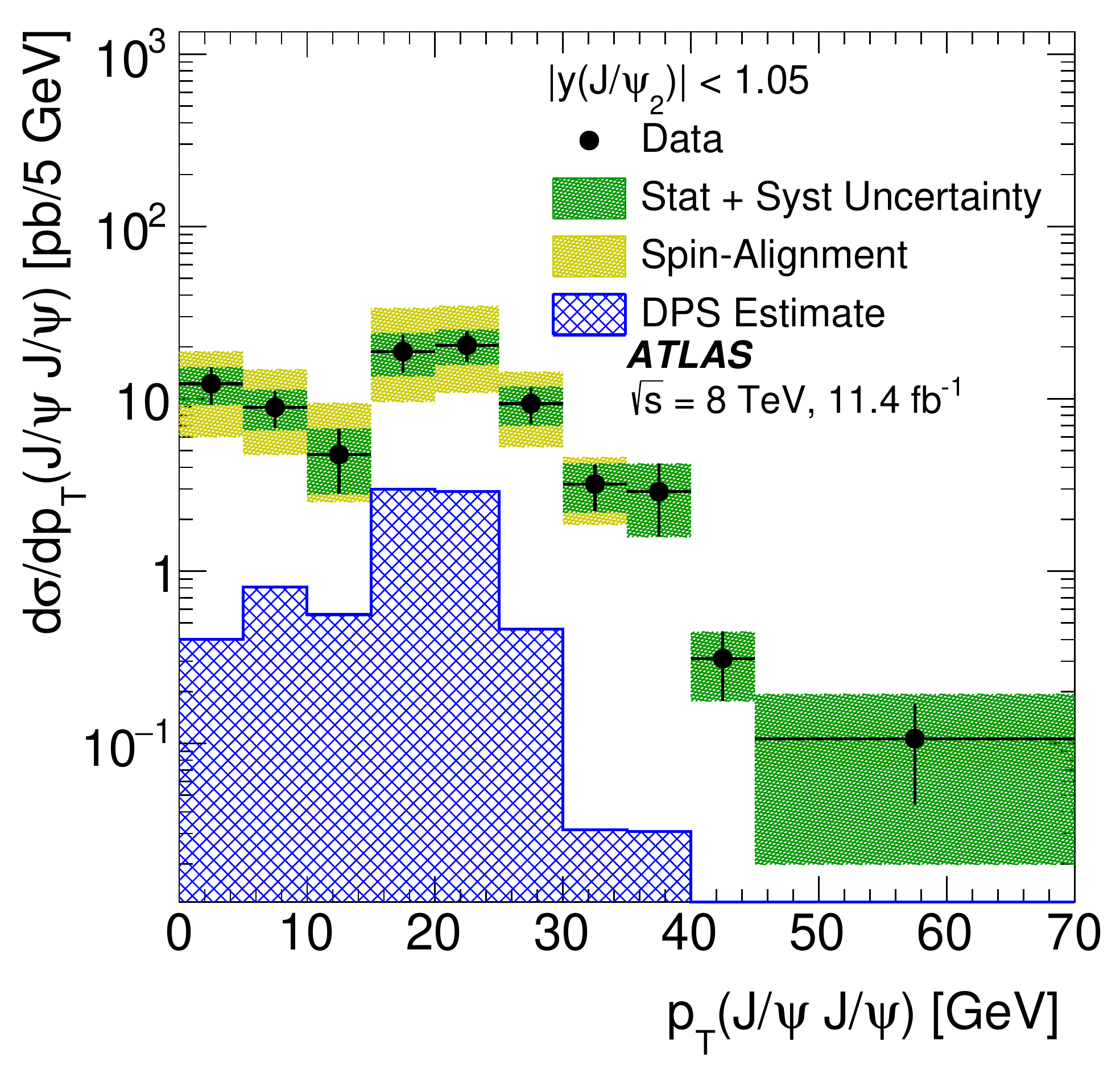}
\label{fig:cp1}
}
\subfloat[]{
\includegraphics*[width=0.45\textwidth]{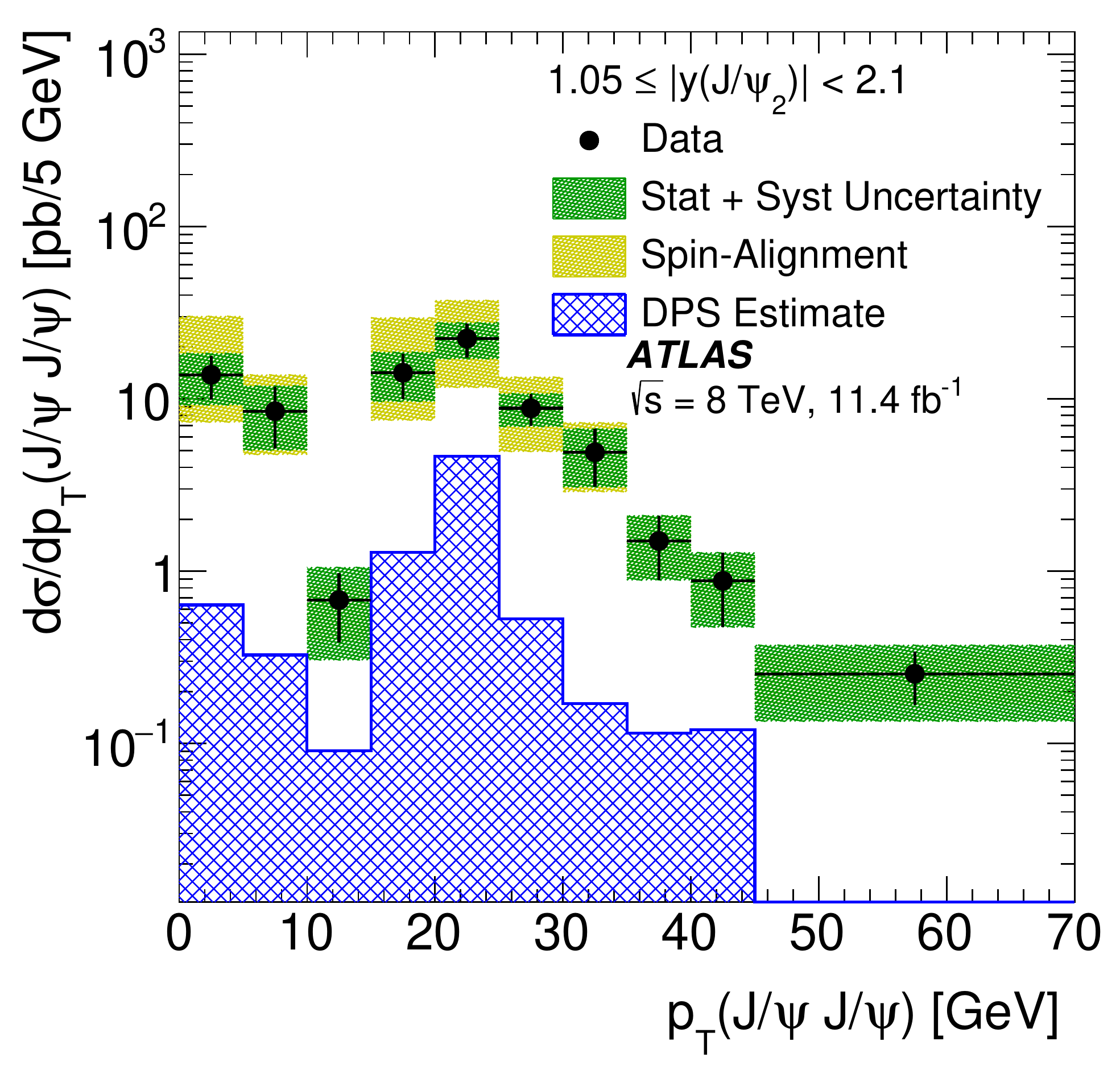}
\label{fig:cp2}
}
\caption{The differential cross-section, d$\sigma$/d$p_{\mathrm{T}}(J/\psi J/\psi)$, in the~\protect\subref{fig:cp1} central and~\protect\subref{fig:cp2} forward rapidity.  The variation due to the choice of $J/\psi$ spin-alignment is shown separately.  The longitudinal polarisation is found to minimise the di-$J/\psi$ cross-section and the positive transverse polarisation is found to maximise the di-$J/\psi$ cross-section.  Also shown is the data-driven DPS distribution derived with the method described in the text.  It is assumed that the DPS weights created within the muon kinematic acceptance can be applied to the acceptance-corrected distributions.  The two peaks at low and high $p_{\mathrm{T}}$ are due to the away and towards event topologies respectively.  The separation is due to the requirement that each $J/\psi$ have $p_{\mathrm{T}}$~$>$~8.5~GeV.}
\label{fig:CrossSyst_DiJpT}
\end{center}
\end{figure}

\begin{figure}[t]
\begin{center}
\subfloat[]{
\includegraphics*[width=0.45\textwidth]{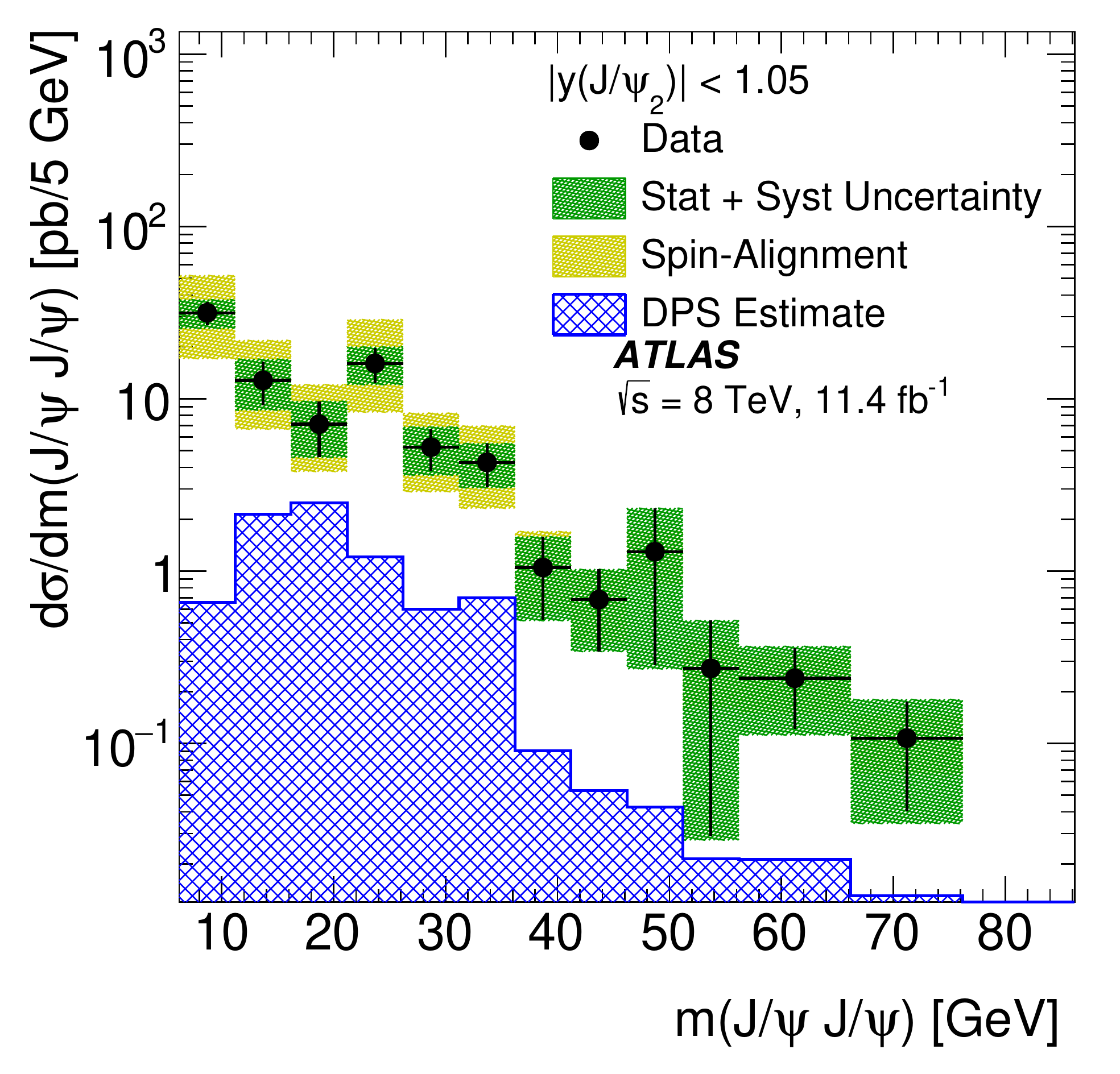}
\label{fig:cm1}
}
\subfloat[]{
\includegraphics*[width=0.45\textwidth]{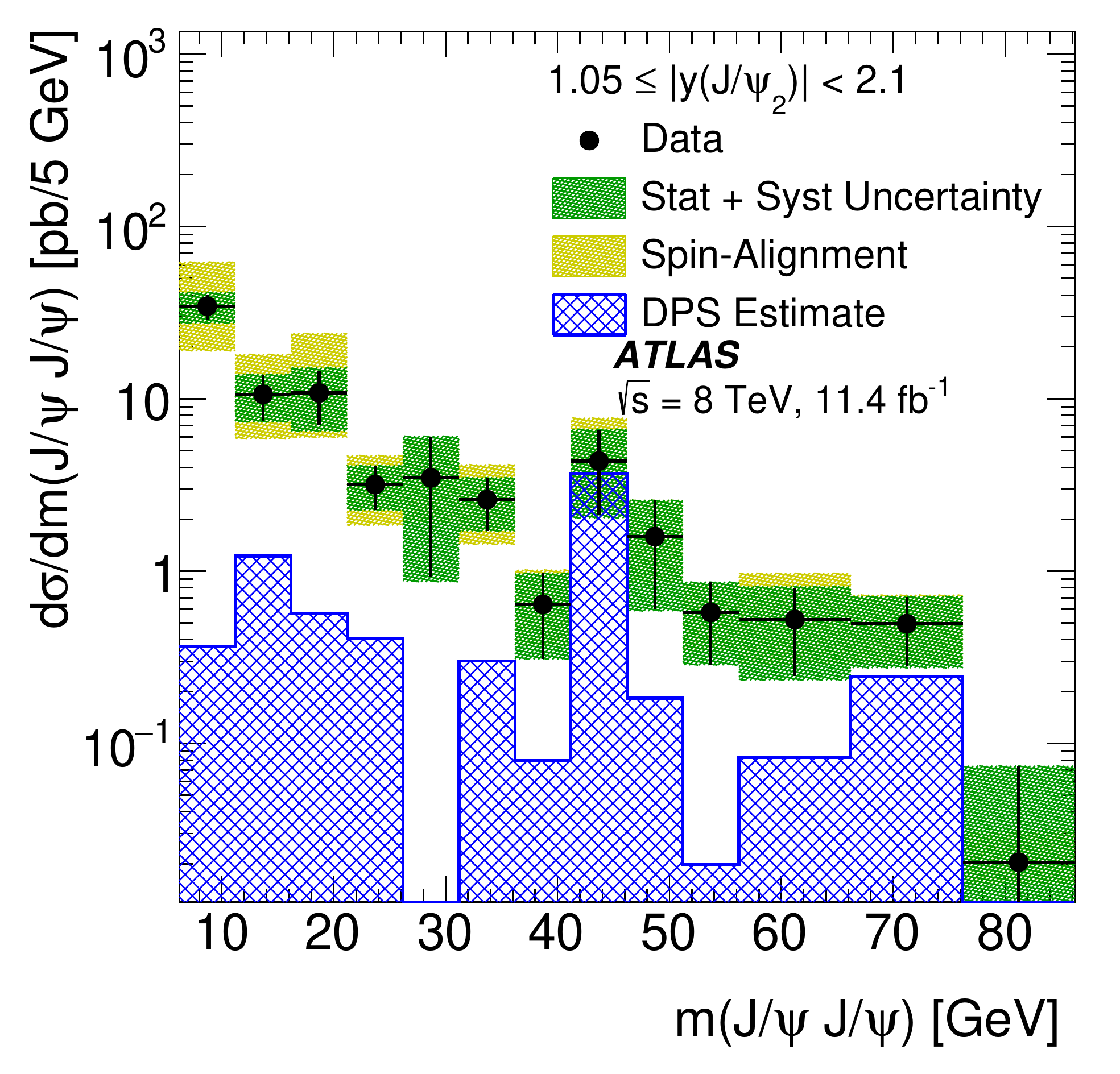}
\label{fig:cm2}
}
\caption{The differential cross-section, d$\sigma$/dm$(J/\psi J/\psi)$, in the~\protect\subref{fig:cm1} central and~\protect\subref{fig:cm2} forward rapidity regions.  The variation due to the choice of $J/\psi$ spin-alignment is shown separately.  The longitudinal polarisation is found to minimise the di-$J/\psi$ cross-section and the positive transverse polarisation is found to maximise the di-$J/\psi$ cross-section.  Also shown is the data-driven DPS distribution derived with the method described in the text.  It is assumed that the DPS weights created within the muon kinematic acceptance can be applied to the acceptance-corrected distributions.}
\label{fig:CrossSyst_DiJM}
\end{center}
\end{figure}

Additionally, the cross-section in the two rapidity regions is measured as a function of the di-$J/\psi$ transverse momentum as well as the invariant mass.  The differential cross-sections as a function of $p_{\mathrm{T}}$($J/\psi$ $J/\psi$) and $m$($J/\psi$ $J/\psi$) are shown in Figure~\ref{fig:CrossSyst_DiJpT} and Figure~\ref{fig:CrossSyst_DiJM} respectively, along with the DPS-weighted distribution.  There are two peaks in the di-$J/\psi$ $p_{\mathrm{T}}$ distribution.  The peak near zero is due to events in which the $J/\psi$ are produced back-to-back in an \textit{away} topology and the peak at higher $p_{\mathrm{T}}$ is due to events that have a \textit{towards} topology in which the two $J/\psi$ are produced in the same direction and are back-to-back with respect to an additional gluon.  The large separation is due to the requirement that each $J/\psi$ have $p_{\mathrm{T}} > $~8.5~GeV.

\subsection{Double parton scattering measurement}
\label{sec:DPSresults}
\begin{figure}[p]
\begin{center}
\begin{tabular}{cc}
\subfloat[]{
\includegraphics*[width=0.45\textwidth]{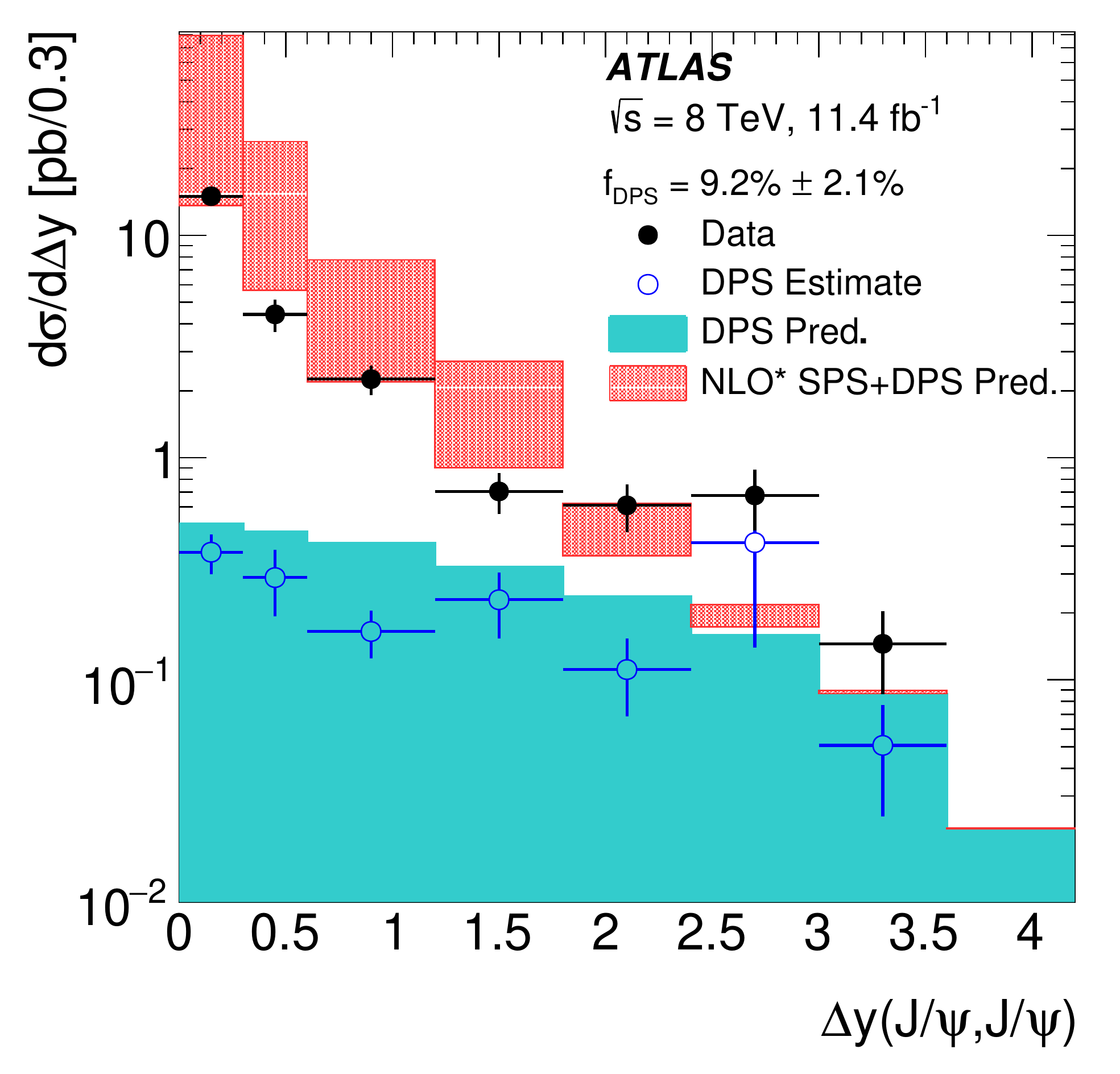}
\label{fig:DPSdy}
}&
\subfloat[]{
\includegraphics*[width=0.45\textwidth]{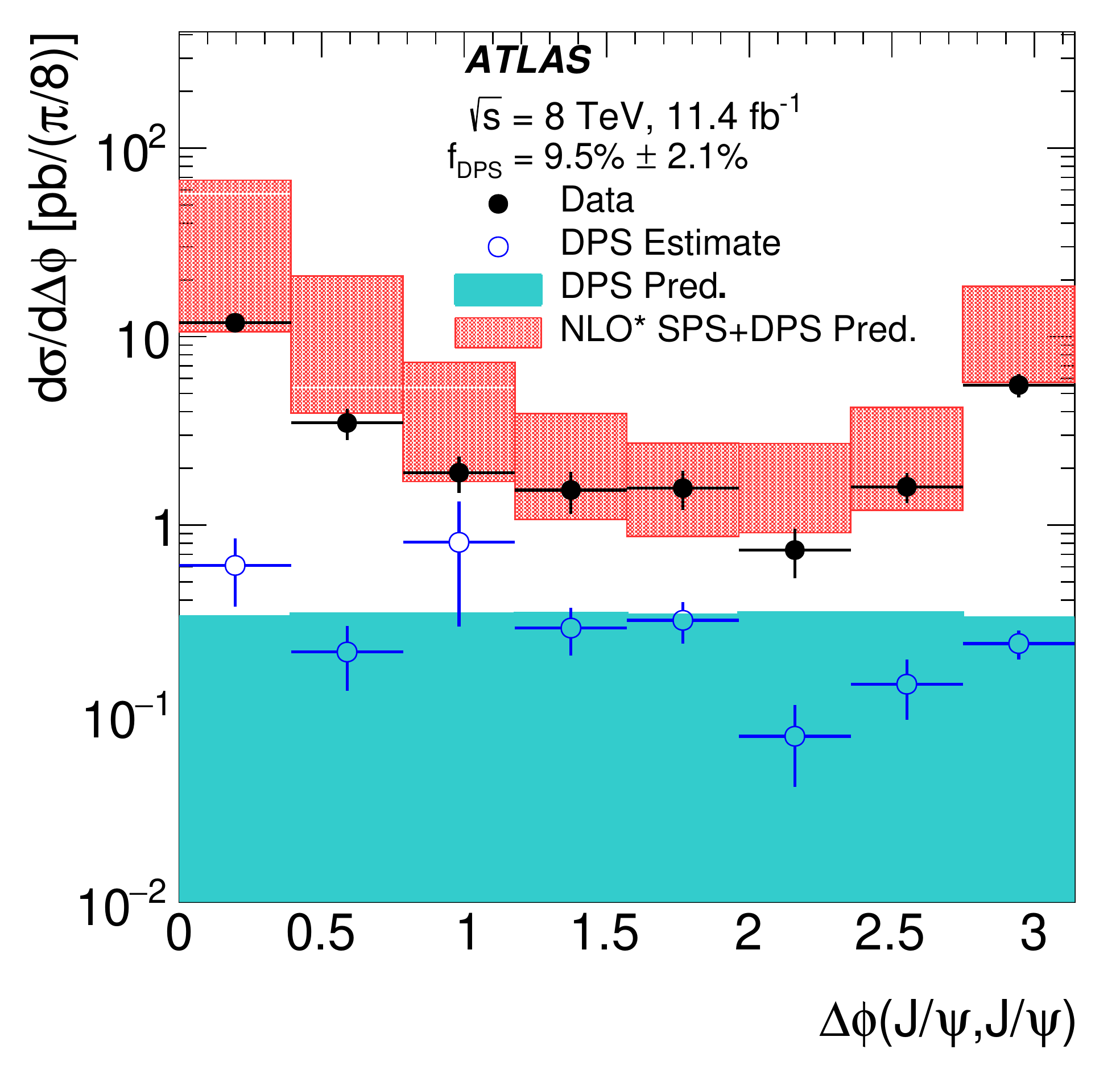}
\label{fig:DPSdphi}
}\\
\subfloat[]{\includegraphics*[width=0.45\textwidth]{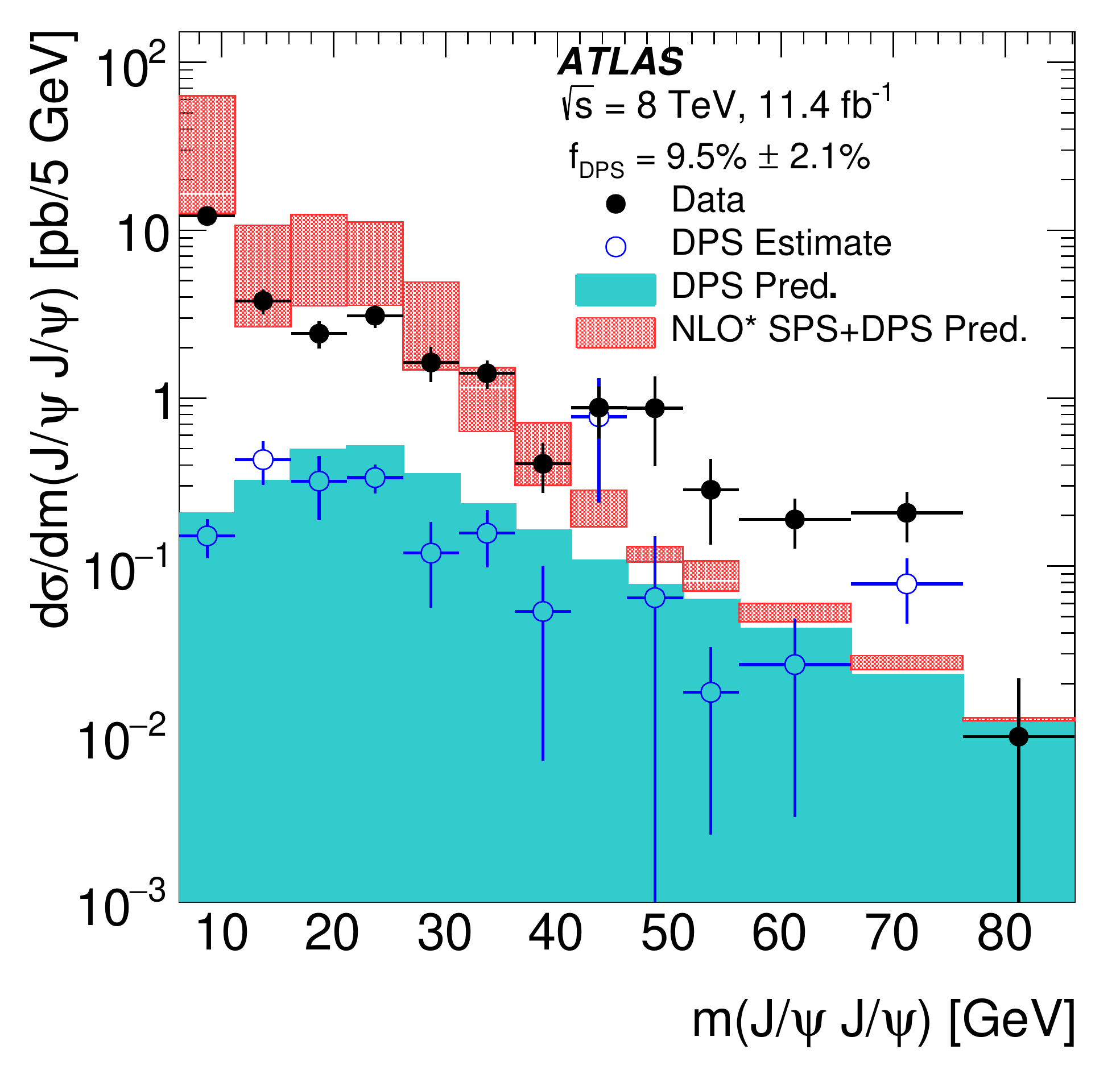}
\label{fig:DPSmass}
}&
\subfloat[]{
\includegraphics*[width=0.45\textwidth]{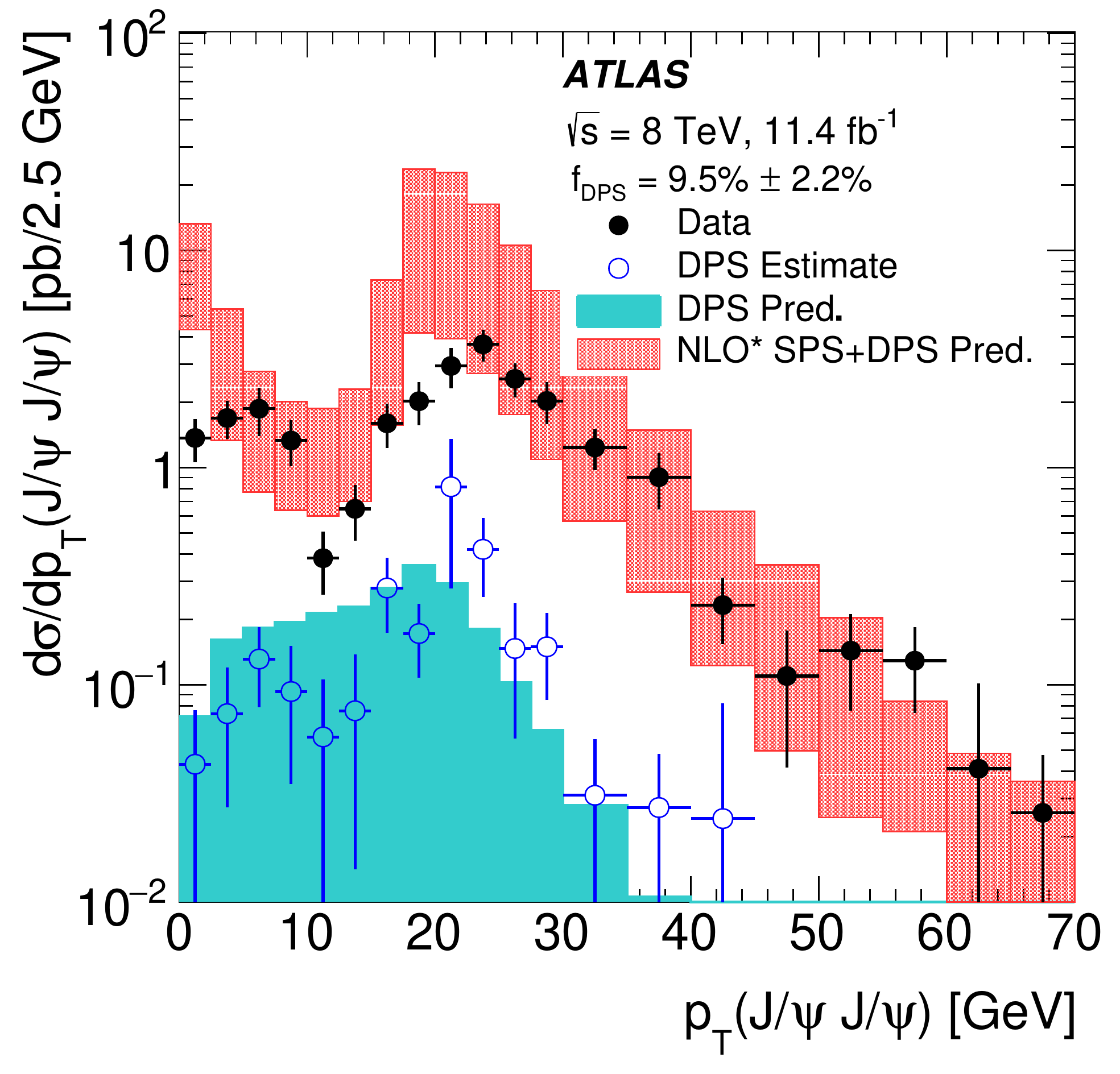}
\label{fig:DPSpt2}
}
\end{tabular}
\caption{The DPS and total differential cross-sections as a function of~\protect\subref{fig:DPSdy} the difference in rapidity between the two $J/\psi$ mesons,~\protect\subref{fig:DPSdphi} the azimuthal angle between the two $J/\psi$ mesons,~\protect\subref{fig:DPSmass} the invariant mass of the di-$J/\psi$,~\protect\subref{fig:DPSpt2} the transverse momentum of the di-$J/\psi$.  Shown are the data as well as the LO DPS~\cite{DPSth2} + NLO* SPS~\cite{SPSth2,SPSth} predictions.  The DPS predictions are normalised to the value of $f_{\mathrm{DPS}}$ found in the data and the NLO* SPS predictions are multiplied by a constant feed-down correction factor.  The data-driven DPS-weighted distribution and the total data distribution are compared to the DPS theory prediction and the total SPS+DPS prediction.}
\label{fig:DPSdist}
\end{center}
\end{figure}

Using the DPS and SPS event weights, which are described in Section~\ref{sec:DPSmodel}, DPS-weighted and SPS-weighted differential distributions are derived.  Due to the limited size of the data set, there are large fluctuations in the acceptance-corrected distributions.  Therefore the muon fiducial volume is used, which does not include the acceptance weight and hence no assumptions about the $J/\psi$ polarisation are made.  Because the SPS and DPS fractions add to unity by construction, only the DPS-weighted distributions are shown in Figure~\ref{fig:DPSdist}.  Since the total distribution and DPS-weighted distribution are shown, the SPS-weighted distribution is understood to be the remainder of the events that are not DPS.

The centre-of-mass energy and fiducial volume requirements of the analysis are applied to the NLO* SPS predictions in Refs.~\cite{SPSth2,SPSth}.  These predictions are generated using HELAC-Onia, which is described in Refs.~\cite{HOnia1,HOnia2}, and used CTEQ6L1 for LO and CTEQ6M~\cite{CTEQ} for NLO* parton distribution functions.  The colour octet contributions and the intrinsic parton transverse momentum are not included in the predictions.  The DPS predictions in Refs.~\cite{SPSth2,SPSth} are based on the models from Refs.~\cite{DPSth,DPSth2}, which assume factorisation of perturbative QCD and use an approximate prompt single-$J/\psi$ matrix element modelled from combined fits of data from multiple experiments.  For comparison, the DPS predictions from Ref.~\cite{DPSth2} are used.  The DPS predictions are created using the MSTW2008 NLO~\cite{MSTW} parton distribution function.  The theory predictions are made in the muon fiducial volume and assume unpolarised $J/\psi$ mesons, and therefore the acceptance correction is not needed for comparison with the predictions.  In Figure~\ref{fig:DPSdist}, the DPS-weighted distribution produced from the event weights of the data-driven method and the total distribution are compared to the LO DPS and sum of the LO DPS and NLO* SPS predictions.  The DPS predictions are normalised to the $f_{\mathrm{DPS}}$ value measured with the data-driven model. For the NLO* SPS predictions, a constant correction factor of 1.85 is applied for feed-down~\cite{SPSth} from $\psi(\mathrm{2S})$ which is present in the data.  Changes in SPS predictions from varying the factorisation and renormalisation scales of perturbative QCD, as well as the mass of the charm quark are assigned as systematic uncertainties.

In each of the plots in Figure~\ref{fig:DPSdist}, the shape of the data-driven DPS distribution approximately agrees with the shape of the DPS predictions.  However, there is disagreement between the total data distribution and the total theory predictions at large $\Delta y$, large invariant mass, and in the low-$p_{\mathrm{T}}$ region that corresponds to di-$J/\psi$ production in an away topology.

The distributions in Figure~\ref{fig:DPSdist} show that a significant fraction of events have a towards topology where the NLO SPS contributions dominate: specifically events in the low-$\Delta \phi$ region of Figure~\ref{fig:DPSdist}\protect\subref{fig:DPSdphi} and the peak of the di-$J/\psi$ $p_{\mathrm{T}}$ distribution in Figure~\ref{fig:DPSdist}\protect\subref{fig:DPSpt2} at around $p_{\mathrm{T}}$~=~22~GeV.  Therefore LO predictions alone, which do not include the towards topology, are not enough to describe PP di-$J/\psi$ production.

Most of the NLO* SPS predictions would appear to require a larger value of $f_{\mathrm{DPS}}$ than the values measured from the data-driven distributions to fit the data.  A possible reason for the discrepancies may be that the NLO* predictions assume a constant correction factor for feed-down from higher-mass charmonium states, which could change the kinematic properties of the SPS distributions.  Requiring $p_{\mathrm{T}}$($J/\psi) >$~8.5~GeV limits feed-down, but a change in the kinematic properties of the feed-down component could lead to a wider SPS tail~\cite{FeedDownDPS,FeedDownDPS2}.  The wide peak at low di-$J/\psi$ $p_{\mathrm{T}}$ could be explained by a large effect due to the inclusion of the intrinsic parton transverse momentum, a non-constant feed-down component, or a combination of the two.

\begin{figure}[ht!b]
\begin{center}
\begin{tabular}{cc}
\subfloat[]{
\includegraphics*[width=0.45\textwidth]{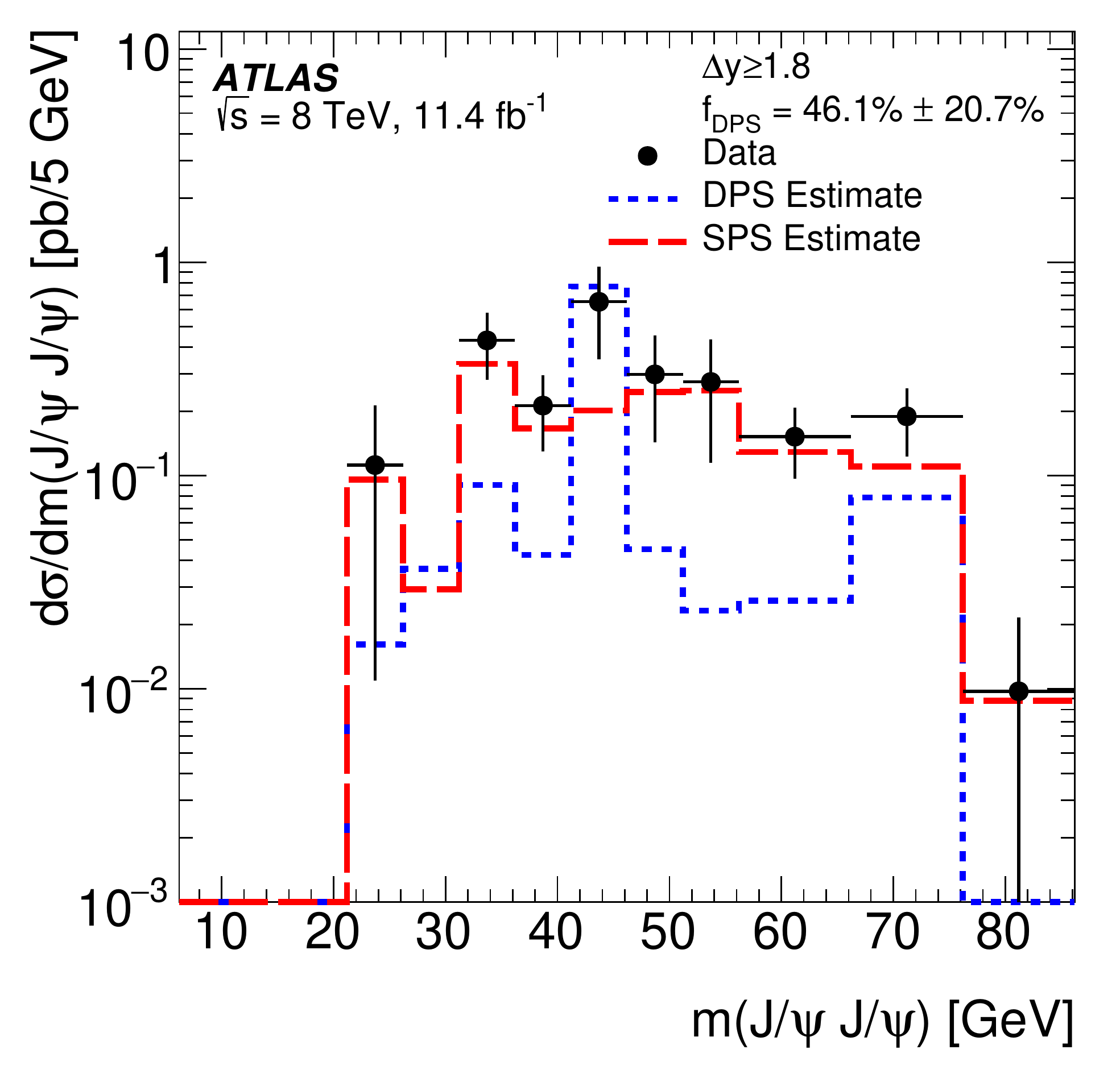}
\label{fig:mdy}
}&
\subfloat[]{
\includegraphics*[width=0.45\textwidth]{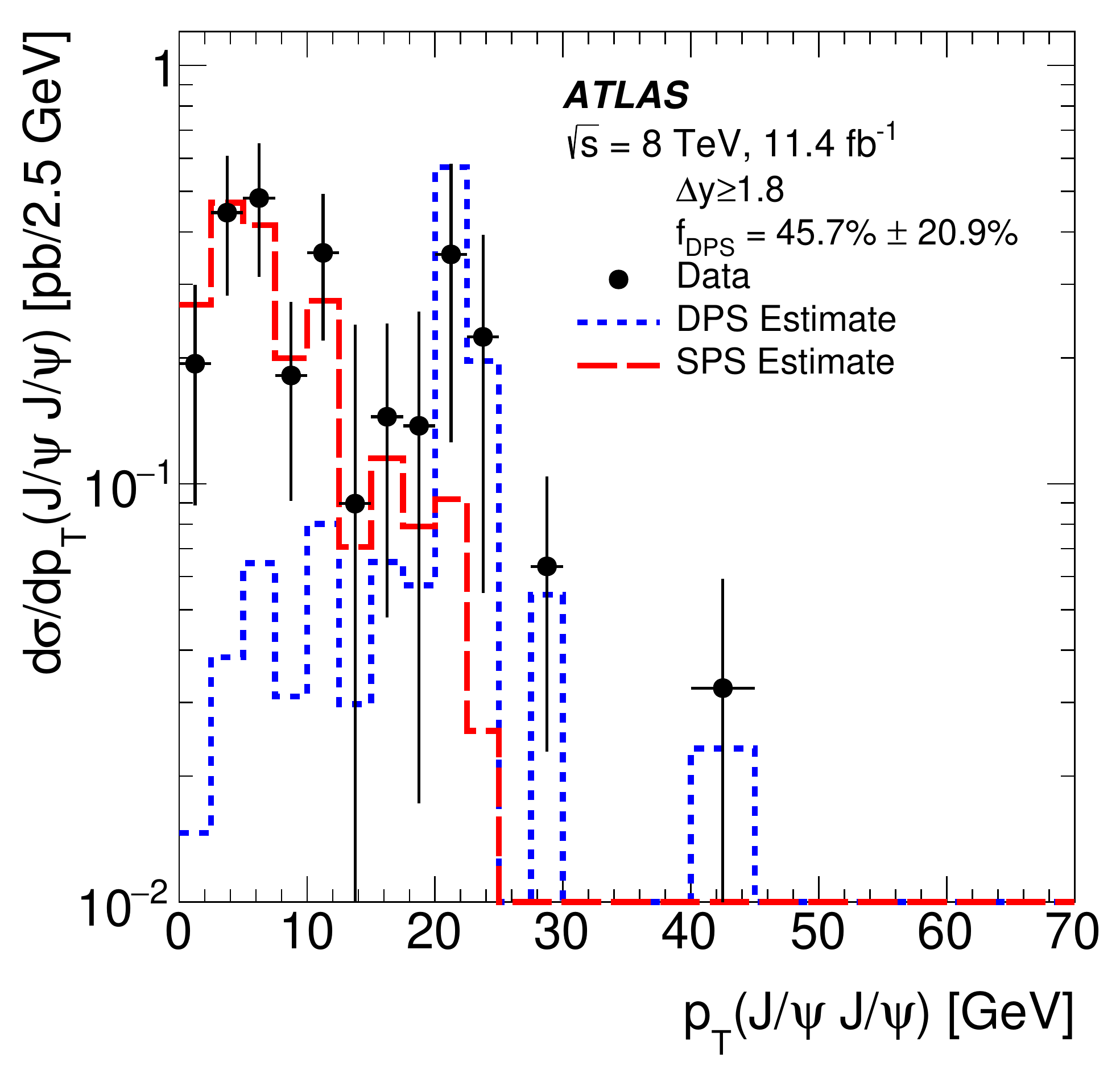}
\label{fig:ptdy}
}\\
\subfloat[]{
\includegraphics*[width=0.45\textwidth]{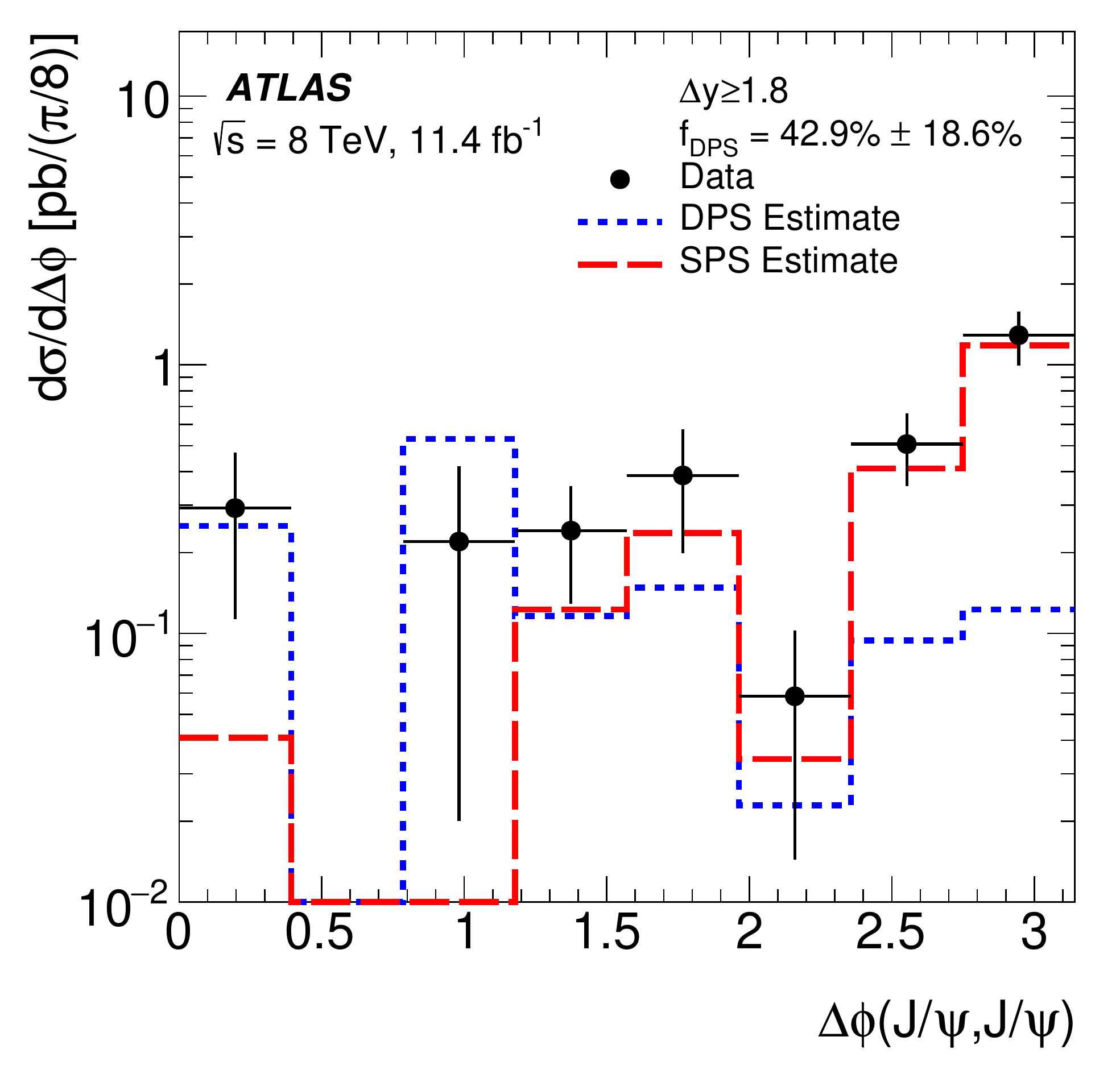}
\label{fig:phdy}
}&
\subfloat[]{
\includegraphics*[width=0.45\textwidth]{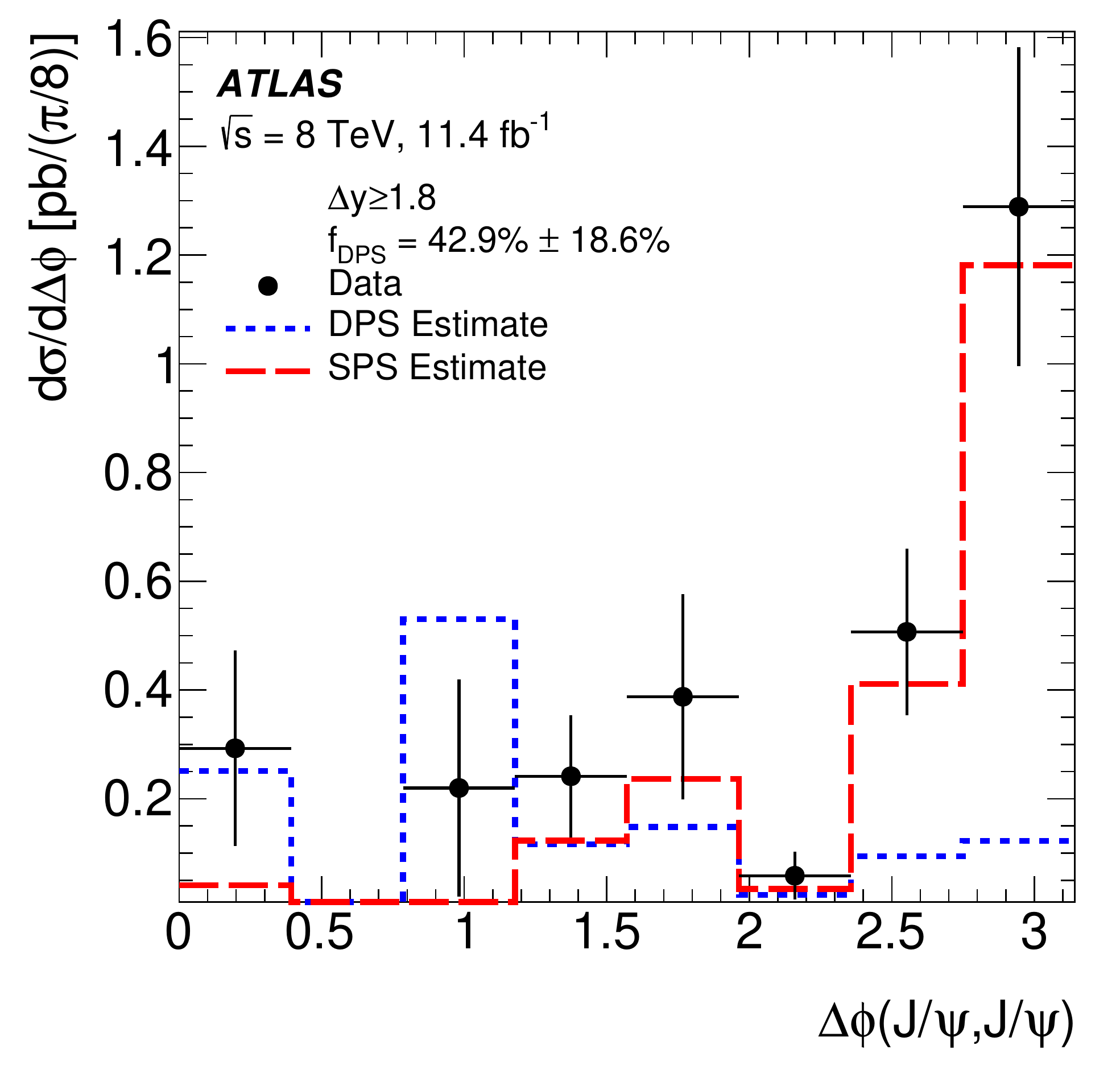}
\label{fig:phdy_lin}
}
\end{tabular}
\caption{The PP, DPS, and SPS total cross-section distributions in the reduced kinematic region of $\Delta y$~$\geq$~1.8 for~\protect\subref{fig:mdy} the di-$J/\psi$ invariant mass,~\protect\subref{fig:ptdy} the di-$J/\psi$ transverse momentum,~\protect\subref{fig:phdy} the azimuthal angle between the two $J/\psi$ mesons and~\protect\subref{fig:phdy_lin} the same distribution as~\protect\subref{fig:phdy} shown with a linear scale on the vertical axis.  The same binning as in Figure~\ref{fig:DPSdist} is used.  Because the SPS and DPS distributions are determined from a data-driven method, the statistics are the same as the data.  Therefore the errors in the SPS and DPS distributions are not included in these figures as they are derived from the data and would obscure the data distribution.}
\label{fig:DPSexcess}
\end{center}
\end{figure}

To study the properties of the discrepancies seen, a requirement of $\Delta y$~$\geq$~1.8 is imposed.  The corresponding distributions are shown in Figure~\ref{fig:DPSexcess}, where for a better comparison the same binning as in Figure~\ref{fig:DPSdist} is used.  Because the SPS and DPS distributions are determined from a data-driven method, the statistics are the same as the data.  For better clarity the errors in the SPS and DPS distributions are not included in these figures.

A comparison of the di-$J/\psi$ invariant mass distribution in Figure~\ref{fig:DPSdist}\protect\subref{fig:DPSmass} with that in Figure~\ref{fig:DPSexcess}\protect\subref{fig:mdy} shows that the events in the region of excess ($\Delta y$~$\geq$~1.8) have large di-$J/\psi$ invariant mass, as expected from the relationship between the invariant mass and the difference in rapidity.  The di-$J/\psi$ transverse momentum, shown in Figure~\ref{fig:DPSexcess}\protect\subref{fig:ptdy}, has an SPS peak near zero and then falls off monotonically while the DPS peaks at a slightly larger $p_{\mathrm{T}}$.  This indicates that the two $J/\psi$ mesons are produced in an away topology.  The $\Delta \phi$ distribution in this region, shown in Figure~\ref{fig:DPSexcess}\protect\subref{fig:phdy}, is not uniform as would be expected for a pure DPS: there is a large SPS peak from the away topology peaked at $\Delta \phi $~=~$\pi$.  To make this comparison easier, the $\Delta \phi$ distribution is also shown on linear vertical scale in Figure~\ref{fig:DPSexcess}\protect\subref{fig:phdy_lin}.  A plausible explanation for the excess of SPS in the distribution is the presence of a non-constant contribution to the di-$J/\psi$ final state from feed-down of back-to-back SPS pair production from excited charmonium states which could change the kinematic properties of the SPS distribution~\cite{FeedDownDPS,FeedDownDPS2}. 

\begin{figure}[ht!b]
\begin{center}
\begin{tabular}{cc}
\subfloat[]{
\includegraphics*[width=0.45\textwidth]{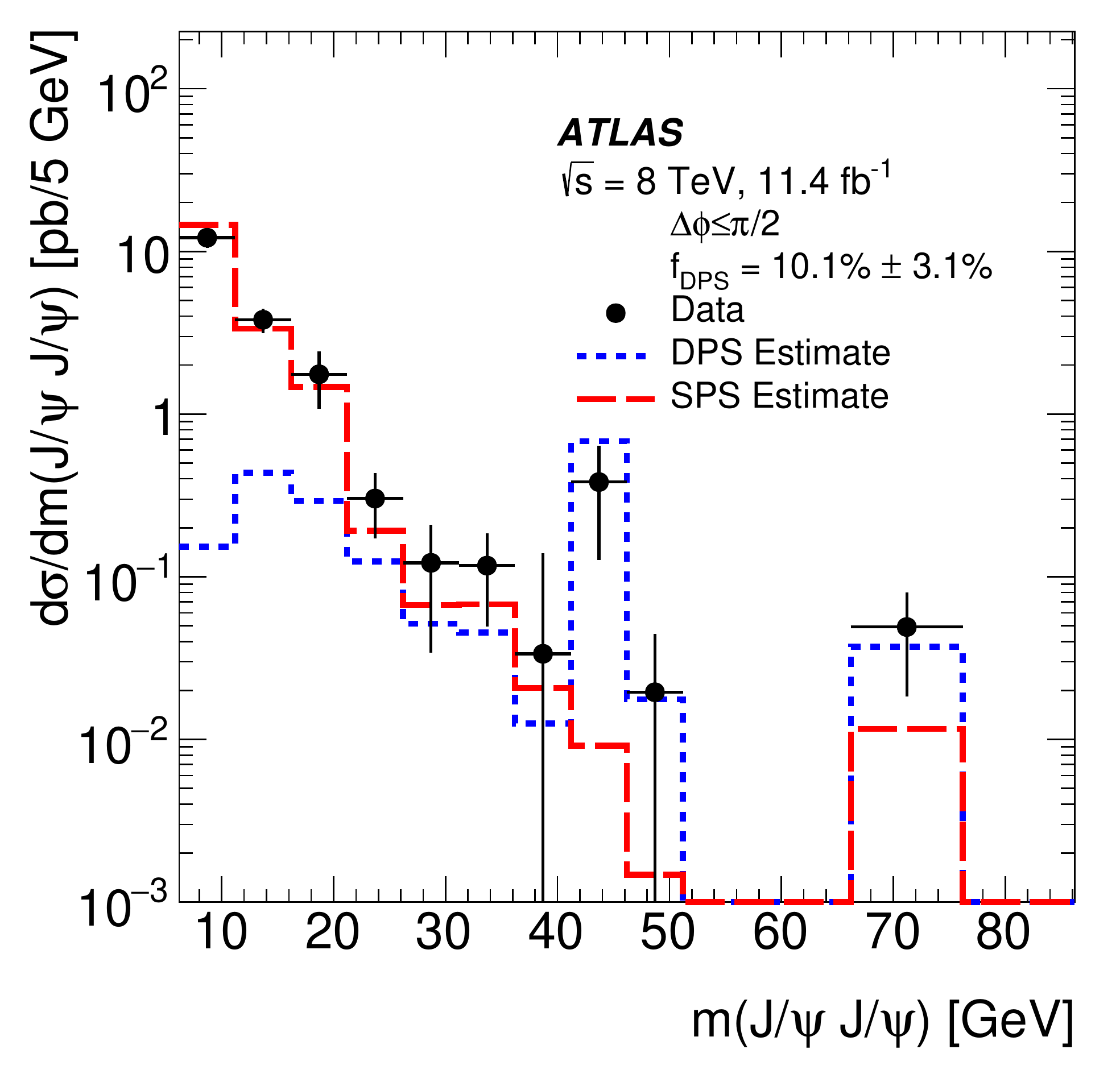}
\label{fig:mdphi}
}&
\subfloat[]{
\includegraphics*[width=0.45\textwidth]{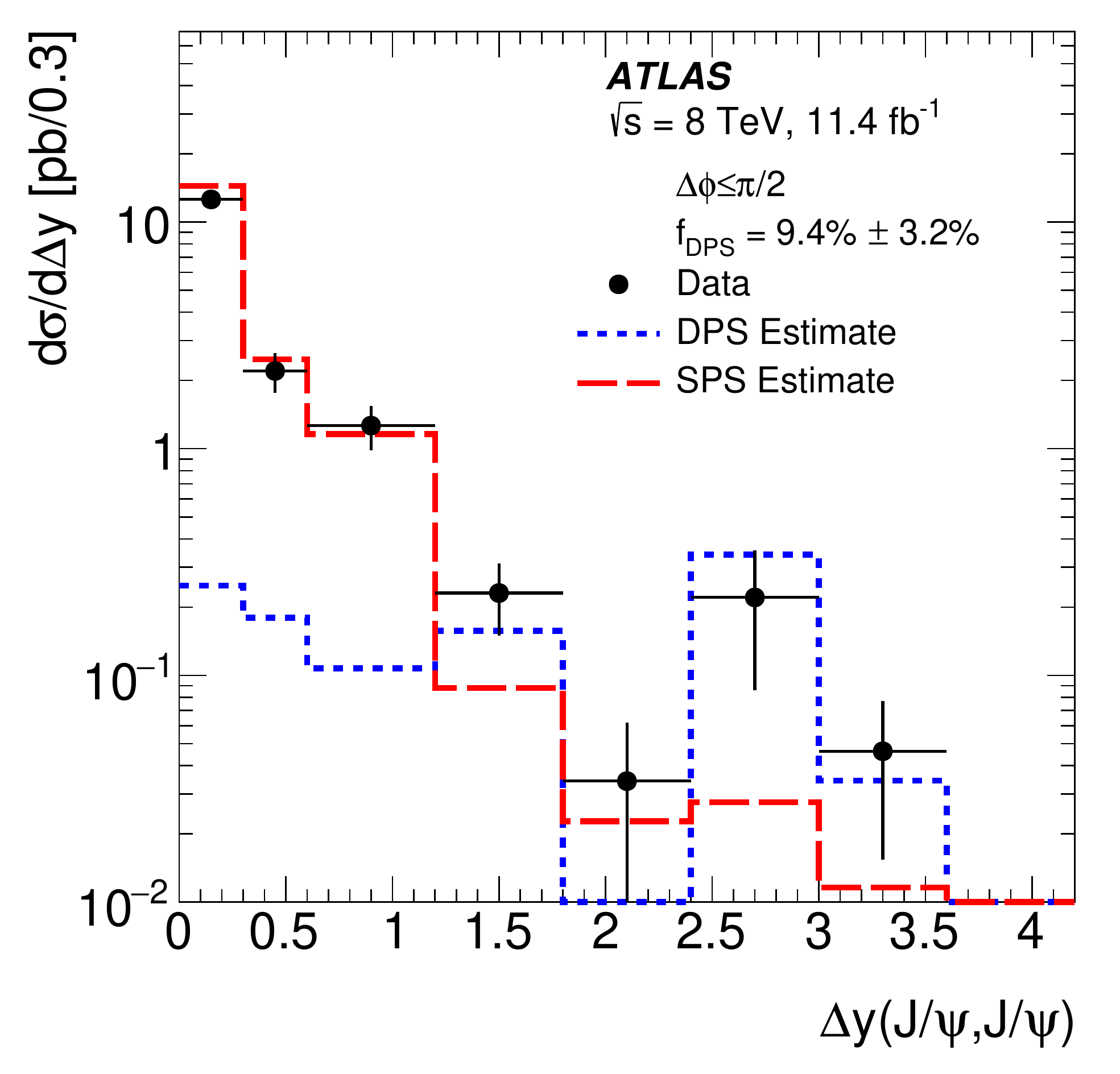}
\label{fig:dydphi}
}
\end{tabular}
\subfloat[]{\includegraphics*[width=0.45\textwidth]{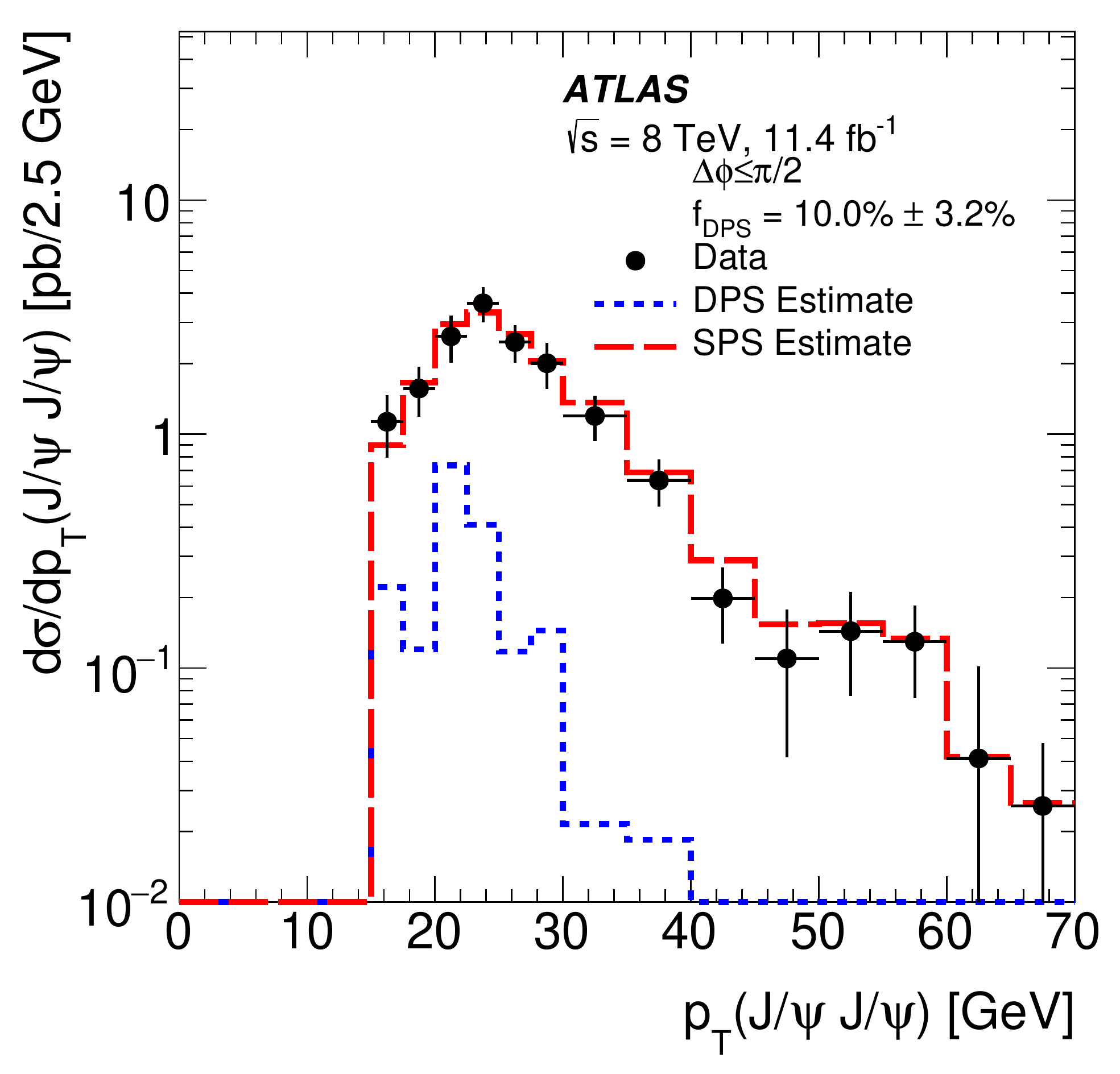}
\label{fig:ptdphi}
}
\caption{The PP, DPS, and SPS total cross-section distributions in the reduced kinematic region of $\Delta \phi$~$\leq$~$\pi$/2 for~\protect\subref{fig:mdphi} the di-$J/\psi$ invariant mass,~\protect\subref{fig:dydphi} the difference in rapidity between the two $J/\psi$ mesons, and~\protect\subref{fig:ptdphi} the di-$J/\psi$ transverse momentum.  The same binning as in Figure~\ref{fig:DPSdist} is used.  Because the SPS and DPS distributions are determined from a data-driven method, the statistics are the same as the data.  Therefore the errors in the SPS and DPS distributions are not included in these figures as they are derived from the data and would obscure the data distribution.}
\label{fig:DPSdphiCut}
\end{center}
\end{figure}

To further understand the relative SPS and DPS composition of events in the normalisation region, the distributions of di-$J/\psi$ invariant mass, $\Delta y$ and di-$J/\psi$ $p_{\mathrm{T}}$ are shown in Figure~\ref{fig:DPSdphiCut}  for events in the kinematic region $\Delta \phi \leq \pi/2$.  There is a clear difference in the shape of the SPS and DPS distributions.  The SPS estimate has a much larger peak at low mass, and the DPS distribution falls off much more quickly as a function of the di-$J/\psi$ $p_{\mathrm{T}}$.  The SPS $\Delta y$ distribution has a large peak near zero and the DPS distribution is flatter.  The different shapes of the distributions, as well as the DPS domination at large $\Delta y$ in this region, further confirms the choice of the normalisation region.

Reference~\cite{SPSth} states that if the data are SPS-dominated, feed-down events should be primarily from LO $\psi$(2S) and $J/\psi$ production and can make up  40$\%$ of the SPS cross-section.  This matches the peaks due to events with an away topology observed in the $\Delta \phi$ and di-$J/\psi$ $p_{\mathrm{T}}$ distributions at large $\Delta y$ in Figure~\ref{fig:DPSexcess}.  Additionally, in Ref.~\cite{SPSth} the $\Delta y$ and di-$J/\psi$ mass from the CMS di-$J/\psi$ measurement~\cite{CMScross} are fit with DPS and NLO* SPS predictions.  The CMS data also show an excess at large $\Delta y$ and di-$J/\psi$ mass.  In Ref.~\cite{FeedDownDPS} a comparison of the predicted NLO* SPS, feed-down, and DPS distributions is shown.  The predicted di-$J/\psi$ mass and $\Delta y$ distributions from feed-down have a wider tail than the NLO* SPS distributions, similar to what is observed in the DPS distributions.  The increase at large di-$J/\psi$ mass and $\Delta y$ is also predicted in Ref.~\cite{FeedDownDPS2}, which studied all possible Fock state contributions to the SPS production of prompt $J/\psi$ meson pairs.  A significant non-constant contribution from feed-down is a possible explanation of the much quicker drop off of the tail in the NLO* SPS predictions relative to the data-driven distribution, seen in Figures~\ref{fig:DPSdy}~--~\ref{fig:DPSpt2}.  Because feed-down can have a distribution similar to DPS, it can explain why the predictions seem to require a larger $f_{\mathrm{DPS}}$ value than measured by the data-driven distribution.  A larger $f_{\mathrm{DPS}}$ would not explain the peak at $\Delta \phi \approx \pi$ for $\Delta y$~$\geq$~1.8, which can be explained by SPS from a non-constant feed-down contribution.  Additionally, the wide peak seen at low di-$J/\psi$ $p_{\mathrm{T}}$ can be explained either by a large effect due to the inclusion of the intrinsic parton transverse momentum, smearing due to a non-constant feed-down component, or a combination of the two. 

\subsubsection{Effective cross-section measurement}

With the measured inclusive di-$J/\psi$ cross-section and the fraction of DPS events as well as the prompt $J/\psi$ cross-section in the corresponding fiducial volume, the effective cross-section can be derived using Eq.~(\ref{eq:sigEffDiJpsi}).  The prompt $J/\psi$ differential cross-section is obtained from measurements in Ref.~\cite{Single} by integrating over $p_{\mathrm{T}}$ and extrapolating to the rapidity acceptance region of this analysis ($|y$($J/\psi)|$~$\leq$~2.1 cf. $|y$($J/\psi)|$~$\leq$~2.0 in Ref.~\cite{Single}). The extrapolation uses a linear fit to the cross-section as a function of the absolute rapidity. The statistical and systematic uncertainties are scaled to keep the relative uncertainties the same before and after extrapolation. The cross-section in the fiducial volume of this analysis is $\sigma_{J/\psi}$~=~429.8~$\pm$~0.1~(stat)~$\pm$~38.6~(syst)~nb.

The value of $f_{\mathrm{DPS}}$ is taken from the $\Delta y$ distribution since it has a well-known DPS distribution, and the other distributions are used as a cross-check.  Using the $\Delta y$ distribution the fraction is measured to be:

$f_{\mathrm{DPS}}$~=~(9.2~$\pm$~2.1~(stat)~$\pm$~0.5~(syst))$\%$.

The DPS cross-section, corrected for the muon acceptance in the full $J/\psi$ rapidity range is measured to be:

$\sigma_{\mathrm{DPS}}^{J/\psi,J/\psi}$~=~14.8~$\pm$~3.5~(stat)~$\pm$~1.5~(syst)~$\pm$~0.2~(BF)~$\pm$~0.3~(lumi)~pb.

A small difference is found between the DPS cross-section measured in the inclusive volume and the cross-section extrapolated from the fiducial volume.  This difference is introduced by fluctuations in the DPS distributions from the acceptance weight which is used to extrapolate to the inclusive volume, and is smaller than the statistical error.  The effective cross-section obtained from these inputs is measured to be:

$\sigma_{\mathrm{eff}}$~=~6.3~$\pm$~1.6~(stat)~$\pm$~1.0~(syst)~$\pm$~0.1~(BF)~$\pm$~0.1~(lumi)~mb.

\begin{figure}[t]
\begin{center}
\includegraphics*[width=0.7\textwidth]{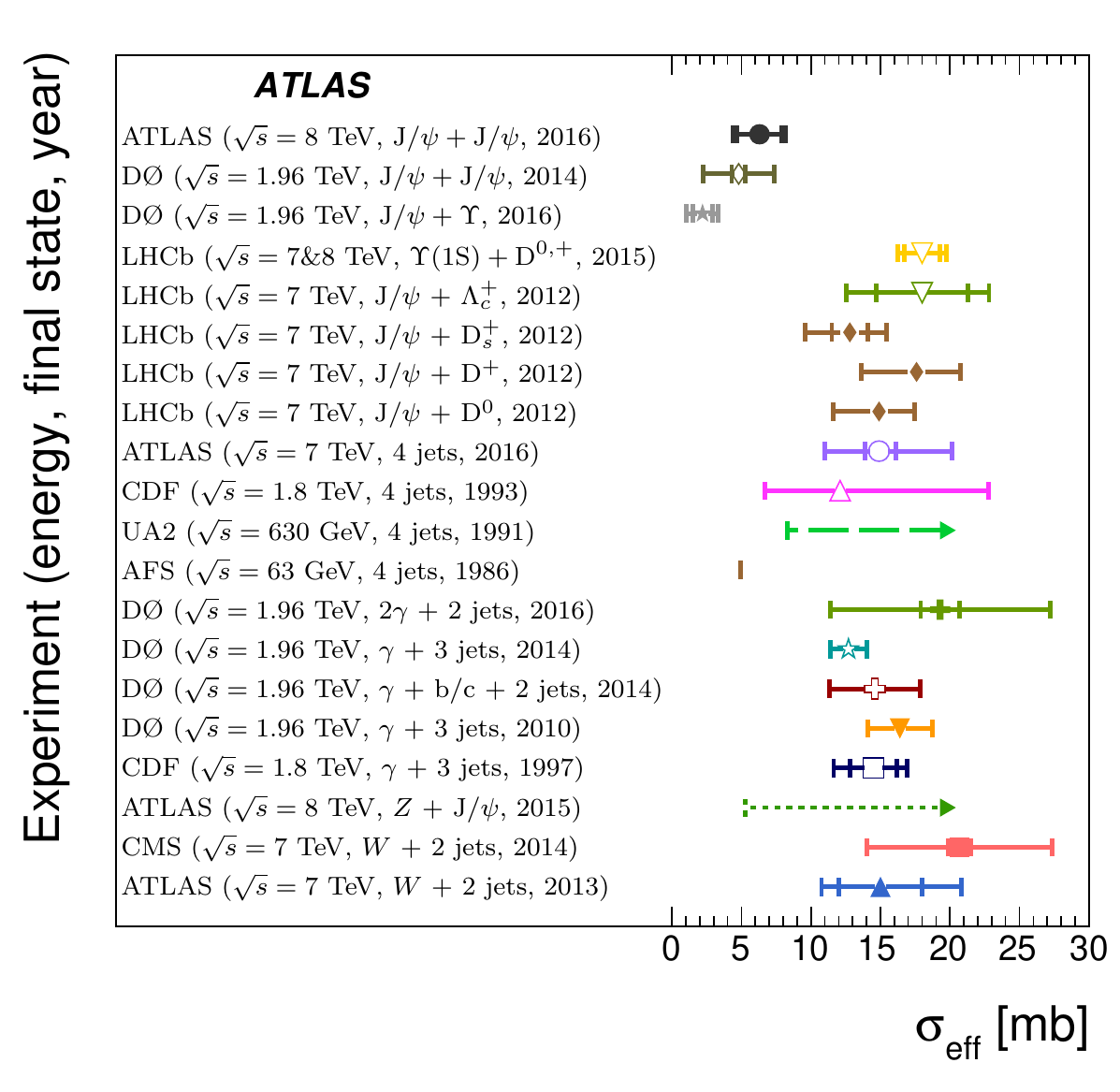}
\caption{The effective cross-section of DPS from different energies and final states measured by the AFS experiment~\cite{AFS}, the UA2 experiment~\cite{UA2}, the CDF experiment~\cite{CDF4j, CDFxF}, the D0 experiment~\cite{gamma, CDF2013, D0DiJ,D0JY,D0Dijet}, the CMS experiment~\cite{CMSW2J}, the LHCb experiment~\cite{LHCbJD, LHCbYD}, and the ATLAS experiment~\cite{WDPS, ATLASZJpsi,ATLAS4jEP}.  The inner error bars represent the statistical uncertainties and the outer error bars represent the sum in quadrature of the statistical and systematic uncertainties. Dashed arrows indicate lower limits and the vertical line represents the AFS measurement published without uncertainties.}
\label{fig:SigmaEff}
\end{center}
\end{figure}

\begin{figure}[ht!b]
\begin{center}
\includegraphics*[trim=0mm 0mm 4mm 5mm,clip,width=0.58\textwidth]{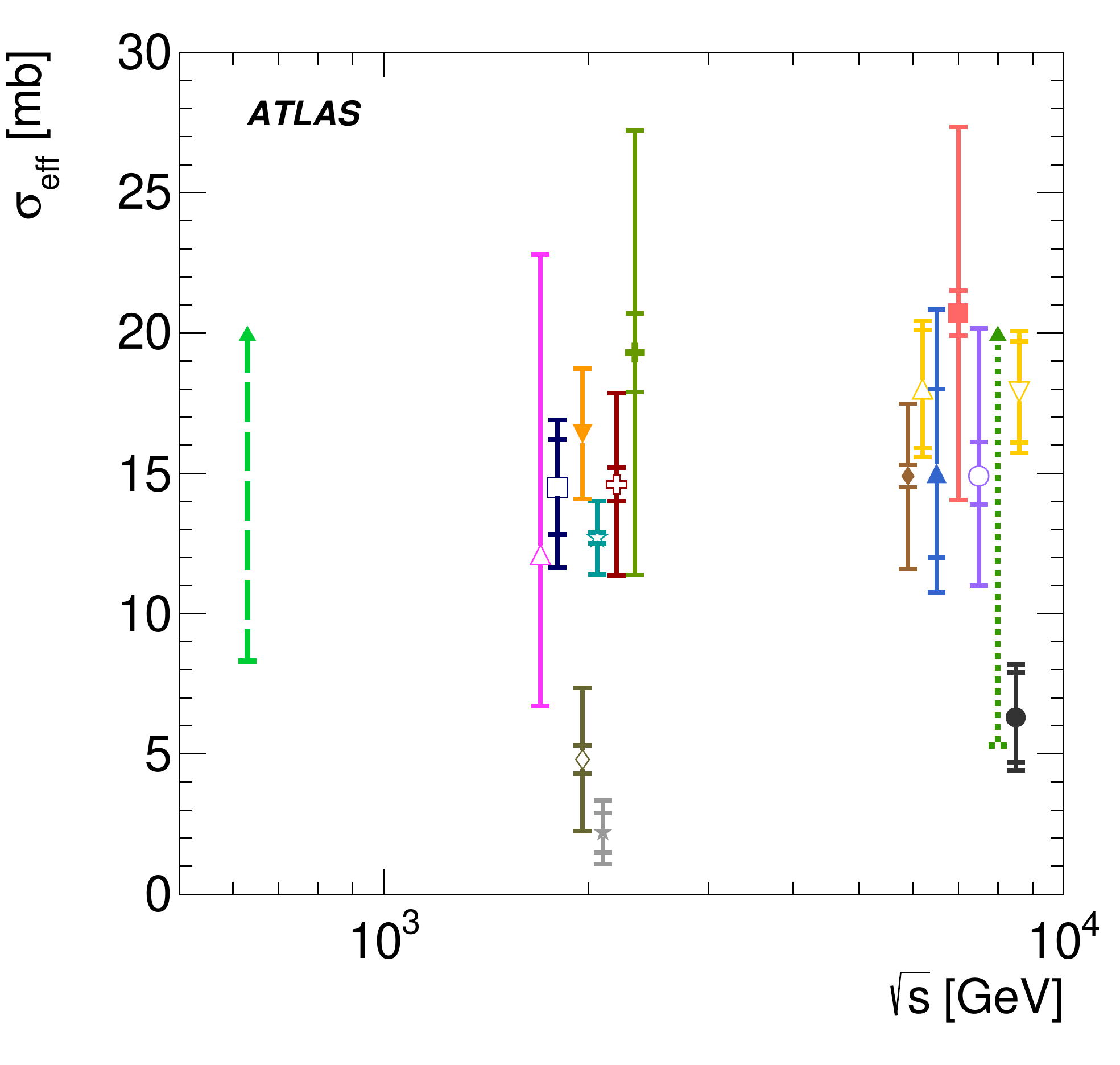}
\includegraphics*[trim=12mm -52mm 10mm 7mm,clip,width=0.4\textwidth]{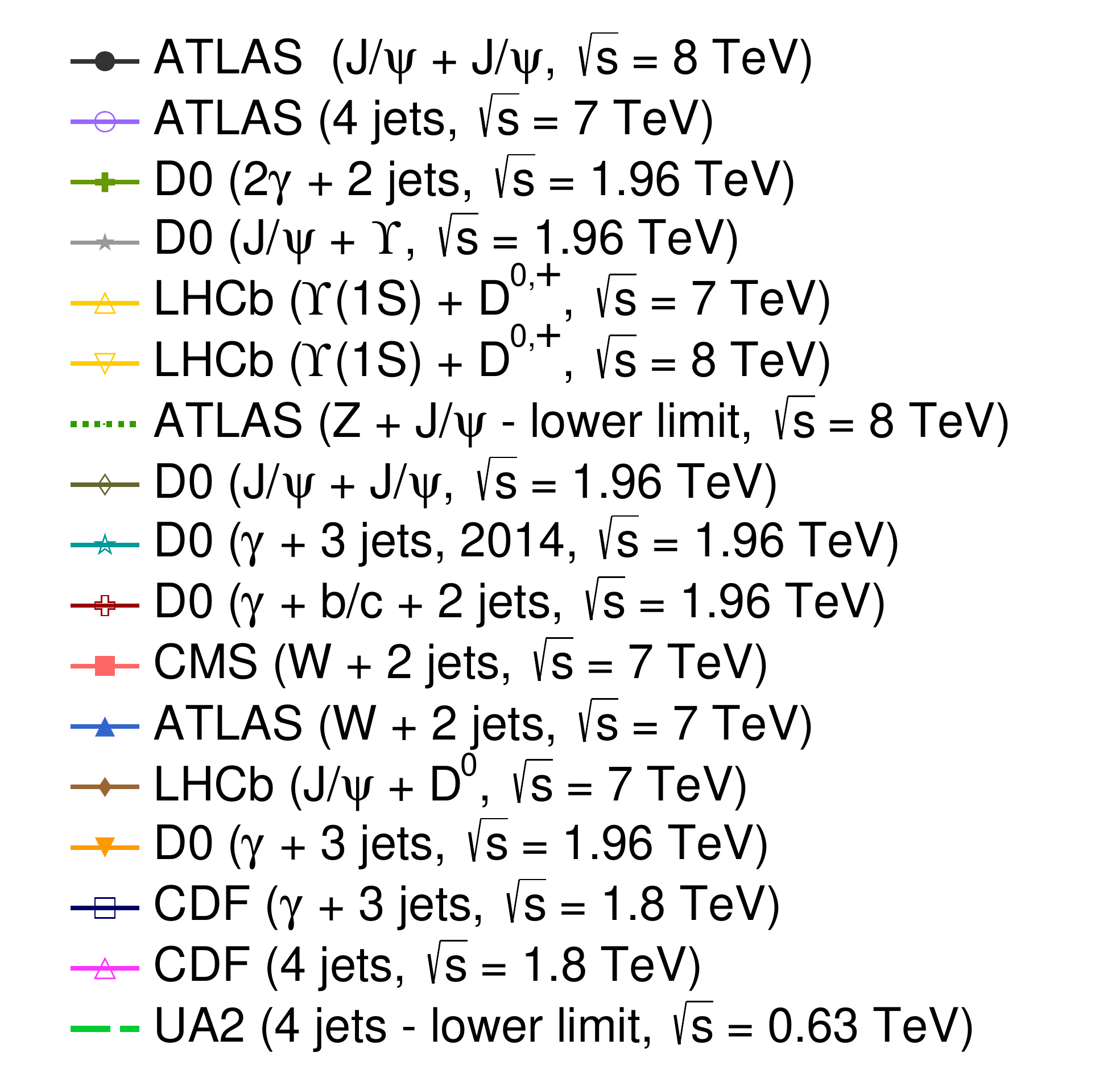}
\caption{The effective cross-section of DPS as a function of the centre-of-mass energy, $\sqrt{s}$, for the UA2 experiment~\cite{UA2}, the CDF experiment~\cite{CDF4j, CDFxF}, the D0 experiment~\cite{gamma, CDF2013, D0DiJ,D0JY,D0Dijet}, the CMS experiment~\cite{CMSW2J}, the LHCb experiment~\cite{LHCbJD, LHCbYD}, and the ATLAS experiment~\cite{WDPS, ATLASZJpsi,ATLAS4jEP}.  The inner error bars represent the statistical uncertainties and the outer error bars represent the sum in quadrature of the statistical and systematic uncertainties. Dashed arrows indicate lower limits.  For clarity, measurements at identical centre-of-mass energies are slightly offset in $\sqrt{s}$.}
\label{fig:SigmaEffSqrtS}
\end{center}
\end{figure}

The effective cross-section measured in this analysis is compared to measurements from other experiments and processes in Figure~\ref{fig:SigmaEff}.  In Figure~\ref{fig:SigmaEffSqrtS} the effective cross-sections are shown as a function of $\sqrt{s}$. In defining the effective cross-section, assumptions are made which lead to process and energy independence although there is no theoretical need for this independence.  More measurements of the effective cross-section at different energies will be helpful to test this assumption.   The ATLAS and D0~\cite{D0DiJ} analyses provide a hint that the effective cross-section measured from the prompt di-$J/\psi$ final state could be lower than that measured for the other final states.  It is interesting to note that the  di-$J/\psi$, $J/\psi$+$\Upsilon$~\cite{D0JY}, and 4-jet~\cite{UA2, CDF4j, ATLAS4jEP} processes are each dominated by gluon interactions and therefore should directly probe the gluon distribution in the proton~\cite{gg,HadroPolar,Jpsi2}.  However other analyses of gluon dominated processes~\cite{LHCbJD, LHCbYD} measured a larger effective cross-sections in these states.  Additional studies could help to learn more about DPS and the dependencies of the effective cross-section.  The pion cloud model~\cite{pion} predicts a smaller average transverse distance between gluons in the nucleon than between quarks. Such a difference could produce a lower effective cross-section for gluon-dominated processes.

 \FloatBarrier

\section{Summary}
\label{sec:conclusion}
In summary, using 11.4~fb$^{-1}$ of $\sqrt{s}$~=~8 TeV proton--proton collision data, the first study of prompt $J/\psi$ pairs from the ATLAS detector at the LHC is presented. The differential cross-section as a function of the sub-leading $J/\psi$ $p_{\mathrm{T}}$, di-$J/\psi$ $p_{\mathrm{T}}$, and di-$J/\psi$ mass are measured for two rapidity regions of the sub-leading $J/\psi$ meson: $|y$($J/\psi_{2})|$~$<$~1.05 and 1.05~$\leq$~$|y$($J/\psi_{2})|$~$<$~2.1.  Integrating over the $p_{\mathrm{T}}$ of the muons, the cross-section is 82.2~$\pm$~8.3~(stat)~$\pm$~6.3~(syst)~$\pm$~0.9~(BF)~$\pm$~1.6~(lumi)~pb in the central region and 78.3~$\pm$~9.2~(stat)~$\pm$~6.6~(syst)~$\pm$~0.9~(BF)~$\pm$~1.5~(lumi)~pb in the forward region.  This measurement assumes unpolarised $J/\psi$ mesons and does not include the $J/\psi$ spin-alignment systematic uncertainty.  In the muon fiducial volume, $p_{\mathrm{T}}(\mu)$~$>$~2.5~GeV, $|\eta(\mu)|$~$<$~2.3, and the triggered $J/\psi$ having both muons with $p_{\mathrm{T}}(\mu)$~$>$~4.0~GeV, the cross-section is 15.6~$\pm$~1.3~(stat)~$\pm$~1.2~(syst)~$\pm$~0.2~(BF)~$\pm$~0.3~(lumi)~pb for $|y$($J/\psi_{2})|$~$<$~1.05 and 13.5~$\pm$~1.3~(stat)~$\pm$~1.1~(syst)~$\pm$~0.2~(BF)~$\pm$~0.3~(lumi)~pb for 1.05~$\leq$~$|y$($J/\psi_{2})|$~$<$~2.1.  No assumptions are made about the $J/\psi$ polarisation in the muon fiducial volume.

Using a data-driven method, the fraction of double parton scattering processes in a single proton--proton collision is measured to be $f_{\mathrm{DPS}}$ = (9.2~$\pm$~2.1~(stat)~$\pm$~0.5~(syst))$\%$ in the muon fiducial volume.  The shapes of the measured double parton scattering distributions are consistent with model predictions.  For single parton scattering, the results are characterised by distributions wider than the next-to-leading-order predictions as seen in the absolute difference between the rapidities of the two $J/\psi$, the absolute difference between the azimuthal angles, the invariant mass of the di-$J/\psi$, and the di-$J/\psi$ transverse momentum.

A significant fraction of events appear to correspond to a topology in which the two colour singlet $J/\psi$ mesons are produced in the same direction and back-to-back with respect to an additional gluon. This topology is only included in next-to-leading-order calculations. A theoretical model based on leading-order DPS plus next-to-leading-order-colour singlet model SPS predictions without loops (NLO*) describes the data well, including in the kinematic regions where NLO contributions dominate.  Possible explanations for the difference between the data and theoretical predictions at large $\Delta y$ and invariant mass might be the need to include a large effect due to the inclusion of the intrinsic parton transverse momentum or a contribution via feed-down from a colour-singlet $\psi$(2S) meson which does not have the same kinematic properties as the NLO* SPS predictions.  The contribution from feed-down can amount to 40$\%$ of the single parton scattering cross-section.  Further studies of the pair production of $J/\psi$ mesons would give an opportunity to further constrain quarkonium production models and provide information on spin physics and heavy ion physics.

From these inputs, the effective cross-section for prompt $J/\psi$ meson pair production at $\sqrt{s}~=~8$~TeV is measured to be $\sigma_{\mathrm{eff}}$~=~6.3~$\pm$~1.6~(stat)~$\pm$~1.0~(syst)~$\pm$~0.1~(BF)~$\pm$~0.1~(lumi)~mb.  The data suggest that the effective cross-section measured from the prompt di-$J/\psi$ final state could be lower than that measured for the other final states.

\section*{Acknowledgements}

We thank CERN for the very successful operation of the LHC, as well as the
support staff from our institutions without whom ATLAS could not be
operated efficiently.

We acknowledge the support of ANPCyT, Argentina; YerPhI, Armenia; ARC, Australia; BMWFW and FWF, Austria; ANAS, Azerbaijan; SSTC, Belarus; CNPq and FAPESP, Brazil; NSERC, NRC and CFI, Canada; CERN; CONICYT, Chile; CAS, MOST and NSFC, China; COLCIENCIAS, Colombia; MSMT CR, MPO CR and VSC CR, Czech Republic; DNRF and DNSRC, Denmark; IN2P3-CNRS, CEA-DSM/IRFU, France; GNSF, Georgia; BMBF, HGF, and MPG, Germany; GSRT, Greece; RGC, Hong Kong SAR, China; ISF, I-CORE and Benoziyo Center, Israel; INFN, Italy; MEXT and JSPS, Japan; CNRST, Morocco; FOM and NWO, Netherlands; RCN, Norway; MNiSW and NCN, Poland; FCT, Portugal; MNE/IFA, Romania; MES of Russia and NRC KI, Russian Federation; JINR; MESTD, Serbia; MSSR, Slovakia; ARRS and MIZ\v{S}, Slovenia; DST/NRF, South Africa; MINECO, Spain; SRC and Wallenberg Foundation, Sweden; SERI, SNSF and Cantons of Bern and Geneva, Switzerland; MOST, Taiwan; TAEK, Turkey; STFC, United Kingdom; DOE and NSF, United States of America. In addition, individual groups and members have received support from BCKDF, the Canada Council, CANARIE, CRC, Compute Canada, FQRNT, and the Ontario Innovation Trust, Canada; EPLANET, ERC, ERDF, FP7, Horizon 2020 and Marie Sk{\l}odowska-Curie Actions, European Union; Investissements d'Avenir Labex and Idex, ANR, R{\'e}gion Auvergne and Fondation Partager le Savoir, France; DFG and AvH Foundation, Germany; Herakleitos, Thales and Aristeia programmes co-financed by EU-ESF and the Greek NSRF; BSF, GIF and Minerva, Israel; BRF, Norway; CERCA Programme Generalitat de Catalunya, Generalitat Valenciana, Spain; the Royal Society and Leverhulme Trust, United Kingdom.

The crucial computing support from all WLCG partners is acknowledged gratefully, in particular from CERN, the ATLAS Tier-1 facilities at TRIUMF (Canada), NDGF (Denmark, Norway, Sweden), CC-IN2P3 (France), KIT/GridKA (Germany), INFN-CNAF (Italy), NL-T1 (Netherlands), PIC (Spain), ASGC (Taiwan), RAL (UK) and BNL (USA), the Tier-2 facilities worldwide and large non-WLCG resource providers. Major contributors of computing resources are listed in Ref.~\cite{ATL-GEN-PUB-2016-002}.

\printbibliography

\newpage 
\begin{flushleft}
{\Large The ATLAS Collaboration}

\bigskip

M.~Aaboud$^\textrm{\scriptsize 137d}$,
G.~Aad$^\textrm{\scriptsize 88}$,
B.~Abbott$^\textrm{\scriptsize 115}$,
J.~Abdallah$^\textrm{\scriptsize 8}$,
O.~Abdinov$^\textrm{\scriptsize 12}$,
B.~Abeloos$^\textrm{\scriptsize 119}$,
R.~Aben$^\textrm{\scriptsize 109}$,
O.S.~AbouZeid$^\textrm{\scriptsize 139}$,
N.L.~Abraham$^\textrm{\scriptsize 151}$,
H.~Abramowicz$^\textrm{\scriptsize 155}$
% ...
\bigskip
\\
% ...
$^{8}$ Department of Physics, The University of Texas at Arlington, Arlington TX, United States of America\\
% ...
$^{12}$ Institute of Physics, Azerbaijan Academy of Sciences, Baku, Azerbaijan\\
% ...
$^{88}$ CPPM, Aix-Marseille Universit{\'e} and CNRS/IN2P3, Marseille, France\\
% ...
$^{109}$ Nikhef National Institute for Subatomic Physics and University of Amsterdam, Amsterdam, Netherlands\\
% ...
$^{115}$ Homer L. Dodge Department of Physics and Astronomy, University of Oklahoma, Norman OK, United States of America\\
% ...
$^{119}$ LAL, Univ. Paris-Sud, CNRS/IN2P3, Universit{\'e} Paris-Saclay, Orsay, France\\
% ...
$^{137}$ $^{(a)}$ Facult{\'e} des Sciences Ain Chock, R{\'e}seau Universitaire de Physique des Hautes Energies - Universit{\'e} Hassan II, Casablanca; $^{(b)}$ Centre National de l'Energie des Sciences Techniques Nucleaires, Rabat; $^{(c)}$ Facult{\'e} des Sciences Semlalia, Universit{\'e} Cadi Ayyad, LPHEA-Marrakech; $^{(d)}$ Facult{\'e} des Sciences, Universit{\'e} Mohamed Premier and LPTPM, Oujda; $^{(e)}$ Facult{\'e} des sciences, Universit{\'e} Mohammed V, Rabat, Morocco\\
% ...
$^{139}$ Santa Cruz Institute for Particle Physics, University of California Santa Cruz, Santa Cruz CA, United States of America\\
% ...
$^{151}$ Department of Physics and Astronomy, University of Sussex, Brighton, United Kingdom\\
% ...
$^{155}$ Raymond and Beverly Sackler School of Physics and Astronomy, Tel Aviv University, Tel Aviv, Israel\\
% ...
$^{a}$ Also at Department of Physics, King's College London, London, United Kingdom\\
$^{b}$ Also at Institute of Physics, Azerbaijan Academy of Sciences, Baku, Azerbaijan\\
$^{c}$ Also at Novosibirsk State University, Novosibirsk, Russia\\
$^{d}$ Also at TRIUMF, Vancouver BC, Canada\\
$^{e}$ Also at Department of Physics {\&} Astronomy, University of Louisville, Louisville, KY, United States of America\\
% ...
\end{flushleft}


\end{document}